\documentclass[%
reprint,
superscriptaddress,
longbibliography,
amsmath,amssymb,
aps,prx
]{revtex4-2}
\usepackage[utf8]{inputenc}
\usepackage[left=2cm,right=2cm,top=2cm,bottom=2cm]{geometry}
\usepackage{amsfonts, amssymb,amsmath,latexsym,graphics,amsthm, graphicx,epsfig,multirow,comment,feyn,slashed,xcolor,afterpage, makecell}
\usepackage{bbm}
\usepackage{bm}

\theoremstyle{definition}
\usepackage [autostyle, english = american]{csquotes}
\usepackage[labelfont=bf,font=small]{caption}
\usepackage{subfig}
\graphicspath{{figures/}}
\MakeOuterQuote{"}
\newcommand{\ket}[1]{|{#1}\rangle}
\newcommand{\bra}[1]{\langle {#1} |}
\renewcommand{\mod}{\mathrm{mod \,}}
\newcommand{\ui}{u^{-1}}
\newcommand\bigzero{{\text{\huge0}}}
\newcommand{\nocontentsline}[3]{}
\newcommand{\tocless}[2]{\bgroup\let\addcontentsline=\nocontentsline#1{#2}\egroup}

\def\equationautorefname~#1\null{Eq. (#1)\null}

\newcommand{\appref}[1]{\hyperref[#1]{App.~\ref*{#1}}}
\renewcommand{\vec}[1]{\bm{#1}}

\newcommand{\subsubsec}[1]{\tocless\subsubsection{#1}}

\newcommand{\gs}[1]{{#1}}
\renewcommand{\comment}[1]{}
\newcommand{\tr}{\mathrm{Tr}}
\newcommand{\iSWAP}{\mathrm{iSWAP}}
\newcommand{\SWAP}{\mathrm{SWAP}}
\usepackage{hyperref}
\begin{document}
\title{Crystalline Quantum Circuits}
\author{Grace M. Sommers}
\affiliation{Department of Physics, Princeton University, Princeton, NJ 08544, USA}   

\author{David A. Huse}
\affiliation{Department of Physics, Princeton University, Princeton, NJ 08544, USA}

\author{Michael J. Gullans}
\affiliation{Joint Center for Quantum Information and Computer Science, NIST/University of Maryland, College Park, Maryland 20742, USA}
\date{\today}

\begin{abstract}
Random quantum circuits continue to inspire a wide range of applications in quantum information science and many-body quantum physics, while remaining analytically tractable through probabilistic methods. Motivated by an interest in deterministic circuits with similar applications, we construct classes of \textit{nonrandom} unitary Clifford circuits by imposing translation invariance in both time and space. Further imposing dual-unitarity, our circuits effectively become crystalline spacetime lattices whose vertices are SWAP or iSWAP two-qubit gates and whose edges may contain one-qubit gates. One can then require invariance under (subgroups of) the crystal's point group. Working on the square and kagome lattices, we use the formalism of Clifford quantum cellular automata to describe operator spreading, entanglement generation, and recurrence times
of these circuits. A full classification on the square lattice reveals, of particular interest, a “nonfractal good scrambling class” with dense operator spreading that generates codes with linear contiguous code distance and high performance under erasure errors at the end of the circuit. We also break unitarity by adding spacetime-translation-invariant measurements and find a class of such circuits with fractal dynamics. 
\end{abstract}
\maketitle
\section{Introduction} 
Random quantum circuits are a model system of many-body quantum physics, in which the degrees of freedom are qubits or qudits and the evolution under a local Hamiltonian is modeled by local unitary gates. Random unitary circuits thus provide a platform for analytic computation of, for example, out-of-time-order correlators and entanglement growth
~\cite{Brown2015,Nahum17,Nahum2017,VonKeyserlingk2018}. They also have numerous applications to quantum complexity theory \cite{Brown2015,Boixo2018,Arute2019a,Bouland2019,Movassagh2018,Hangleiter22}, tomography \cite{Huang2020,Elben2022}, benchmarking \cite{Emerson2005,Liu2021}, and circuit complexity bounds \cite{Brandao2021,Haferkamp22}. {A particular motivation for this work comes from the field of quantum error correction, where random circuits have also played an important role~\cite{Brown2012,Brown13}. For example, random finite-rate stabilizer codes have linear code distance and reach channel capacity, and their performance under erasure errors can be modeled by random matrix theory~\cite{Gullans21}.} \comment{For example, in the presence of projective measurements at a rate $p$, there is an encoding/decoding transition at $p_c$ between a well-encoded volume-law{-entangled} phase at low $p$ 
and a {poorly encoding, area-law} phase at large $p$~\cite{Li2018,Skinner2018,Li2019,Gullans2020}.} Randomness has also proven useful for improving the error threshold and logical error rates of surface codes under biased noise, through random Clifford-gate deformations~\cite{Dua2022}.

While randomness is a valuable theoretical tool for studying quantum circuit dynamics, ultimately, there is a need for deterministic circuits with similar applications. 
For example, the behavior of practically relevant algorithms may not be well captured by random circuits. Indeed, in the case of the variational quantum eigensolver (VQE), initializing the solver with random circuits leads to  barren plateaus in the gradient \cite{McClean2018,Wang2021}.  Nonrandom circuits are likely to be more natural for many applications and avoid these barren plateaus. In the context of quantum simulation algorithms, one may question whether generic Hamiltonian evolution displays the same phenomena as random circuits.  The growth of quantum circuit complexity with evolution time is not understood outside random circuits  \cite{Bulchandani21}.  In addition, specific circuit families with more identifiable structure have been necessary to boost the performance of gate-set tomography in practical use cases \cite{Nielsen20}, and are likely to play a crucial role in the efficient  verification of quantum advantage on near-term devices \cite{Aaronson2022}.  Even addressing these questions from a conceptual point of view or providing a route towards future progress can be useful.  From a theoretical computer science perspective, this research avenue has echoes of ``derandomization''.  In classical complexity theory, this term refers to the process of turning probabilistic algorithms into deterministic ones as part of the quest to prove that the latter are just as powerful (i.e., $BPP=P$)~\cite{Aaronson2013}.   Similarly, in the theory of expander graphs and error-correcting codes, derandomization refers to the art of finding explicit constructions for objects only known to exist from  probabilistic arguments  \cite{Hoory06}.

Here we take a less formal, more physical view of the problem by analyzing a class of deterministic circuits with ``translational'' invariance in both time and space. These spacetime translation-invariant (STTI) circuits are endowed with two special features that enable an analytic treatment while still allowing for ergodic dynamics.
First, all the gates are dual-unitary, namely, unitary when viewed in the spatial direction as well as the usual time direction. As a nontrivial  model of quantum chaos with certain exactly solvable correlation functions, dual-unitary circuits are the subject of a rich, rapidly developing literature on which we build~\cite{Gopalakrishnan2019,Bertini2019,Piroli2019,Bertini2020,Bertini2021, Aravinda2021,Lerose2021,Borsi2022,Claeys2022a,Claeys2022,Kasim2022,Masanes2023}. Second, the gates in our circuits are Clifford. {Clifford circuits hold appeal because they can be classically simulated in polynomial time~\cite{Gottesman1998,Aaronson2004}, yet are physically relevant in the sense that the $n$-qudit uniform Clifford ensemble is a unitary 2-design for the $n$-qudit Haar ensemble~\cite{DiVincenzo2002} if the qudit dimension is a prime power~\cite{Zhu2017} (and in fact a 3-design if the qudit dimension is a power of 2~\cite{Zhu2017,Webb2016}). Analytically, Clifford circuits with spacetime randomness obey effective hydrodynamic equations~\cite{Nahum17,Nahum2017,Richter2022}, while spatially random Floquet Clifford circuits can exhibit strong localization in 1+1D~\cite{Chandran2015,Farshi2022d1,Farshi2022}.} In the present work, with spacetime translation invariance, our circuits can be interpreted as quantum cellular automata (QCA)~\cite{Farrelly2019,Arrighi2019}, and restricting to Clifford gates allows us complement the exact methods for treating dual-unitary circuits with the tools of symplectic cellular automata~\cite{Schlingemann2008,Gutschow2010solo,Gutschow2010long}.

Clifford quantum cellular automata (CQCA) \gs{on prime-dimensional qudits} with spatial period $a=1$ have received a thorough treatment in earlier work, but to our knowledge there is no systematic classification of automata with $a>1$ and beyond. Our primary focus in this work is on brickwork dual-unitary Clifford circuits, \gs{which naturally are expressed as qubit CQCA with $a=2$} and exhibit richer behavior than $a=1$. We highlight several physical properties of these circuits that can be gleaned from the symplectic automaton representation, including fractality in operator spreading and recurrence times. In addition to classifying and situating these circuits within the broader context of CQCA, we extend the concept of "self-dual-unitary" gates---gates such as the SWAP gate whose spacetime rotation is not only unitary, but in fact invariant~\cite{Bertini2019,Rather2022}---to all the point group symmetries of the lattice, associated with dual-unitarity, time reversal, and reflection. We further generalize to (self-) tri-unitary~\cite{Jonay2021} automata using the kagome lattice, for which $a=4$ and we can define 3 axes of time with unitary evolution. 

{On the quantum information side, we focus in this work on the applications to quantum error correction. We highlight a class of CQCA on the square lattice in which initially local operators scramble and spread densely within the lightcone, which can serve as encoding circuits for finite-rate codes with high performance under erasure errors and whose quasicyclic structure~\cite{Lally2001,Guneri2020} could provide a path toward efficient decoding under more general noise~\cite{Grassl1999,Feng1989,Semenov2012,Zeh2014,Mitchell2014}. More broadly, our results on these specific classes of quantum dynamics have potential applications in the same areas as random circuits, including benchmarking, quantum chaos, and complexity theory.}

{\subsection{Outline}}
The paper proceeds as follows. \autoref{sect:overview} provides a high-level overview of our results. As a case study in the most novel class of circuits discovered in our work,~\autoref{sect:circuit8} details the behavior of the "dense good scrambling class" on the square lattice~\autoref{sect:circuit8}. Taking a step back, in~\autoref{sect:symm}, we define the general models in detail and demonstrate how the symmetry transformations are enacted at the level of the one- and two-qubit gates. To gain greater insight into these symmetries, we introduce the CQCA formalism and show how the corresponding matrices transform under rotations and reflections of the lattice, in~\autoref{sect:cqca}.~\autoref{sect:classes} specializes to the square lattice, classifying the SWAP-core and iSWAP-core $a=2$ automata including the nonfractal good scrambling class. In~\autoref{sect:kagome}, we turn to the kagome lattice, where the circuits are described by $a=4$ CQCA. Returning to the square lattice, in~\autoref{sect:measurements} we describe the fractal structure that arises when we introduce projective measurements. Finally, we conclude in~\autoref{sect:conclude} with a discussion of future research avenues. 

\section{Overview}\label{sect:overview}
Before presenting our methods and results in detail, we begin with an overview of our findings. 
The two common features of the STTI circuits considered in this work---dual-unitarity and Cliffordness---provide complementary avenues for study. 

\subsection{Symmetries, dual-unitarity, and tri-unitarity}
The circuits we consider are all crystalline lattices, in which vertices correspond to gates and edges correspond to qubits, possibly dressed with single-qubit gates. Focusing our attention on two-qubit gates, we choose lattices with coordination number $z=4$. In addition to spacetime translation invariance, the bare lattices are invariant under the rotations and reflections that comprise their point group~\cite{Ashcroft1976}. In the circuit perspective, however, vertices are no longer pointlike objects, and edges have a directionality imposed by the single-qubit gates. We can therefore ask which of the symmetries of the \textit{lattice} are also symmetries of the \textit{circuit}.

\begin{figure}[t]
    \centering
    \includegraphics[width=\linewidth]{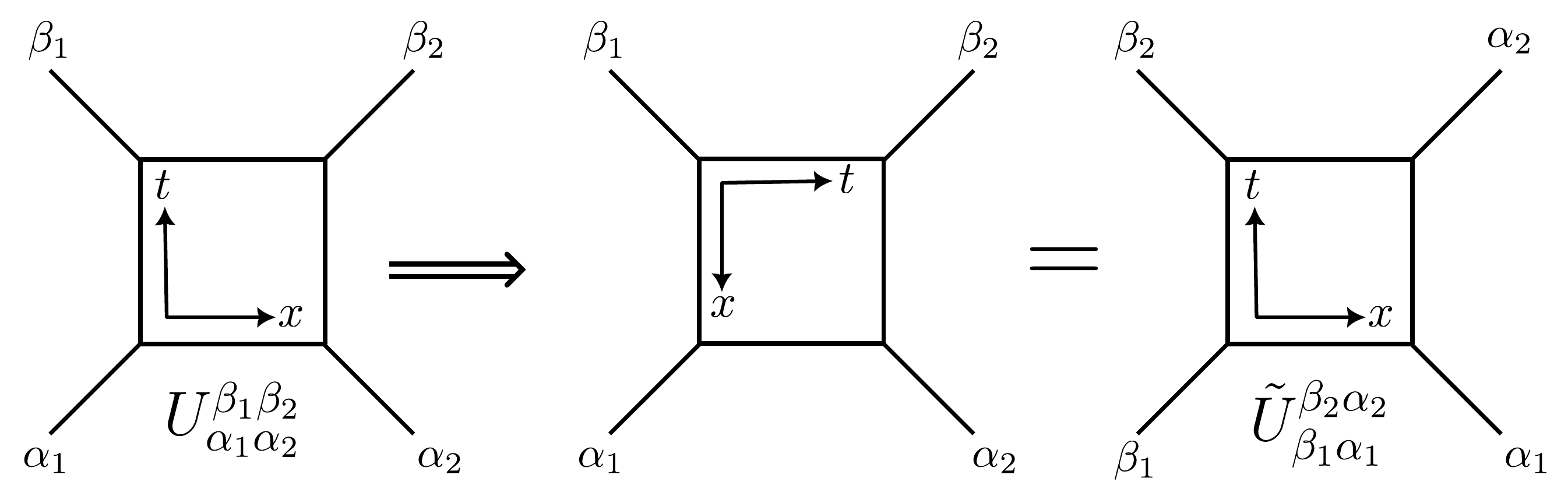}
    \caption{The convention for the dual gate used in this paper~\cite{Bertini2021,Borsi2022}: given a unitary gate $U^{\beta_1 \beta_2}_{\alpha_1 \alpha_2}$ (left), we rotate the spacetime axes by $\pi/2$ (center) to obtain the dual $\tilde{U}^{\beta_2 \alpha_2}_{\beta_1 \alpha_1}$ (right). If $U$ is dual-unitary, then $\tilde{U}$ is unitary (as is the spacetime rotation in the opposite direction).}
    \label{fig:dual}
\end{figure}

One main thrust of this work is organizing and classifying these symmetries for two such lattices, square and kagome. Implicit in this analysis is that the transformed gates are unitary. For two-qubit gates, this imposes dual-unitarity: rotating the gate by $\pi/2$ in spacetime yields another unitary gate (\autoref{fig:dual}). 
From the parameterization of dual-unitary gates in Ref.~\cite{Bertini2019}, restricting to the Clifford group, the dual-unitary operator $U$ implemented by the gate can be written as either a SWAP core (non-entangling) or iSWAP core (maximally entangling), with single-qubit Clifford gates on each of the four legs. 

\begin{figure}[t]
    \includegraphics[width=\linewidth]{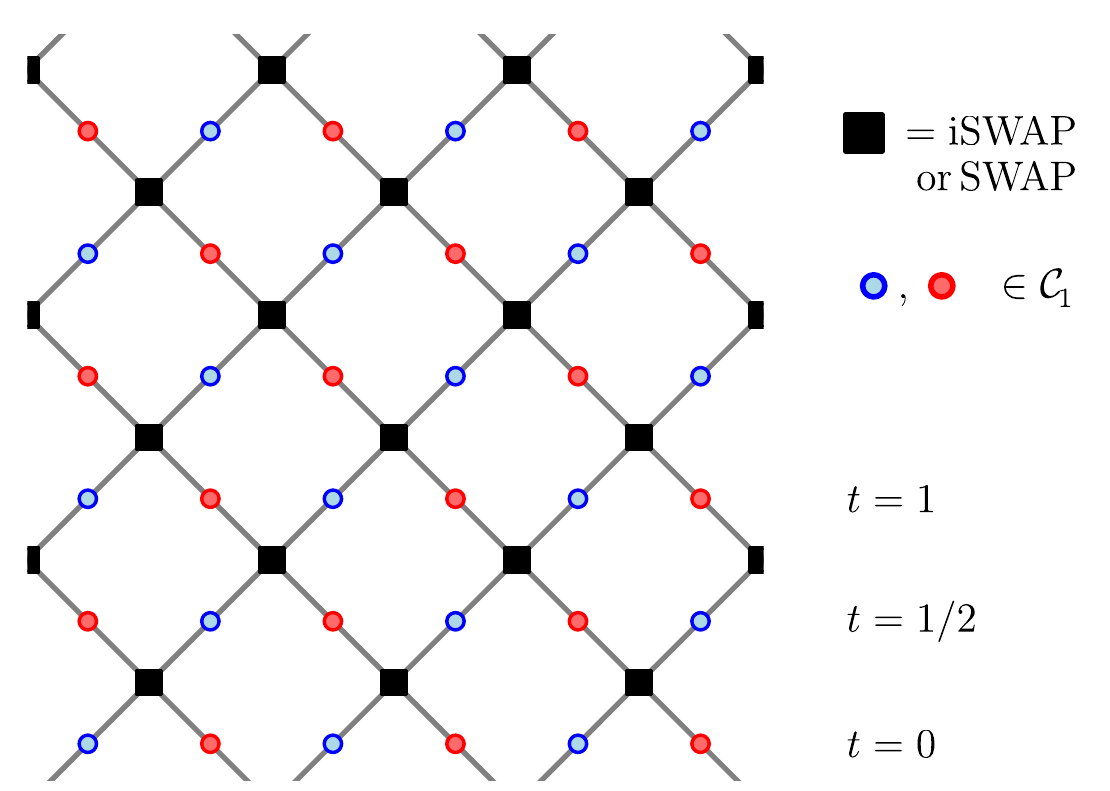}
    \caption{STTI dual-unitary brickwork circuit represented as a rotated square lattice. Black squares are (i)SWAP cores. Edges are decorated with single-qubit Clifford gates, represented as red and blue circles. One time step is defined as two layers of the brickwork circuit.}
    \label{fig:lattice}
\end{figure}

Our main model is the brickwork circuit shown in~\autoref{fig:lattice}, a square lattice of SWAP or iSWAP cores, with single-qubit gates on each edge. The bare SWAP and iSWAP cores are "self-octa-unitary" since they are invariant under all eight point group transformations of the square. With the inclusion of single-qubit gates, the resulting STTI circuit can have some, all, or none of these symmetries. {This is the focus of~\autoref{sect:symm}.}

On the kagome lattice, whose point group is $D_6$ instead of $D_4$, we can define three axes (six arrows) of time, making these circuits (self)-tri-unitary. {In Ref.~\cite{Jonay2021}, where tri-unitarity is first introduced, tri-unitary \textit{gates} are defined on three qubits and tiled on a triangular lattice. However, as the authors note, the family of tri-unitary gates considered in that paper can be decomposed into three two-qubit gates, and the resulting circuit can then be expressed on the kagome lattice.} The three axes of time restrict the two-point correlations between traceless one-site operators averaged over all states to vanish except at $x_1-x_2=0$ and at $|x_1-x_2|=v|t_1-t_2|$ where $v$ is the velocity of the lightcone.

\subsection{Classification of CQCA}
Because our circuits are both STTI and Clifford, we can represent them as Clifford quantum cellular automata (CQCA), which is the primary analytic technique used in this work. For a more detailed introduction to the CQCA formalism, the reader is referred to~\autoref{sect:cqca} \gs{and to Refs.~\cite{Schlingemann2008,Gutschow2010long,Gutschow2010solo}.}

The circuit in~\autoref{fig:lattice} is translation-invariant with a unit cell of $T=1/2$, $a=2$, composed with a shift by 1 site, \gs{so it can be treated as an "$a=2$ automaton."}
In~\autoref{sect:classes}, we classify all iSWAP-core automata on the square lattice into six classes, where members of each class are related by a reflection about the center of the gate, and/or a change of basis. The point group transformations exchange members of the same class. {A similar classification scheme can be applied on the kagome lattice, where $a=4$, but in~\autoref{sect:kagome} we focus our attention on those with a high amount of symmetry, the "self-tri-unitary" circuits.} 

\gs{Since the Clifford group normalizes the Pauli group, the dynamics under a Clifford circuit with spatial period $a$ is fully encoded (modulo phases) by the image of $X_i$ and $Z_i$ on each site $i=1,...,a$ of the unit cell. Leveraging this translation invariance, a CQCA with a unit cell containing $a$ qudits is described by a $2a\times 2a$ matrix $M$, whose entries are Laurent polynomials in the variable $u$ which labels the unit cell~\cite{Berenstein2021}.}

\gs{We adapt and extend to $a>1$ the techniques presented in foundational works~\cite{Schlingemann2008,Gutschow2010solo,Gutschow2010long}, which focus on prime $q$, $a=1$ automata}~\footnote{\label{foot:qudits}\gs{An STTI Clifford circuit with spatial period $a$ can be recast as a circuit with period $1$ acting on qudits with dimension $q^a$~\cite{Ketkar2005,Gottesman2014,Zeng2020}. However, such a circuit will not in general admit a representation as an $a=1$ symplectic cellular automaton. This is because as defined in Ref.~\cite{Gottesman2014}, the "true" $\mathbb{F}_{q^a}$ Clifford group on $n$ $q^a$-dimensional qudits---comprised of gates which admit a symplectic representation over the finite field $\mathbb{F}_{q^a}$---is only a subset of the full $n$-qudit Clifford group. Said differently, a general quantum stabilizer code on $n$ qudits of dimension $q^a$ corresponds to a length-$2na$ linear classical code over $\mathbb{F}_q$, but as a length-$2n$ classical code over $\mathbb{F}_{q^a}$, it is only an \textit{additive} code, and thus not amenable to the matrix techniques of this paper~\cite{Zeng2020}.}}. $a=1$ CQCA have determinant $u^{2d}$ where $d\in \mathbb{Z}$. Factoring out a shift of $u^d \mathbbm{1}$ makes a centered symplectic cellular automaton (CSCA) with determinant 1, whose characteristic polynomial is uniquely determined by $\tr(M)$~\cite{Gutschow2010solo,Gutschow2010long}:{
\begin{equation}\label{eq:char-poly-a1}
\chi_M(y) = y^2 + \tr(M) y + 1.
\end{equation}}
\gs{While this simple relationship between $\tr(M)$ and $\chi_M(y)$ no longer holds for $a>1$, the characteristic polynomial remains inextricably linked to three related properties of the automaton: entanglement generation, operator spreading, and the recurrence time in a finite system.} 

The recurrence time of the unitary, up to a phase, on a system of $L$ qubits, or $m=L/a$ unit cells (with periodic boundary conditions) is denoted $\tau(m)$, the minimum power such that $M^\tau = \mathbbm{1} \, \mod \, (u^m-1)$ up to global shifts. Under the evolution of the automaton, any stabilizer group, mixed or pure, repeats modulo signs and shifts after an interval that divides $\tau(m)$. The scaling of $\tau(m)$ divides the six square lattice classes into two groups: three for which $\tau(m)\leq 3m$ for all $m$, and three for which $\tau(m)$ is linear in $m$ for $m=2^k$, but grows much faster for generic $m$. We also demonstrate a sharp distinction between these two groups with respect to the entanglement generation for a random initial product state. The first group consists of "poor scramblers," for which the resulting Page curve~\cite{Page1993} has a slope less than 1, i.e. the total entropy of a subsystem of length $|A|<L/2$ is $f|A|$, where $0<f<1$. This submaximal entanglement generation can be attributed, at least in part, to the presence of conserved $Z$ charges, or "gliders." In particular, we find a close connection between the "bare iSWAP class" (all single-qubit gates are the identity) and the standard glider automaton with $a=1$~\cite{Gutschow2010long}.

\subsection{Fractality, dense operator spreading, and quantum error correction}
The second group of iSWAP-core automata on the square lattice is comprised of "good scramblers", which, {when acting on random initial product states, generate Page curves of slope 1 at times away from} the recurrences. The three classes within this group exhibit different fractal behavior. {The fractal in question is the footprint of an initially local Pauli operator which spreads within the lightcone. We define the fractal dimension through the scaling of the cumulative number of non-identity sites within this footprint vs. the depth of the circuit, so that $d_f\leq 2$ for CQCA defined in 1+1D. \gs{In the limit of infinite time, the fractal structure of the footprint depends only on the minimal polynomial $\mu_M$ of the automaton $M$~\cite{Gutschow2010fractal}. The minimal polynomial is the lowest-degree monic polynomial $\mu_M$ for which $\mu_M(M)=0$, thus encoding a recursion relation for $M$.}

We refer to one class as the self-dual kicked Ising (SDKI) class, a representative of which maps to the SDKI model via a ``boundary'' circuit~\cite{Bertini2019}. Without invoking this direct mapping at the level of gates, the connection to SDKI is clear from the automata, which both have the minimal polynomial $\mu(y)=y^2 + (\ui + 1 + u)y + 1$. Initially local operators {spread in this class of circuits} with a fractal dimension $d_f = \log_2[(3+\sqrt{17})/2]= 1.8325...$~\cite{Gutschow2010fractal}. A second good scrambling class has fractal dimension $d_f \cong 1.9$, a pattern not seen in $a=1$ automata~\footnote{We have not analytically derived an exact expression for $d_f$ in this case, but leave it as a challenge for the reader to adapt the methods of Ref.~\cite{Gutschow2010fractal}.}.

Special attention is paid to the third "good scrambling" class, {the subject of a case study in~\autoref{sect:circuit8}. We describe its operator spreading as "nonfractal" or "dense", because the number of $X$, $Y$, and $Z$ sites within a spreading operator are all a finite fraction of the lightcone volume ($d_f=2$).} On one hand, as with all of these dual-unitary CQCA, this nonfractal class has large amounts of structure not seen in random Clifford circuits. In fact, a representative of this class, which has $\pi/2$ $X$ rotations on each leg, is self-octa-unitary.
On the other hand, it shares important features with random circuits, including dense operator spreading. It also has promise for error correction. Namely, when a random initial product state with nonzero entropy density is fed into this circuit, the logical operators spread linearly in time, so that at late times the contiguous length of the shortest logical operator---the contiguous code distance~\cite{Bravyi2009}---is linear in $m$. {Since operators also spread densely, we expect their weight to scale proportionally to their length, which then implies a linear code distance. Indeed, quasicyclic codes generated from initial periodic product states perform well under erasure errors applied at the end of the circuit. Under more general noise, the crystalline symmetries of the encoding circuit could be beneficial for finding efficient optimal decoders.} Note that we have not addressed the overhead needed to make these codes or the circuits fault-tolerant, which we leave as a problem for future work.

\subsection{Adding measurements}
Finally, in~\autoref{sect:measurements} we break unitarity by adding measurements in a STTI fashion. With one measurement per doubled spacetime unit cell of the square lattice, in most cases an initial fully mixed state reaches a steady state (mixed or pure) after $O(1)$ time steps, but for the $d_f\cong 1.9$ good scrambling class in the appropriate measurement basis, a fully mixed initial state purifies in $m$ time steps for $m=2^k$. {During the initial transient, the state acquires volume-law entanglement, but loses it before reaching the steady state, which has zero entanglement.} A perturbation to this product steady state spreads as a Sierpinski gasket, a pattern not seen on the square-lattice dual-unitary circuits without measurements. {We present this as just one example of the rich menagerie of hybrid STTI circuits, deferring an extended discussion of the hierarchical classification of such circuits, including those whose steady state is a high-performing finite-rate code, to a future paper~\cite{Sommers2023a}.}

\section{Case study of the dense good scrambling class}\label{sect:circuit8}
As motivation for the broader classification program undertaken in the rest of this paper, consider a realization of~\autoref{fig:lattice} in which all of the two-qubit gates (black squares) are the iSWAP gate:
\begin{equation}\label{eq:U-iswap}
    \iSWAP = e^{-i\frac{\pi}{4}\left(XX + YY\right)} = \begin{pmatrix} 1 & 0 & 0 & 0 \\
    0 & 0 & -i & 0 \\
    0 & -i & 0 & 0 \\
    0 & 0 & 0 & 1 \end{pmatrix}
\end{equation}
and all of the single-qubit gates (red and blue circles) are rotations by $\pi/2$ about the $X$ axis on the Bloch sphere:
\begin{equation}\label{eq:circuit8-gate}
    R_X[\pi/2] = e^{-i \frac{\pi}{4} X}.
\end{equation}

This circuit is a Clifford quantum cellular automaton (CQCA) with unit cell $a=2$ composed solely of dual-unitary gates, thus lending it a high degree of structure. In fact, in addition to being spacetime translation-invariant (STTI), the class to which this $(R_X[\pi/2], R_X[\pi/2])$ circuit belongs is the only one, besides the "bare iSWAP class" (in which all the single-qubit gates are the identity), that contains circuits left invariant under the 8 rotations and reflections of the unit cell of the square lattice. We call this property "strong self-octa-unitarity" and define it formally in~\autoref{sect:symm}.

On the other hand, the dynamics under this circuit is in many ways reminiscent of random Clifford circuits, with local operators spreading densely rather than as fractals, and with initial product states evolving to volume-law-entangled states whose Page curve has slope 1. In this section, we explore this dichotomy between structure and scrambling and discuss the application of these circuits to developing codes with linear distance. {We will revisit these concepts in more general settings throughout the paper.} 

\subsection{Recurrence times}
An immediate difference from random circuits is the presence of recurrences: since the dynamics are Floquet, Clifford, and unitary, any initial state on a finite system must eventually repeat under the action of the circuit. To wit, there are $\prod_{k=0}^{L-1} (2^{L-k}+1) = O(2^{cL^2})$ unique stabilizer groups (modulo signs) on $L$ qubits~\cite{Aaronson2004}, which places an upper bound on the recurrence time. 

In fact, for all $m$, where $m=L/a$ is the number of unit cells with periodic boundary conditions, the recurrence time $\tau(m)$ is well below this bound. Of special note are system sizes $m=2^k$, for which $\tau(m)$ grows linearly. This linear trend in $\tau(m)$ for STTI circuits has been proven for $m=2^k$ in $a=1$ CQCA over qubits~\cite{VonKeyserlingk2018,Stephen2019} as well as for $m=q^k$ in a class of dual-unitary circuits known as perfect permutation maps, where the odd prime $q$ is the dimension of the qudits~\cite{Borsi2022}. 

What distinguishes this circuit and the other good scrambling classes from the "poor scrambling" classes discussed in~\autoref{sect:poor} is the trend in $m\neq 2^k$. As the example of Ref.~\cite{Borsi2022} indicates, the sensitivity in our good scrambling circuits to the power of 2 is related to the onsite Hilbert space dimension $q=2$. As shown in~\autoref{fig:recurrence-times}, $\tau(m)$ is strongly nonmonotonic in $m$. {A curious trend, left for the interested reader to ponder, is that if we write $m = j 2^k$, then $\tau(m)/2^k$ is either $2^p+2$ or $2^p-2$ for some $p$, where $p$ is a function of $j$ alone. If this trend holds for all $m$, then $\tau(29)\geq 2^{24}-2$ (indicated as the lower bound on an error bar in~\autoref{fig:recurrence-times}).} Speculatively, the upper envelope of $\tau(m)$ grows exponentially in $m$ but no faster than $O(2^m)$ (gray line), which is still exponentially smaller than the generic upper bound {of $O(2^{cm^2})$}.

\begin{figure}[t]
    \centering
    \includegraphics[width=0.9\linewidth]{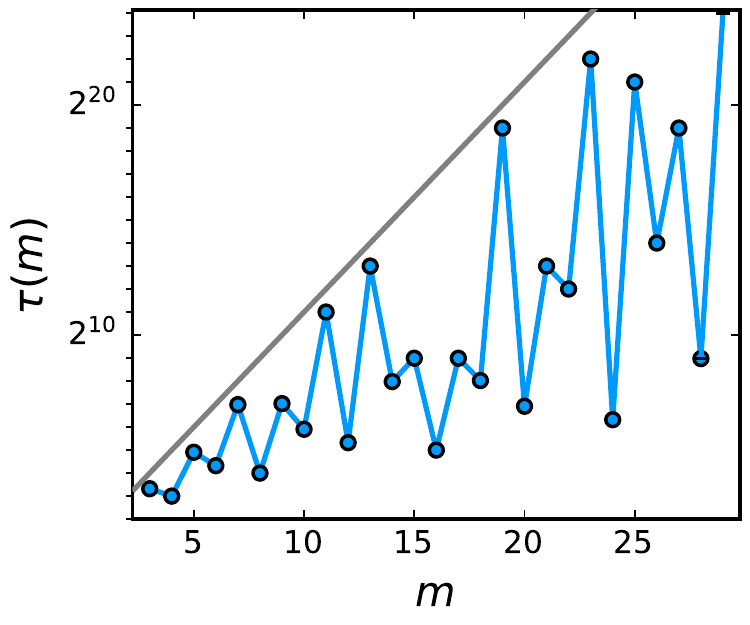}
    \caption{Recurrence time $\tau(m)$ of the unitary, modulo signs and shifts, for a brickwork circuit of iSWAP cores and $\pi/2$ $X$ rotations (\autoref{eq:U-iswap} and~\autoref{eq:circuit8-gate}), acting on $m=L/2$ unit cells with periodic boundary conditions. Gray line is $\tau = 2^{m+1}$, which appears to be an upper bound on $\tau(m)$.}
    \label{fig:recurrence-times}
\end{figure}
\subsection{Entanglement generation for pure product states}
The second defining feature of this class, along with the other good scrambling classes, is in the generation of entanglement for initial pure product states. In this aspect it behaves like a random circuit: starting from a random product state, the subsystem entropy averaged over all contiguous regions of the same length increases linearly in time before saturating at a near-Page curve with slope 1~(\autoref{fig:page})~\cite{Nahum17}. However, 
the initial product state does eventually recur. Since $\tau(m)$ is linear in $m$ for $m=2^k$, on those system sizes, the system spends a finite fraction of its evolution in a state of suppressed entanglement. For the time evolution on $m=64$ unit cells shown in~\autoref{fig:page}, the initial product state recurs (modulo signs) with a period of $\tau(m)=128$, but the state returns to area-law entanglement twice per period.  For generic large $m$, the recurrence time generally satisfies $\tau(m)\gg m$, so the state spends most of its time near-maximally entangled.

\begin{figure}[t]
    \centering
    \includegraphics[width=\linewidth]{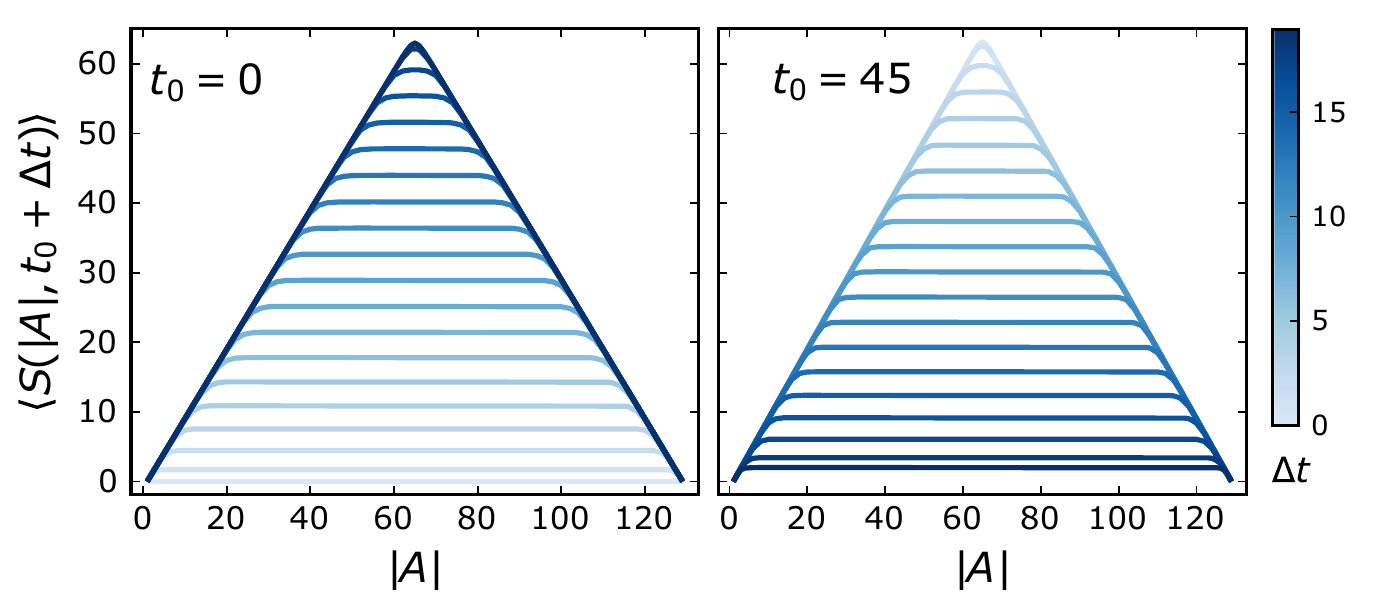}
    \caption{Entanglement generation on a random pure product state on $L=128$ qubits, or $m=64$ unit cells, for a brickwork circuit of iSWAP cores and $\pi/2$ $X$ rotations (\autoref{eq:U-iswap} and~\autoref{eq:circuit8-gate}). For $t<20$, the subsystem entropy increases at a near-maximal rate before reaching a Page curve with slope 1 (left). The state remains near-maximally entangled until $t\cong 45$, before the subsystem entropy starts to decrease until reaching an area-law state at $t=64$ (right). In both panels, the entropy $\langle S(|A|,t_0+\Delta t) \rangle$ is averaged over all contiguous regions of length $|A|$ with periodic boundary conditions, with darker (lighter) curves corresponding to later (earlier) times $\Delta t$ with respect to $t_0$.}
    \label{fig:page}
\end{figure}

\subsection{Operator content}
The two above properties---superlinear recurrence times for generic $m$ and generation of slope-1 Page curves starting from a pure product state---are also seen in two other classes of good scrambling automata, discussed in~\autoref{sect:classes}. What makes this class unique among all those studied in this work is that, whereas Pauli strings spread as fractals in the other classes, in this class all initial local operators spread densely, i.e. with fractal dimension 2 (\autoref{fig:paulis-circuit8}). {Dense operator spreading sets this class of circuits outside the range of possible behavior of $a=1$ CQCA~\cite{Gutschow2010long}, where fractal operator spreading (as diagnosed by the out-of-time-order commutator) has been interpreted as evidence of quantum scarring, i.e. weak ergodicity breaking \cite{Kent2023}. Thus, the absence of fractals in this class suggests a stronger form of ergodicity than that found in other CQCA.}

As quantitative evidence for $d_f=2$, the cumulative number of appearances of the Pauli $\sigma$ within the light cone for times $t'=0,...,t$ is shown for the initial string $Z_1$ in~\autoref{fig:paulis-cumul}. {The cumulative count of each Pauli scales as $t^2$, albeit with a larger prefactor for the pair $I$ and $Z$, compared to the pair $X$ and $Y$. This asymmetry in the frequency of the two pairs of Paulis, which depends on the initial string, is one indication that in spite of the dense spreading, the substructure of the operator content in the bulk is still distinguishable from that of a random circuit. It is also in contrast to the "Pauli mixing" behavior---proximity to a uniform distribution on the Paulis---of operator spreading in \textit{random Floquet} Clifford circuits, proven to hold within the lightcone for large-dimensional qudits, and also observed in the interior of localized operators in qubit circuits~\cite{Farshi2022d1}.} 
\begin{figure}[t]
    \subfloat[]{
    \includegraphics[width=\linewidth]{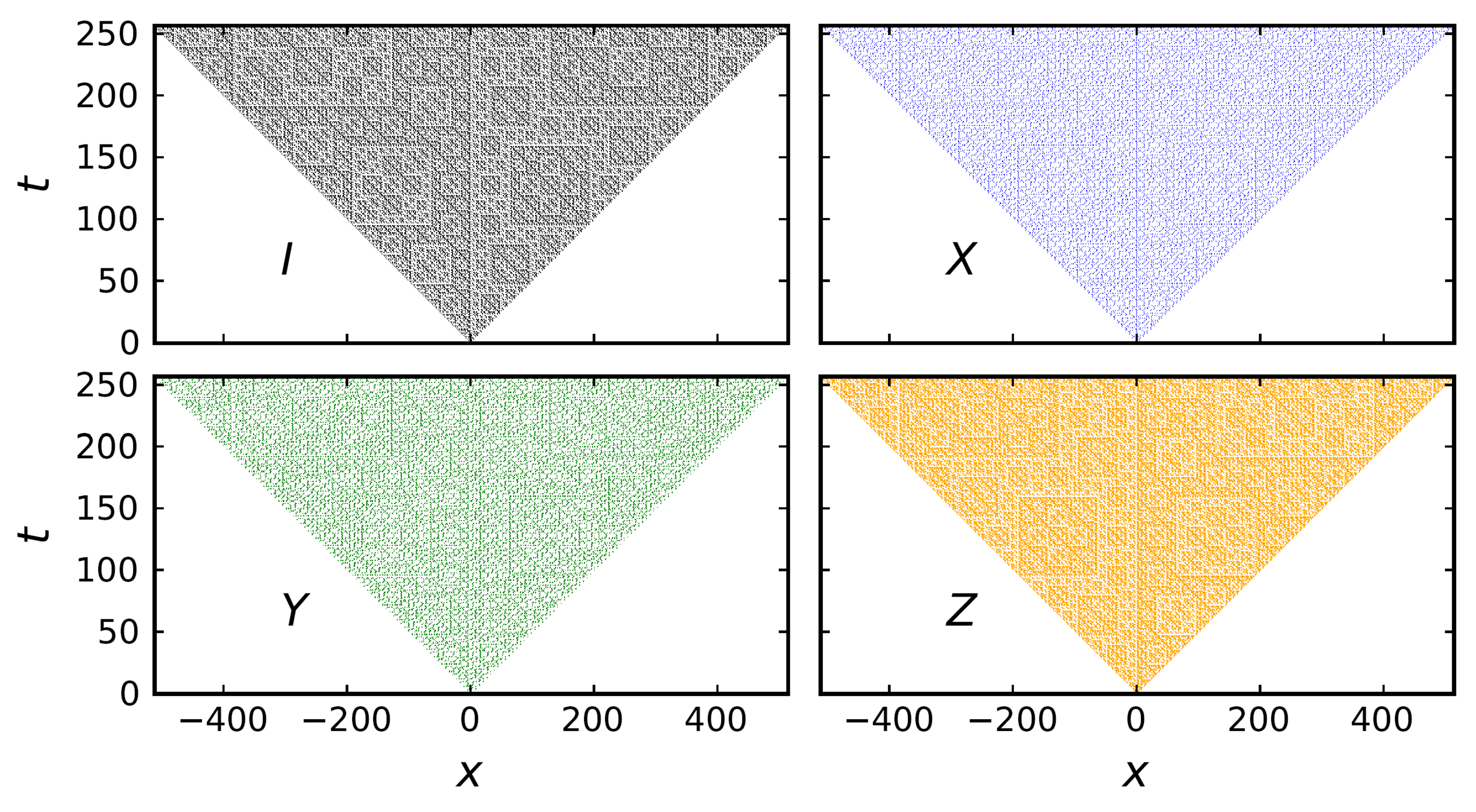}\label{fig:paulis-circuit8}
    } \\
    \subfloat[]{
    \includegraphics[width=\linewidth]{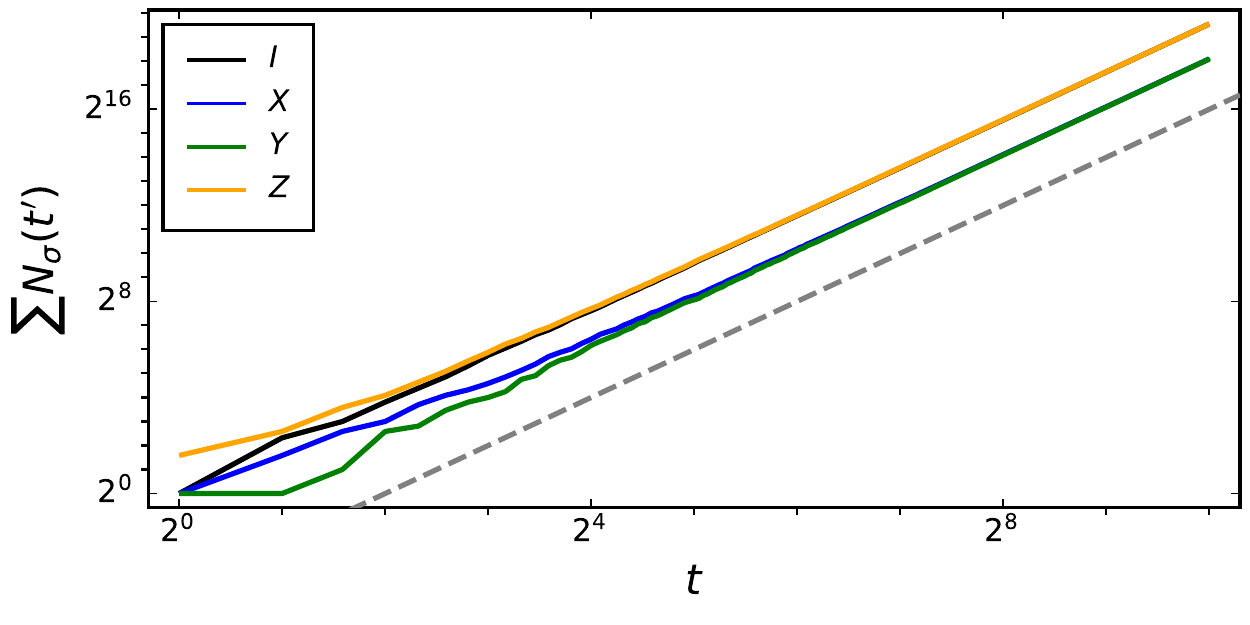}\label{fig:paulis-cumul}
    }
    \caption{(a) Image of the initial string $Z_1$ up to time $t=256$ under the action of the circuit defined by~\autoref{eq:U-iswap} and~\autoref{eq:circuit8-gate}, separated into identity (black), $X$ (blue), $Y$ (green), and $Z$ (orange) sites within the lightcone. (b) Cumulative number of appearances, $\sum_{t'=0}^t N_\sigma(t')$, of $\sigma=I,X,Y,Z$ within the lightcone $|x|\leq t$ for $Z_1(t)$ up to $t=1024$. Gray dashed line shows $\sum N(t') \propto t^2$.}
\end{figure}

\subsection{Code length and code distance}\label{sect:code}
Viewing the CQCA as an encoding circuit for a stabilizer code, the nonfractal spreading of Pauli strings gives this class strong potential for quantum error correction. 

{One figure of merit in describing quantum codes is the code distance $d$, the number of non-identity Paulis in the support of the lowest-weight logical operator~\cite{nielsen2010}. This property relates to the operator spreading in the encoding circuit in the following way. Consider a stabilizer code generated by running the circuit for $O(m)$ layers on an initial mixed product state with a finite entropy density $s$. The stabilizer group $\mathcal{S}$ is generated by $(1-s)L$ stabilizer generators, which can initially be chosen to live on single sites, while $s L$ logical pairs live on the unstabilized sites. Under the action of a dense good scrambling circuit, both the stabilizer generators and the logicals spread nonfractally within the lightcone, eventually saturating at $O(m)$ weight.}

{Because the code distance is the minimum weight across all logical representatives---elements of the normalizer of $\mathcal{S}$ that act nontrivially on the codespace, which can potentially lower their weight through multiplication by elements of $\mathcal{S}$---the growth of a \textit{single} operator in isolation only provides an upper bound on the code distance. Since minimizing the weight overall all logical representatives has exponential complexity, we use the \textit{contiguous} code distance, defined as the length $d_1$ of the shortest contiguous region (with periodic boundary conditions) that contains a logical operator~\cite{Bravyi2009}, as an efficiently computable proxy for $d$. $d_1$ is only an upper bound on $d$, but in circuits with dense operator spreading where the weight of an operator is proportional to its contiguous length, it is a reasonable stand-in for determining the scaling of $d$ with system size, and has been used previously to characterize codes produced by geometrically local, random monitored circuits~\cite{Gullans2020,Ippoliti2020,Li2021a,Li2021}.} 

For concreteness we choose $s=1/4$ and take an initial state with randomly-placed single-site stabilizers on $\lceil 3L/4 \rceil$ sites. As shown in~\autoref{fig:d1}, starting from $d_1=1$ in the product state, the contiguous code distance increases linearly before reaching a maximum slightly below the quantum Singleton bound of $d_{max} = 1+ (1-s)L/2 \rightarrow 3L/8$~\cite{Knill1997,Cerf1997}. \gs{As with the half-cut entanglement entropy}, for $m=2^k$, $d_1$ returns to $O(1)$ twice per period, whereas for other $m$ the extensive-code-length plateau in $d_1$ persists long past the duration of the run owing to the superlinear recurrence time.

\begin{figure}[t]
    \centering
    \includegraphics[width=\linewidth]{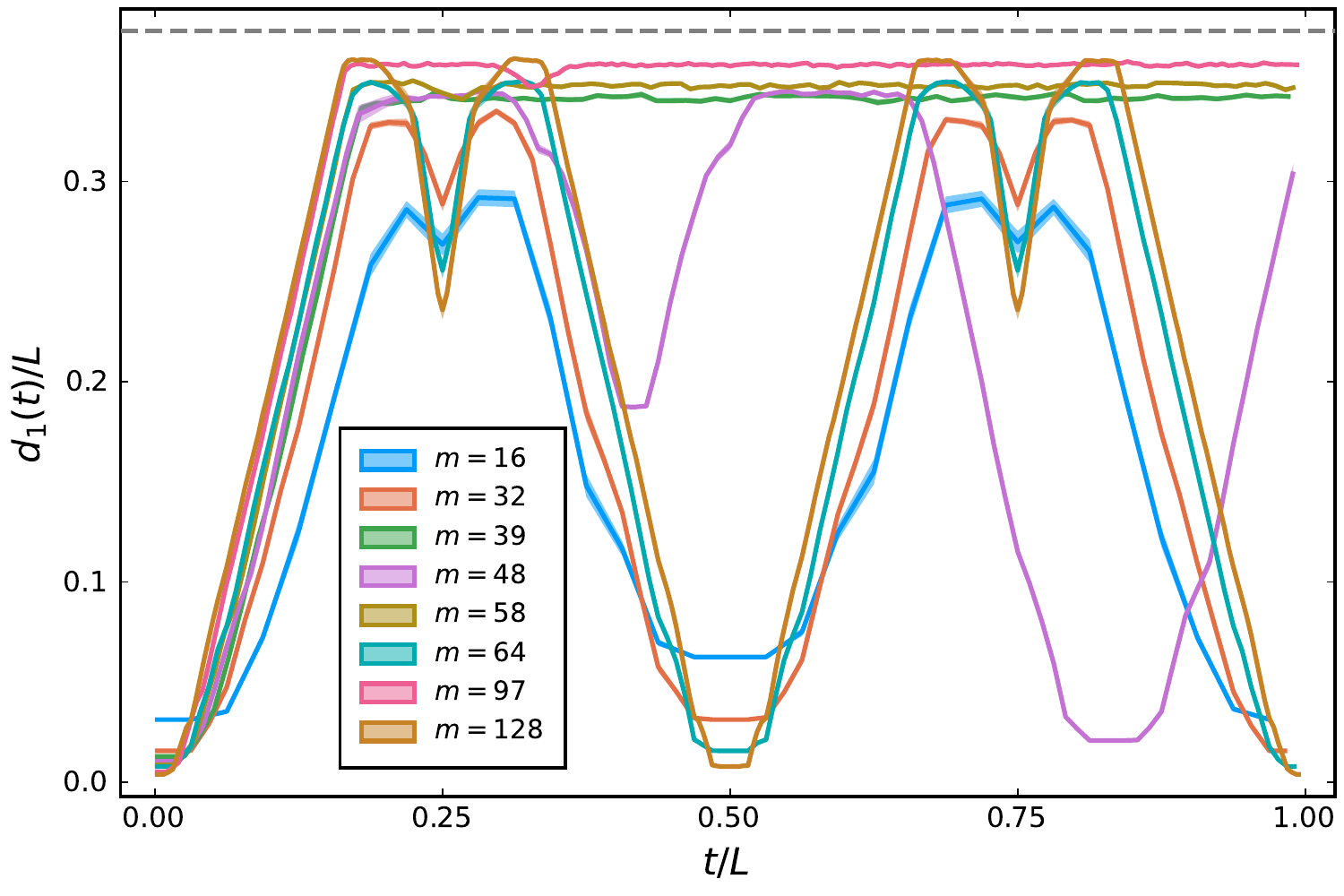}
    \caption{Code length vs. time averaged over 100 random samples for system sizes $m=L/2=$ 16, 32, 39, 48, 58, 64, 97, 128, for the circuit defined by~\autoref{eq:U-iswap} and~\autoref{eq:circuit8-gate} Gray dashed line shows the quantum Singleton bound $d_{max}/L=0.375$.}
    \label{fig:d1}
\end{figure}

{One potential benefit of codes generated by crystalline circuits, as opposed to random encoding circuits, is that their added structure could aid in finding efficient decoding algorithms. To take full advantage of this symmetry, in~\autoref{sect:good} we study codes generated by good scrambling circuits acting on translation-invariant initial states. To assess their performance beyond the heuristics provided by the contiguous code distance, we subject these codes to  erasures, for which an efficient optimal decoder is known~\cite{Delfosse2020,Gullans21}, and obtain recovery probabilities comparable to random codes for a range of system sizes.}

\section{Model and symmetries}\label{sect:symm}
The circuit described in the previous section is just one of many that can be constructed by tiling a crystalline lattice with unitary gates. The building blocks of our spacetime translation-invariant circuits are dual-unitary two-qubit gates, which admit the parameterization~\cite{Bertini2019}:
\begin{equation}\label{eq:du-gate}
    (u_1 \otimes u_2) V[J] (v_1 \otimes v_2)
\end{equation}
where
\begin{equation}
    V[J] = \exp[-i(\pi/4 (XX + YY) + J ZZ)]
\end{equation}
and $u_1, u_2, v_1, v_2$ are single-qubit gates.

Restricting to Clifford unitaries, which map elements of the Pauli group to elements of the Pauli group~\cite{Gottesman1998}, our only choices for $V[J]$ are the SWAP gate ($J=\pi/4$, up to an overall phase) and the iSWAP gate ($J=0$). The latter gate, per standard convention, selects $Z$ as a special axis, as in~\autoref{eq:U-iswap}. A consequence of this convention is that while a generic separable state of two qubits becomes entangled under the action of the iSWAP, product states in the computational ($Z$) basis remain product states. 

A two-qubit gate can naturally be represented as a four-leg tensor, with two incoming and two outgoing legs, as in~\autoref{fig:dual}. {Viewed as a four-qubit state via the operator-state correspondence, a 2-qubit unitary gate corresponds to a state with maximal entanglement of the bipartition into "incoming" and "outgoing" legs, while dual-unitarity also imposes maximal entanglement between the "left" and "right" bipartitions~\cite{Borsi2022}.} We can also interpret this tensor as a geometric object, which has $D_4$ symmetry: the four-legged square is invariant under four-fold rotations, as well as reflections about the horizontal, vertical, and two diagonal axes passing through the center of the square. The corresponding gate need not have these symmetries; thus, our objective is to determine which circuits possess the symmetries of their underlying lattice. 

\subsection{Symmetry of SWAP and iSWAP cores} 

{One motivation for focusing on circuits where the two-site gates on the vertices of the lattice are all dual-unitary is that under any point-group transformation, the circuit remains unitary. In fact, these dual-unitary "cores"---SWAP and iSWAP---are more than just dual-unitary: they possess the full $D_4$ symmetry of the square.} Thus, we can treat the black vertices in the lattice representation (\autoref{fig:lattice} and~\autoref{fig:kagome-lattice}) as "just squares" and focus on the effect of the point group transformations on the edges, which are dressed by single-qubit gates. 

{As depicted in~\autoref{fig:dual}, the spacetime dual of a two-qubit unitary gate is the operator resulting from the $\pi/2$ rotation of its legs. In matrix form,
\begin{equation}
\tilde{U}^{\beta_2 \alpha_2}_{\beta_1 \alpha_1} = U^{\beta_1\beta_2}_{\alpha_1 \alpha_2}.
\end{equation}
Therefore the SWAP gate is self-dual (as was previously noted in Ref.~\cite{Bertini2019}), as is the iSWAP gate, which can be explicitly verified from~\autoref{eq:U-iswap}.}

{The $D_4$ point group can be generated by composing $\pi/2$ rotations with any reflection. Again this just amounts to a reshuffling of matrix indices. Reflection about the horizontal corresponds to time reversal, which is implemented by the taking the transpose}~\footnote{{While in other contexts time reversal corresponds to the \textit{conjugate} transpose, here the ordinary transpose is the most natural consequence of performing a reflection in spacetime~\cite{Mestyan2022}}}.

{SWAP and iSWAP are both symmetric matrices, and hence are invariant under time reversal. Combined with invariance under $\pi/2$ rotations, both gates can be said to be $D_4$-symmetric, or self-octa-unitary.}

{Note that a generic two-site Clifford gate can be written in terms of one-site gates dressing a SWAP, iSWAP, identity, or CNOT core. The latter two gates act as (non-unitary) projectors when rotated by $\pi/2$. Translation-invariant CNOT-core circuits do exhibit nontrivial scrambling behavior, which we have fully classified on the square lattice (see~\autoref{sect:dual-unitary}), but the range of behavior is actually a subset of what we find in dual- and tri-unitary circuits.}


\subsection{Symmetries on the square lattice}
What becomes of the $D_4$ symmetry when we include single-qubit gates? In the brickwork geometry of~\autoref{fig:lattice}, each single-qubit gate is represented as a red or blue circle on the edges between the black (i)SWAP cores. {Each unit cell contains one core, one blue gate, and one red gate, but to make the symmetry explicit, we can consider the enlarged "vertex" comprised of one core + one-site gates on all four legs.
Then, since the core is invariant under these operations, it is sufficient to impose the point group symmetry at the level of the four legs~\footnote{This is also a necessary condition, because we impose the symmetry at each point in the unit cell, i.e. rather than just demanding that $V = (r\otimes b) U_{core} (b\otimes r)$ is invariant under the transformation, we require the one-site gates and core to be individually invariant.}.}

Since the transpose operation implements time reversal, reversing the direction of a leg corresponds to taking the transpose of the single-qubit gate on that leg. Labeling each vertex by the single-qubit gates on the incoming legs, the "standard vertex" is denoted $(b,r)$ (upper left of~\autoref{fig:point-group}(a)).
\begin{figure}[t]
    \centering
    \includegraphics[width=\linewidth]{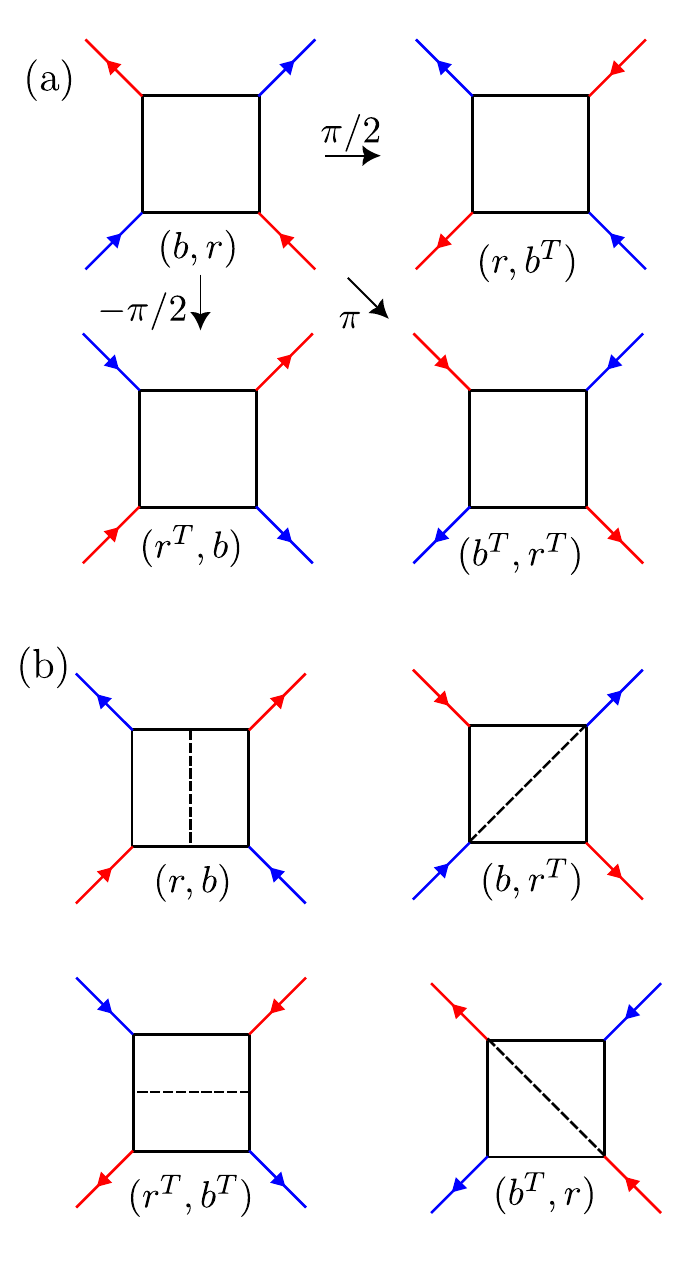}
    \caption{Point group operations on a SWAP or iSWAP vertex. Starting from the diagram in the upper left, the rest are produced by (a) rotations by the indicated angle and (b) reflections about the dashed axis.}
    \label{fig:point-group}
\end{figure}
By comparing the label of the standard vertex to that of the transformed vertex, we can read off the symmetries of each class of circuits. For example, since time reversal (bottom left of~\autoref{fig:point-group}(b)) sends $(b,r)\rightarrow (r^T, b^T)$, $b=r^T$ is a necessary and sufficient condition for time reversal symmetry.

We return to this in more detail in~\autoref{sect:classes}, where the formalism of symplectic cellular automata described in the next section provides a complementary framework for interpreting these symmetries.

\subsection{Symmetries on the kagome lattice}\label{sect:self-tri}
Tri-unitarity was introduced in Ref.~\cite{Jonay2021} as an extension of dual-unitarity in which gates are unitary under three distinct arrows of time. In that work, tri-unitarity is imposed at the level of individual three-qubit \textit{gates}, with $K=6$, which can then be tiled on the triangular lattice to produce a tri-unitary STTI circuit. Our construction instead uses the kagome lattice, which has the same point group as the triangular lattice but, since its coordination number is 4 instead of 6, corresponds to a circuit with two-qubit gates~\footnote{This construction was already anticipated by the particular subfamily of tri-unitary gates considered in Ref.~\cite{Jonay2021}. Each left- or right-facing triangle in the kagome lattice then corresponds to a single six-coordinated vertex of the triangular lattice. A more general triangular-lattice circuit would also include an irreducible three-qubit interaction, which our kagome lattice construction does not allow.}. For the \textit{circuit} to be tri-unitary, the two-qubit gates must be dual-unitary, so restricting to Clifford gates yields a lattice of (i)SWAP cores with single-qubit Cliffords on each edge~(\autoref{fig:kagome-lattice}), similarly to the square lattice. We focus on the case where each core is an iSWAP, since that allows for interacting dynamics.

{In our analysis, the symmetry imposed is that of the lattice, requiring that the full circuit be invariant under (a subgroup of) its corresponding lattice's point group. In this sense our approach differs from Ref.~\cite{Mestyan2022}, in which the full symmetry is imposed on the individual gates, which have $K\geq 4$ legs. These spatially symmetric gates are included under the umbrella of "multi-directional unitary operators", which encompasses families of gates including dual-unitary ($K=4$), tri-unitary ($K=6$), and ternary unitary ($K=8$)~\cite{Milbradt2023}.}

\begin{figure}[t]
    \centering
    \includegraphics[width=\linewidth]{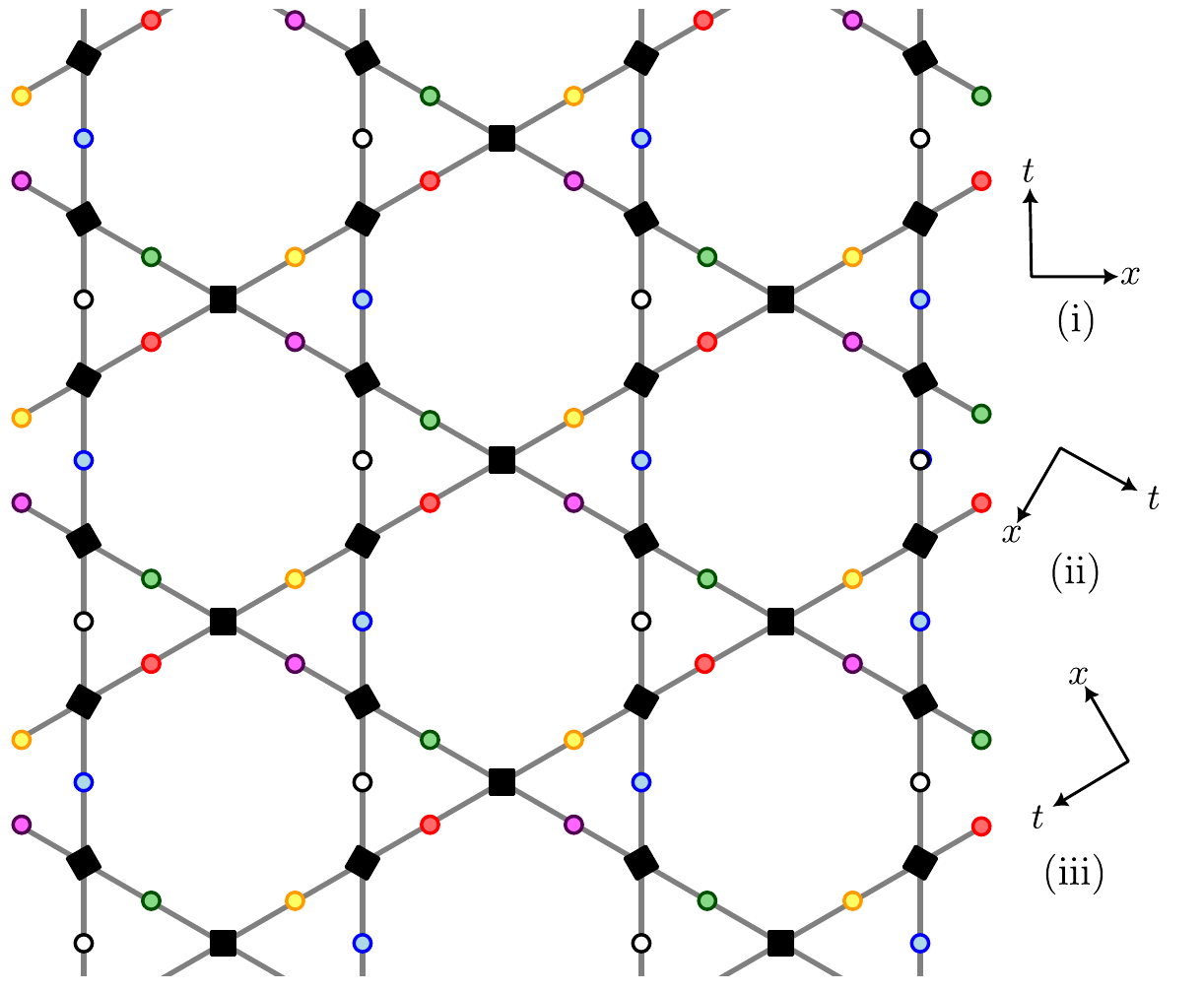}
    \caption{Tri-unitary circuit on a kagome lattice. Black squares are iSWAP cores. The six colors of circles correspond to the six single-qubit gates that populate one unit cell. Three sets of spacetime axes are shown; each axis could also be reversed to give a total of six possible time directions.}
    \label{fig:kagome-lattice}
\end{figure}

\begin{figure}[hbtp]
    \centering
    \includegraphics[width=\linewidth]{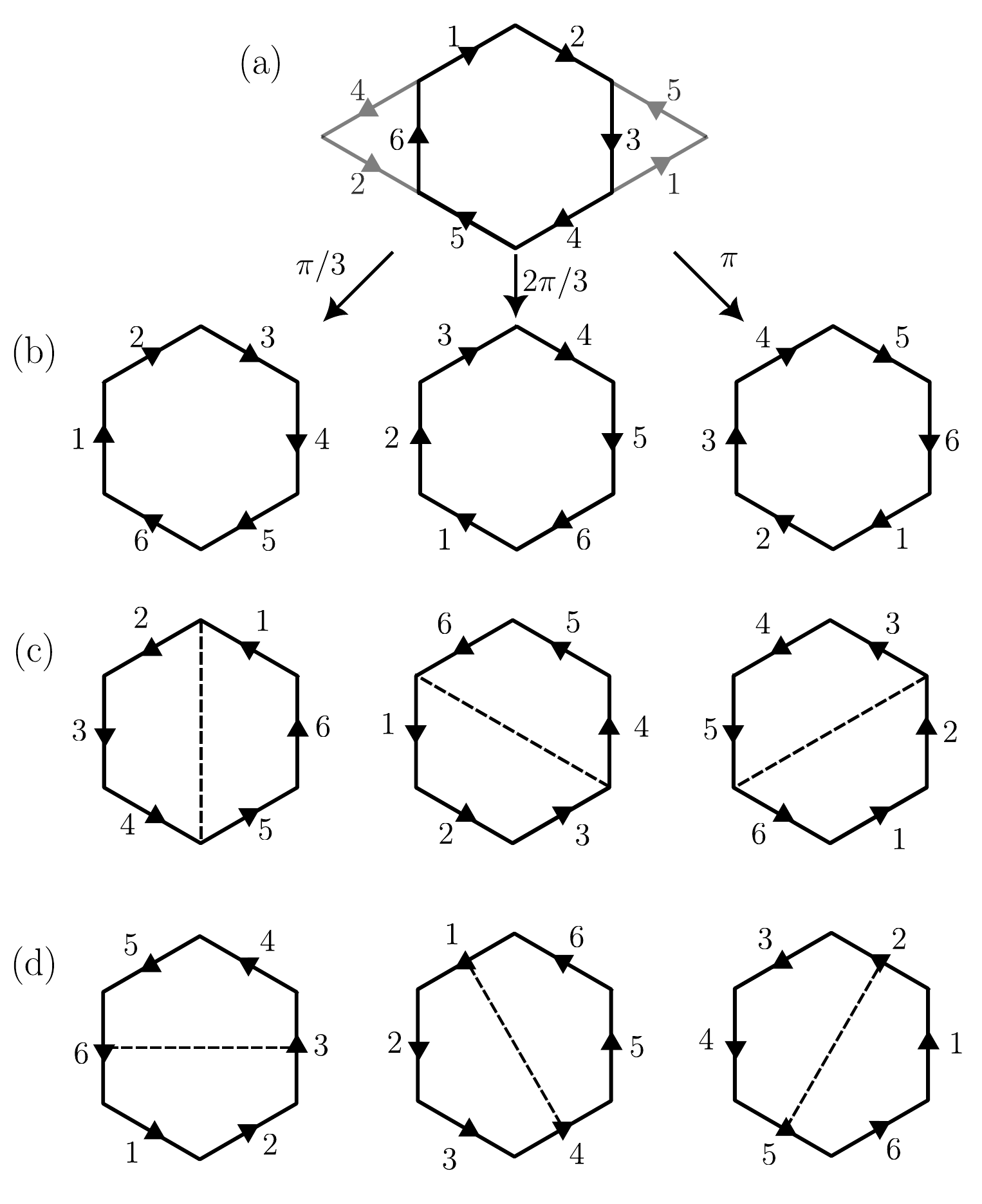}
    \caption{Point group operations on a unit cell of the kagome lattice. (a) The unit cell consists of a hexagon and two adjacent triangles. The black edges labeled 1-6 are treated as "belonging" to the unit cell, while the gray edges belong to adjacent cells. (b) Rotations by $\pi/3$, $2\pi/3$, and $\pi$ ($-\pi/3$, $-2\pi/3$ not shown). (c) Reflections about the three axes connecting opposite vertices, indicated with dashed lines. (d) Reflections about the three axes (dashed) connecting opposite edges.}
    \label{fig:kagome}
\end{figure}

The kagome lattice has the unit cell shown in~\autoref{fig:kagome}(a). The space group of the lattice factors into symmetry under translation by a unit cell and the point group $D_6$, which consists of the transformations shown in~\autoref{fig:kagome}(b-d). Since the iSWAP core is invariant under these rotations and reflections, it suffices to consider the effect of the transformations on the single-qubit gates decorating the edges, as with the square lattice. This can be determined by assigning each edge a direction and label; as above, reversing the direction of the edge corresponds to taking the transpose of the gate.

Demanding the full symmetry of the kagome lattice yields the condition $1=2=...=6$ from invariance under $\pi/3$ rotations (left panel of~\autoref{fig:kagome}(a)), and $1=1^T$ from invariance under any of the reflections. An immediate example is the bare iSWAP circuit, in which all single qubit gates are identities. This is one of the circuits analyzed in~\autoref{sect:kagome}.

The group of six-fold rotations, $C_6$, contains $C_2$ and $C_3$ as subgroups. The symmetry group $C_3$ is of particular interest since rotation by $2\pi/3$ corresponds to changing from one arrow of time to another. Thus, a circuit left invariant under this rotation, which imposes $1=3=5$ and $2=4=6$, can be called "self-tri-unitary." Time reversal symmetry along each of these arrows of time would further impose the symmetries in~\autoref{fig:kagome}(d).

\subsection{Strong and weak self-duality of correlations}\label{sect:corr}
In the previous subsections we have defined a strong form of self-duality: applying the given point group transformation leaves the circuit strictly invariant. In the ensuing analysis, we will also see examples of a weaker form of invariance, wherein the transformed circuit is related to the original circuit by a change of basis. 

What distinguishes strong and weak self-duality? One difference is in the symmetries of the two-point correlations of one-site Pauli operators at infinite temperature, at a spacetime displacement of $(x,t)$. In a dual-unitary circuit, these correlations are nonvanishing only on the edges of the lightcone, $x=\pm vt$, where $v$ is the lightcone velocity. Hence, any correlation function can be decomposed in terms of left- and right-moving quantum channels $\mathcal{M}_\pm$~\cite{Bertini2019}. In a tri-unitary circuit, owing to the existence of three axes of time (\autoref{fig:kagome-lattice}), correlations can also be nonzero along the "static wordline", $x=0$, with the associated quantum channel $\mathcal{M}_0$~\cite{Jonay2021}. Thus, in both cases the analytic tractability of two-point correlations provides a simple way to probe circuit symmetries. Loosely speaking, invariance under a given point group transformation manifests as an equality between correlations at displacements related by that transformation. If the correlations are only equal after a change of basis, then the circuit only possesses a weak form of the symmetry. We leave a more detailed treatment of this topic to~\appref{app:corr}.

\section{Clifford quantum cellular automata}\label{sect:cqca}
Now we introduce the main analytical tool used in the rest of the paper: Clifford quantum cellular automata (CQCA). After presenting the formalism, we write down the general form for the automaton on the square lattice and show how it is transformed under the point group operations described in the previous section.

\subsection{Matrix representation}
By definition, Clifford gates transform single Paulis into single Paulis, rather than superpositions of many Paulis. {As a result, the action of a Clifford unitary is defined by the images of $X$ and $Z$. This property forms the bedrock of the stabilizer tableau representation, by which Clifford circuits can be simulated classically with quadratic complexity in the number of qubits. The uninitiated reader is referred to Refs.~\cite{Gottesman1998,Aaronson2004} for a detailed discussion of this approach. The essence of the tableau representation is a shift in perspective: to understand how a (mixed or pure) stabilizer state evolves, it suffices to track the evolution (in the Schrodinger picture) of the generators of the stabilizer group, comprised of the operators with expectation value $+1$ in the state. The stabilizer tableau gives an efficient means of tracking phases on these operators~\cite{Aaronson2004}, but these will not be relevant to our study of how operators scramble and spread. We will therefore represent $X$ and $Z$ as the binary vectors $\vec{\xi}(X) = \begin{pmatrix} 1 & 0 \end{pmatrix}^T$ and $\vec{\xi}(Z)=\begin{pmatrix}
0 & 1
\end{pmatrix}^T$, which implies $\vec{\xi}(I)=\begin{pmatrix} 0 & 0 \end{pmatrix}^T$ and $\vec{\xi}(Y) = \begin{pmatrix} 1 & 1 \end{pmatrix}^T$.} A single-qubit gate can then be expressed as:
\begin{equation}\label{eq:1-qub}
C_{1 \, \mathrm{qb}} = \begin{pmatrix}
\vec{\xi}(UXU^\dag) & \vec{\xi}(U Z U^\dag)
\end{pmatrix}
\end{equation}

As written, $C$ is a matrix over the binary field $\mathbb{F}_2$. To handle Pauli strings that spread beyond one unit cell, let
\begin{equation}\label{eq:xi-x}
\vec{\xi}(x) = \begin{pmatrix} \xi_X(x) & \xi_Z(x) \end{pmatrix}^T.
\end{equation} 
$\vec{\xi}(x)$ is a function of the lattice position $x$, whose value at $x$ is the two-component binary vector representing the Pauli operator on that site.

{When the circuit in question is translation-invariant with unit cell $a$, it is useful to express it as a Clifford quantum cellular automaton (CQCA). Here we describe the straightforward generalization of the formalism in Refs.~\cite{Schlingemann2008,Gutschow2010long,Gutschow2010solo}, which is written for $a=1$, to general $a$.}
Exploiting translation invariance, we transform $\vec{\xi}(x)$ to a $2a$-component vector over the Laurent polynomial ring $\mathbb{F}_2[u,\ui]$ via the algebraic Fourier transform. Explicitly, letting $x=an+j$ denote the coordinates of the $j$th site in the $n$th unit cell, we define:
\begin{align}\label{eq:ft}
\vec{\xi}(u) &= (\vec{\xi}^{(1)}(u), \vec{\xi}^{(2)}(u),...,\vec{\xi}^{(a)}(u)) \notag \\
\mathrm{where} \quad
    \vec{\xi}^{(j)}(u) &=\sum_{n\in \mathbbm{Z}} u^n \vec{\xi}(na+j).
\end{align} 
{The argument $u$ of the Fourier-transformed vector is defined implicitly through~\autoref{eq:ft}, where for ease of notation, we use the same variable, $\vec{\xi}$, to denote the original vector-valued function $\vec{\xi}(x)$ and its algebraic Fourier transform $\vec{\xi}(u)$, distinguishing them by their arguments.} 
A CQCA with unit cell $a$ can then be expressed as a $2a \times 2a$ matrix $M$ over $\mathbbm{F}_2[u,\ui]$, \gs{i.e. $M \in \mathcal{M}_{2a}(\mathbb{F}_2[u,\ui])$.}

The form of $M$ is constrained by the fact that CQCA preserve the Pauli commutation relations. In the Fourier-transformed representation, these commutation relations are encapsulated in the symplectic form~\cite{Schlingemann2008}: 
\begin{equation}
    \hat{\sigma}(\vec{\xi},\vec{\eta}) = \sum_{j=1}^a (\overline{\vec{\xi}}^{(j)}_X \vec{\eta}^{(j)}_Z - \overline{\vec{\xi}}^{(j)}_Z \vec{\eta}^{(j)}_X)
\end{equation}
where $\overline{f}(u) = f(u\rightarrow \ui)$. Then $M$ is a valid CQCA (also referred to as a symplectic cellular automaton (SCA)) if and only if~\cite{Schlingemann2008}:
\begin{equation}\label{eq:symplectic}
    \hat{\sigma}(M\vec{\xi}, M\vec{\eta}) = \hat{\sigma}(\vec{\xi},\vec{\eta}).
\end{equation} 

Taking the algebraic Fourier transform allows us to compactly represent the action of the CQCA on an infinite system, but sometimes it is useful to consider the behavior on finite chains with periodic boundary conditions. For a system of $m$ unit cells, a shift by $u^m$ is equivalent to the identity, so we take the entries of $M$ to belong to the residue ring $\mathbbm{F}_2[u,\ui] / \langle u^m - 1 \rangle$. We define the recurrence time of the unitary whose CQCA is given by $M$, denoted $\tau(m)$, as the minimum power such that $M^\tau = u^d\mathbbm{1}$ modulo $u^m-1$, for some $d\in \mathbb{Z}$. Allowing $d\neq 0$ accounts for the case where $U$ repeats up to an overall shift by an integer number of unit cells. Under the evolution of the automaton, any stabilizer group on $m$ unit cells, mixed or pure, repeats modulo signs and shifts after an interval that divides $\tau(m)$.

\subsection{Review of $a=1$ automata}

Before turning to the square and kagome lattice, whose automata have $a=2$ and $a=4$ respectively, it will be useful to recall some facts about $a=1$ CQCA \gs{over qudits with prime dimension $q$}. For a more thorough treatment complete with proofs, the reader is referred to Refs.~\cite{Schlingemann2008,Gutschow2010long,Gutschow2010solo}. 

For $a=1$, an automaton with local Hilbert space dimension $q$ is an element $M \in \mathcal{M}_2(\mathbbm{F}_q[u,\ui])$. From the symplectic condition one can prove that $M$ is an SCA if and only if~\cite{Schlingemann2008}:
\begin{enumerate}
    \item Each element $f(u)$ of $M$ is reflection-invariant with respect to the same lattice point $d \in \mathbb{Z}$, \gs{that is, $u^{2d}\overline{f}(u) = f(u)$}.
    \item $\det M = u^{2d}$.
\end{enumerate}
A third condition, which is often stated separately~\cite{Gutschow2010long,Gutschow2010solo} but actually follows from the above two, is that the images of $X$ and $Z$, i.e. the column vectors of $M$, are coprime. 

Due to condition 2, we can always "center" the automaton by factoring out $u^{d}$. This simply expresses that the shift automaton $u^d \mathbbm{1}$, which acts by shifting all operators to the right by $d$ units, commutes with all other automata. Then, it suffices to consider centered symplectic cellular automata (CSCA) whose entries are symmetric Laurent polynomials~\cite{Schlingemann2008,Gutschow2010long,Gutschow2010solo}.

Centered symplectic cellular automata with $a=1$ can be classified into three groups based on their trace~\cite{Gutschow2010long,Gutschow2010solo}: {those with $\tr(M)=$ constant belong to the periodic class, those with $\tr(M)=u^n + u^{-n}$ for some $n\in \mathbb{N}$ belong to the glider class, and all others belong to the fractal class.} This simple classification stems from the fact that the characteristic polynomial of a 2x2 matrix is determined by its trace and determinant, the latter being 1 for CSCA (\autoref{eq:char-poly-a1}):
\begin{equation}\label{eq:char-poly}
\chi_M(y) = y^2 + \tr(M) y + \mathrm{det}(M).
\end{equation}
{By the Cayley-Hamilton theorem, $M$ satisfies its characteristic equation, so for $a=1$ CSCA,
\begin{equation}\label{eq:cayley}
M^2 = -\tr(M) M - \mathbbm{1}.
\end{equation}}

This recursion relation for $M$ underlies several related properties. First, the asymptotic generation rate of bipartite entanglement on a translation-invariant pure state is $\deg(\tr(M))$; that is, the bipartite entanglement across a cut of the infinite system grows linearly for glider and fractal automata but oscillates about a constant for periodic automata~\cite{Gutschow2010long,Gutschow2010solo}. {The behavior of periodic automata is particularly simple for $q=2$ (qubits): straightforward application of~\autoref{eq:cayley} implies that a non-identity CSCA with $\tr(M)=c \in \mathbb{F}_2$ repeats with period $c+2$, thus explaining its designation as periodic.} Of the three classes, only the periodic automata admit (1) pure stationary translation-invariant stabilizer states on an infinite chain and (2) stationary product states, of any entropy density below 1~\cite{Gutschow2010long}.

Members of the glider class earn their name because they have eigenvectors $\vec{\xi}_\pm$ with eigenvalues $u^{\pm n}$~\footnote{{That $\tr(M)=u^n + u^{-n}$ is a sufficient condition for $M$ to have eigenvalues $u^{\pm n}$ follows straightforwardly from~\autoref{eq:char-poly}; the necessity of this condition is proven in Prop. II.8 of Ref.~\cite{Gutschow2010long}.}}. These so-called "gliders" are operators that shift but do not spread under the action of the automaton, corresponding to conserved charges and resulting in a recurrence time $\tau(m)\leq m$ on a system with periodic boundary conditions~\cite{Stephen2019}. In contrast, the recurrence time for fractal CSCA is exponentially large for generic $m$, but from the recursion relation~\autoref{eq:cayley}, one can prove that for all CSCA, including those in the fractal class,   $\tau(m) \leq 3m/2$ for $m=2^k$~\cite{Stephen2019}~\footnote{Ref.~\cite{VonKeyserlingk2018} quotes a weaker upper bound of $12m$, but this for the recurrence of $U$ itself, not $M$, i.e. taking signs into account.}. We will see that the a linear bound on $\tau(m)$ for $m=2^k$ also holds for $a>1$.

For $a>1$, the characteristic polynomial remains important for characterizing $M$, although it is no longer solely determined by the trace. More precisely, we will be interested in the minimal polynomial---the monic polynomial $\mu_M$ of least degree for which $\mu_M(M)=0$---which always divides $\chi_M$. In Ref.~\cite{Gutschow2010fractal}, it is demonstrated that for any linear cellular automaton over an abelian group, a broad class of automata that includes CQCA with generic $a$, one can construct a sequence of "colored spacetime diagrams" which depict the evolution of an initial string (in our case a Pauli operator) under the action of the automaton, as time $t\rightarrow\infty$. For a given initial string, the spacetime diagram converges in the limit of infinite time, and in particular automata with the same minimal polynomial produce evolutions with similar fractal structure. \gs{This link between the minimal polynomial and operator spreading is not unique to $a=1$, and rests on the fact that $\mu_M$ implies a recursion relation for $M$.}

\gs{The discerning reader may question why we do not recast our $a>1$ qubit CQCA as $a=1$ quantum cellular automata acting on $2^a$-dimensional qudits. However, representation as an element of $\mathcal{M}_2(\mathbb{F}_{2^a}[u,\ui])$ does not readily follow; see footnote [64] for more details. Our $a>1$ qubit CQCA should also be contrasted with the $a=1$ CQCA studied in Ref.~\cite{Kent2023} with local Hilbert space dimension $N$ (not necessarily prime) endowed with a generalized Clifford algebra, which are described by elements of $\mathcal{M}_2(\mathbb{Z}_{N}[u,\ui])$ and for which $N\rightarrow\infty$ is the semiclassical limit.}
\subsection{Decomposition of dual-unitary CQCA}
Expressing our STTI Clifford circuits as SCA, we now compute the matrix form for the time evolution of one unit cell of the circuit. In full generality, the evolution consists of three fundamental elements: the SWAP/iSWAP cores, single-qubit gates, and optionally, a spatial shift between successive time steps.

\subsubsection{Shift}
Concretely, let's consider the circuit on the square lattice. Although the brickwork only repeats after two layers of the circuit, we can use a smaller unit cell, $T=1/2, a=2$, by also including a spatial shift of $d=1$ between time steps. This simply expresses that the square lattice is translation-invariant under translations by $t=1/2,d=1$.

A generic CQCA on unit cell $a=2$ takes the form:
\begin{equation}
    M = \begin{pmatrix} M_{X_1\rightarrow X_1} & M_{Z_1 \rightarrow X_1} & M_{X_2 \rightarrow X_1} & M_{Z_2 \rightarrow X_1} \\
    M_{X_1\rightarrow Z_1} & M_{Z_1 \rightarrow Z_1} & M_{X_2 \rightarrow Z_1} & M_{Z_2 \rightarrow Z_1} \\
    M_{X_1\rightarrow X_2} & M_{Z_1 \rightarrow X_2} & M_{X_2 \rightarrow X_2} & M_{Z_2 \rightarrow X_2} \\
    M_{X_1\rightarrow Z_2} & M_{Z_1 \rightarrow Z_2} & M_{X_2 \rightarrow Z_2} & M_{Z_2 \rightarrow Z_2}
    \end{pmatrix}
\end{equation}
i.e., the columns are the images of $X_1, Z_1, X_2, Z_2$.

This means that a shift by 1 site to the right, $M_{shift}$, takes the block-off-diagonal form:
\begin{equation}\label{eq:shift}
M_{shift} = \begin{pmatrix}
0 & u \mathbbm{1} \\
\mathbbm{1} & 0 
\end{pmatrix}
\end{equation}
This equation can be straightforwardly generalized to shifts by $j=1,2,...,a-1$ in a unit cell of size $a$.

Note that $M_{shift}^2 = u\mathbbm{1}$, i.e. a shift by one full unit cell, which can be factored out to center the automaton as in the $a=1$ case.  Formally, we could account for this by writing
\begin{equation}\label{eq:center}
    \tilde{M}_{shift} = \begin{pmatrix}
0 & u^{1/2} \mathbbm{1} \\
u^{-1/2}\mathbbm{1} & 0 
\end{pmatrix}
\end{equation}
although of course $u^{1/2}$ is not an element of the Laurent polynomial ring. 

It is sometimes useful to consider the automaton with a larger unit cell, $T=1,a=2$, since after two layers the brickwork circuit repeats without a shift. The centered automaton is:
\begin{equation}\label{eq:one-step}
    \tilde{M} \equiv M_{shift}^{-2} M^2 = \ui M^2.
\end{equation}

\subsubsection{(i)SWAP cores}
The two-qubit gates naturally act on a unit cell of $a=2$. If the circuit is translation-invariant with a larger unit cell, as on the kagome lattice, we can just take a tensor product with the matrices corresponding to the other gates in that layer.

For the SWAP gate, the automaton is:
\begin{equation}\label{eq:swap}
M_{\SWAP} = \begin{pmatrix}
0 & \mathbbm{1} \\
\mathbbm{1} & 0
\end{pmatrix}
\end{equation}
while for the iSWAP:
\begin{equation}\label{eq:iswap}
M_{\iSWAP} = \begin{pmatrix}
\mathbf{a} & \mathbf{b} \\
\mathbf{b} & \mathbf{a}
\end{pmatrix}
\end{equation}
where
\begin{equation}\label{eq:iswap-ab}
\mathbf{a} = \begin{pmatrix}
0 & 0 \\
1 & 0
\end{pmatrix}, \qquad
\mathbf{b} = \begin{pmatrix}
1 & 0 \\
1 & 1
\end{pmatrix}.
\end{equation}

\subsubsection{Single-qubit gates}
The final ingredient in our circuits is the single-qubit gates. Out of the 24 elements of the single-qubit Clifford group, we consider two gates to be equivalent if they differ by only a Pauli operator, since that only affects the signs on the stabilizers.

The six remaining unique elements fall into three groups~\cite{Crooks2022}. As in~\autoref{eq:1-qub}, these gates can be expressed as 2x2 matrices over $\mathbb{F}_2$, which if promoted to matrices over $\mathbb{F}_2[u,\ui]$ (i.e., we imagine applying the same gate to each qubit) would be $a=1$ CQCA in the periodic class: single-qubit gates alone cannot generate any entanglement. Yet, when incorporated into circuits with iSWAP cores, these different groups of gates produce qualitatively different classes of behavior as described in~\autoref{sect:classes}. {{This is a manifestation of the broader point that although circuits with the same core are locally unitarily equivalent, the mixing properties are sensitive to the local (one-site) gates~\cite{Aravinda2021}.}} The three groups are:
\begin{enumerate}
    \item Identity, which trivially has period 1.
    \item $\pi/2$ rotation about $X,Y,$ or $Z$, which preserves the Pauli along the axis of rotation and exchanges the other two. As 2x2 matrices:
    \begin{subequations}
    \begin{align}
    M_{R_X[\pi/2]} &= \begin{pmatrix}
    1 & 1 \\
    0 & 1
\end{pmatrix} \\
M_{R_Y[\pi/2]} &= \begin{pmatrix}
0 & 1 \\
1 & 0
\end{pmatrix} \\ M_{R_Z[\pi/2]} &= \begin{pmatrix}
1 & 0 \\
1 & 1
\end{pmatrix}
    \end{align}
    \end{subequations}
As CQCA, these are all period 2 automata. This reflects the fact that up to a Pauli, a counterclockwise rotation by $\pi/2$ is equivalent to a clockwise rotation about the same axis. Explicitly,
    \begin{equation}\label{eq:rotation}
        (R_{\sigma}[\pi/2])^2 = R_{\sigma}[\pi] = -i\sigma \simeq \mathbbm{1}
    \end{equation}
    where $\sigma = X, Y,Z$, and $\simeq$ is used to denote "equal up to a Pauli."
    \item $\pm 2\pi/3$ rotation about the axis $(1,1,1)$ on the Bloch sphere, which implements a cyclic permutation of $X$, $Y$, and $Z$ modulo signs. Explicitly, the clockwise rotation sends $X\rightarrow Z \rightarrow Y \rightarrow X$, while the counterclockwise rotation sends $X \rightarrow Y \rightarrow Z \rightarrow X$:
    \begin{subequations}
\begin{align}
    M_{R_{(1,1,1)}[2\pi/3]} &= \begin{pmatrix}
    1 & 1 \\
    1 & 0
    \end{pmatrix} \\
    M_{R_{(1,1,1)}[-2\pi/3]} &= \begin{pmatrix}
    0 & 1 \\
    1 & 1
    \end{pmatrix}
    \end{align}
    \end{subequations}
    which are period 3 automata. From the matrix form we can also immediately see that
    \begin{align}\label{eq:permute-3}
        M_{R_{(1,1,1)}[\pm 2\pi/3]}^2 &= M_{R_{(1,1,1)}[\pm 2\pi/3]}^{-1} \notag \\
        &= M_{R_{(1,1,1)}[\mp 2\pi/3]}.
    \end{align}
\end{enumerate}

\subsubsection{Decomposition on the square lattice}
In a brickwork circuit, we can simplify matters by noting that while a generic dual-unitary gate has the parameterization~\autoref{eq:du-gate}, with single-qubit gates before and after the core, in the context of a full circuit the gates on the outgoing legs can be absorbed into the incoming legs of the next layer. We choose to cut the links in such a way that the single-qubit gates come before the core:
\begin{equation}\label{eq:v-first}
    U = V[J] (v_+ \otimes v_-)
\end{equation}

With this convention, the automaton for one time step decomposes as:
\begin{equation}\label{eq:M-decompose}
M= M_{shift} M_{core} \begin{pmatrix}
M_{v_+} & 0 \\
0 & M_{v_-}
\end{pmatrix}
\end{equation}
where $M_{core}$ is the matrix for the SWAP (\autoref{eq:swap}) or iSWAP (\autoref{eq:iswap}) core.

Three alternative conventions are shown in~\autoref{fig:square-cores}. It is straightforward to prove that all four conventions have the same characteristic and minimal polynomials, consistent with the fact that they represent the same physical circuit~\footnote{{We treat the blue and red gates as indivisible, and thus do not include conventions in which they are split between adjacent layers, which would result in (not necessarily Clifford) gates on three or more legs.}}.
\captionsetup[subfigure]{labelformat=empty}
\begin{figure}[t]
    \centering
    \subfloat[$M_{core} \begin{pmatrix} M_{v_+} & 0 \\ 0 & M_{v_-}\end{pmatrix}$]{
    \includegraphics[width=0.45\linewidth]{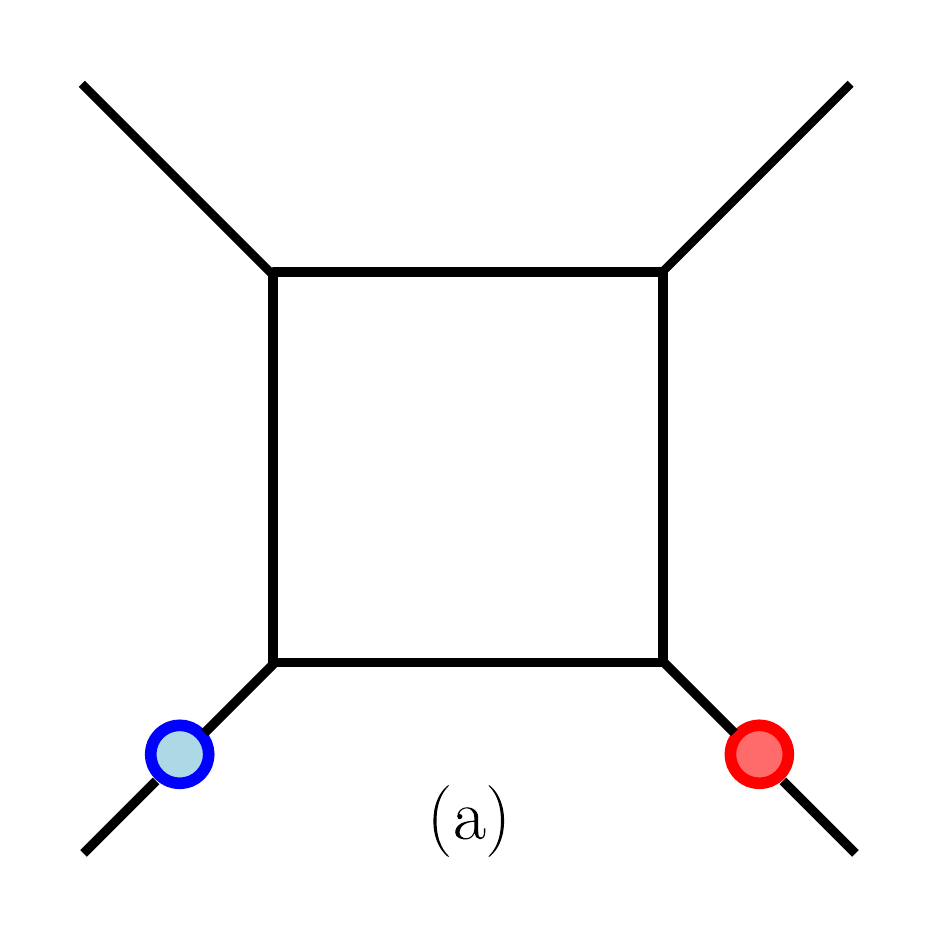}
    }
        \subfloat[$\begin{pmatrix} M_{v_-} & 0 \\ 0 & M_{v_+}\end{pmatrix} M_{core}$]{
    \includegraphics[width=0.45\linewidth]{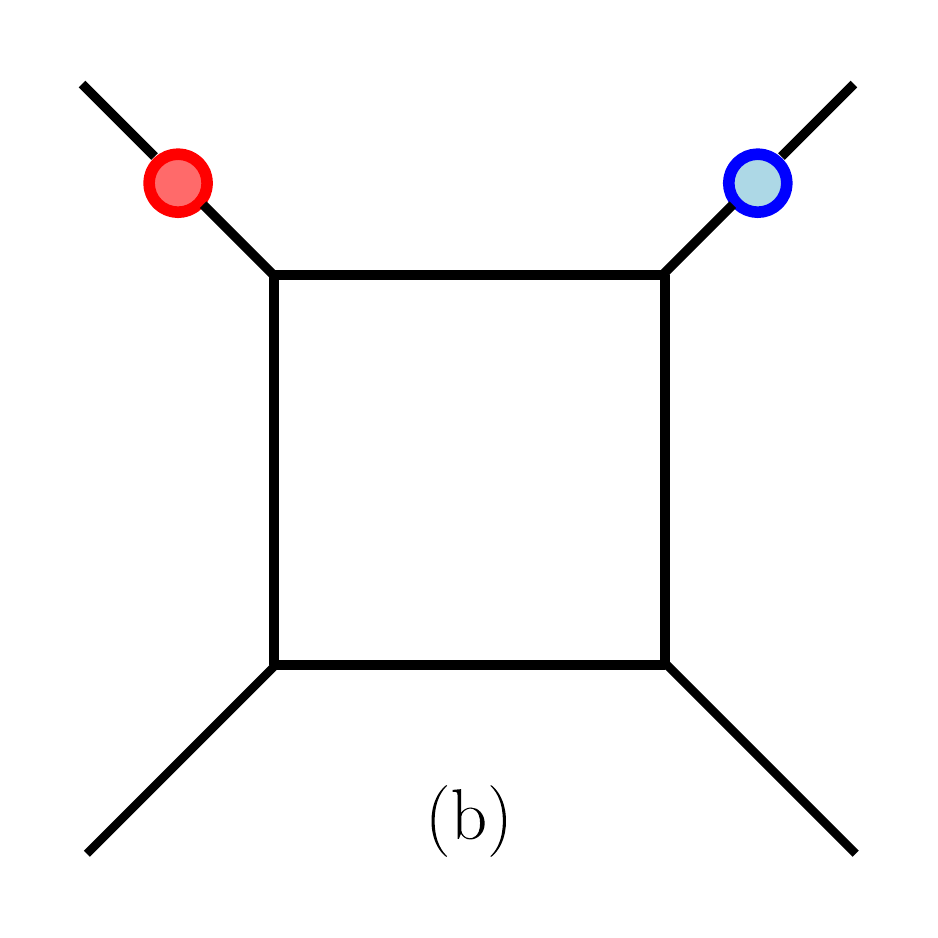}
    }\\
        \subfloat[$\begin{pmatrix} M_{v_-} & 0 \\ 0 & \mathbbm{1}\end{pmatrix} M_{core} \begin{pmatrix} M_{v_+} & 0 \\ 0 & \mathbbm{1}\end{pmatrix}$]{
    \includegraphics[width=0.45\linewidth]{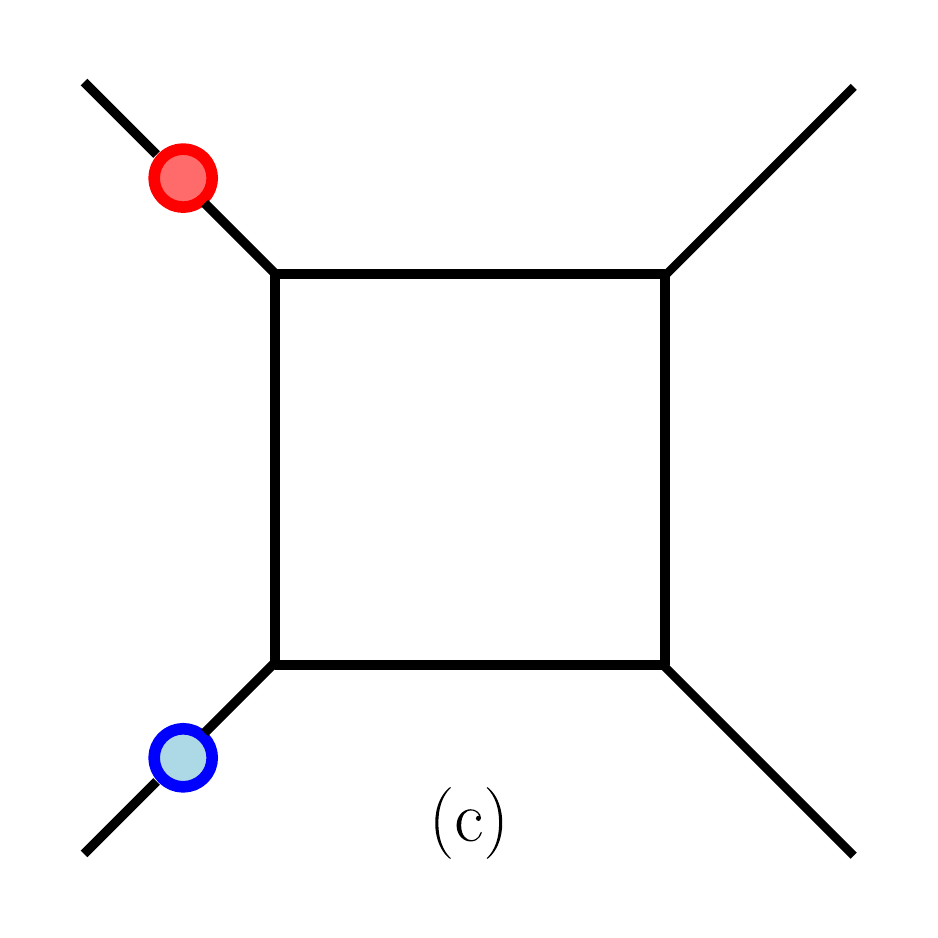}
    }
        \subfloat[$\begin{pmatrix} \mathbbm{1} & 0 \\ 0 & M_{v_+}\end{pmatrix} M_{core} \begin{pmatrix} \mathbbm{1} & 0 \\ 0 & M_{v_-}\end{pmatrix}$]{
    \includegraphics[width=0.45\linewidth]{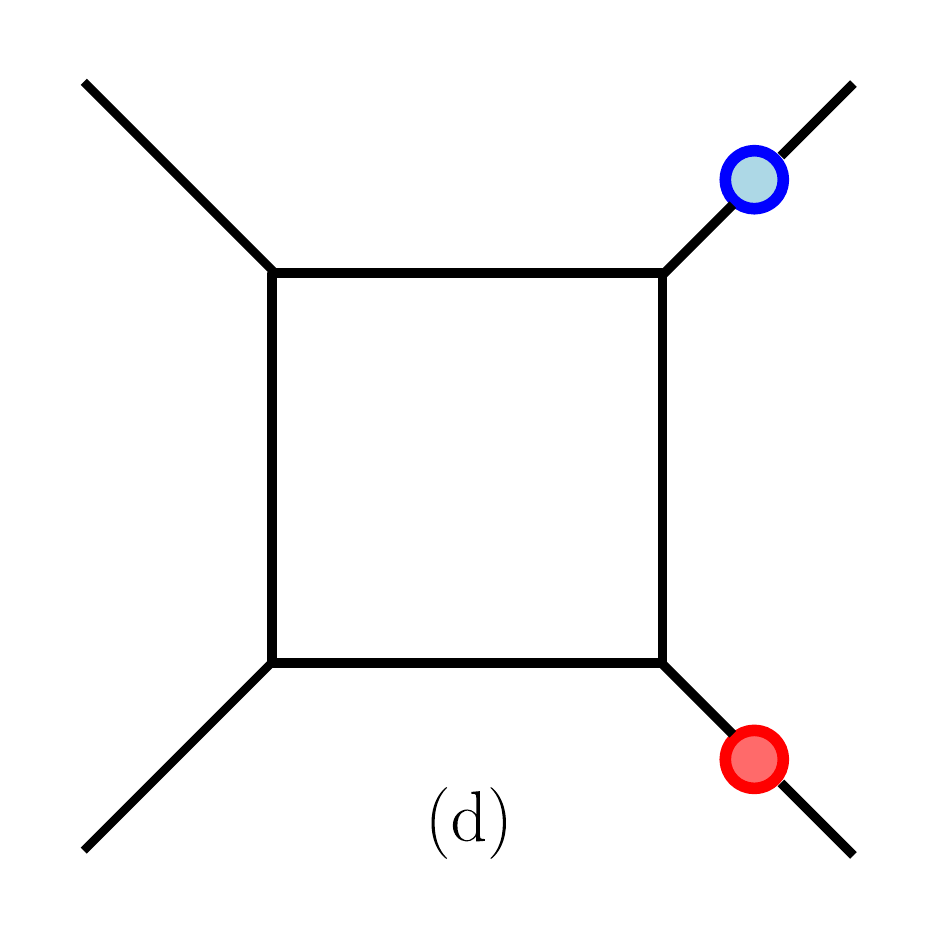}
    }
    \caption{Four conventions for the CQCA on the square lattice. Each is followed by a shift by 1 site, $M_{shift}$. We use the top left convention, which places both single-qubit Cliffords before the core.}
    \label{fig:square-cores}
\end{figure}
\captionsetup[subfigure]{labelformat=parens}

\subsection{Symmetries and similarity transformations}
Now we can analyze the point group symmetries by asking how the automata transform under rotations and reflections of the lattice. 

{An important caveat is that since the unit cell contains only two one-site gates, i.e. in each of the four conventions shown in~\autoref{fig:square-cores} only two of the four legs are decorated with gates, none of these conventions have the full $D_4$ symmetry (unless both gates are identities). This is in contrast with~\autoref{fig:point-group} and the surrounding discussion, where the "expanded vertex" contains a gate on each leg. Thus, when asking if a QCA has a given symmetry, we must compare the transformed automaton to the version of $M$ in the convention $\alpha\in \{a,b,c,d\}$ with the appropriate placement of one-site gates relative to the core.} {All four conventions yield automata with the same characteristic polynomial, so a necessary (but not sufficient) condition for symmetry is that the characteristic polynomial be left invariant under the transformation.}

The eight point group transformations of the square can be expressed as the composition of left-right reflection and the spacetime dual (rotation by $\pi/2$)~\cite{Mestyan2022}. We discuss these, along with time reversal (reflection about the horizontal), in turn.

\subsubsec{Left-right reflections}
For a unit cell of size $a$, reflection about the center of the unit cell is expressed as:

\begin{equation}\label{eq:reflection}
M_{j \leftrightarrow a+1-j} = \begin{pmatrix}
0 & & \mathbbm{1} \\
& \reflectbox{$\ddots$} & \\
\mathbbm{1} & & 0
\end{pmatrix} \overline{M} \begin{pmatrix}
0 & & \mathbbm{1} \\
& \reflectbox{$\ddots$} & \\
\mathbbm{1} & & 0
\end{pmatrix}
\end{equation}
where each $\mathbbm{1}$ is a 2x2 matrix. Explicitly, for $a=2$ this simplifies to:
\begin{equation}\label{eq:1->2}
    M_{1 \leftrightarrow 2} = \begin{pmatrix}
    0 & \mathbbm{1} \\
    \mathbbm{1} & 0
    \end{pmatrix} \overline{M} \begin{pmatrix}
    0 & \mathbbm{1} \\
    \mathbbm{1} & 0
    \end{pmatrix}
\end{equation}

{The resulting transformation of the characteristic polynomial is 
\begin{equation}\label{eq:char-poly-LR}
\chi_M(y)\rightarrow \chi_M(y; u\rightarrow \ui).
\end{equation}}
$M_{core}$ is manifestly invariant under~\autoref{eq:1->2}, while $M_{shift}\rightarrow \ui M_{shift}$, an overall shift that can be removed by "centering" $M_{shift}$ as in~\autoref{eq:center}. Thus, the net result of the transformation is just to exchange $M_{v_+}$ and $M_{v_-}$, as expected:
\begin{equation}
    M^{1\leftrightarrow 2}_a(v_+, v_-) = \ui M_a(v_-,v_+)
\end{equation}
where we have introduced the notation $M_\alpha(v_+,v_-)$ to denote the automaton with convention $\alpha = a,b,c,d$ and single-qubit gates $v_+, v_-$. Note, though, that in imposing this symmetry we do not actually require $v_+=v_-$ as was implied at the level of the unitary in~\autoref{fig:point-group}. Since the symplectic cellular automaton does not include signs on the stabilizers, $M'= M$ (up to a global shift) just imposes that the corresponding unitaries are equal up to a Pauli.

\subsubsec{Time reversal}
The time-reversed automaton is $M'=M^{-1}$, {with characteristic polynomial
\begin{equation}\label{eq:char-poly-TR}
    \chi_{M^{-1}}(y) = \frac{y^{2a}}{\det(M)} \chi_M(1/y) = y^{2a} \chi_M(1/y),
\end{equation}
where the second equality holds for CSCA, for which $\det(M)=1$.}
The automaton transforms as
\begin{align}
    M^{-1}_a(v_+, v_-) &= \begin{pmatrix} M_{v_+}^{-1} & 0 \\ 0 & M_{v_-^{-1}} \end{pmatrix} M_{core} M_{shift}^{-1} \notag \\
    &= M_{shift}^{-1} M_b(v_-^T, v_+^T) M_{shift}^{-1} \notag \\
    &\simeq \ui M_b(v_-^T,v_+^T).
\end{align}
In the last line, $\simeq$ denotes that while the two automata are not equal, they represent the same circuit, since the half-unit-cell shifts can be absorbed into the subsequent layers at the expense of an overall shift by one unit cell.

Up to multiplication by Paulis, we draw the same conclusion as in ~\autoref{fig:point-group}: a square lattice CQCA is time-reversal symmetric if $v_+ \simeq v_-^T$ and $v_- \simeq v_+^T$.

\subsubsec{Spacetime dual}
The transformation of the CQCA $M$ under a counterclockwise $\pi/2$ rotation can in general be written by looking at the action on a complete basis of stabilizers and solving a set of linear equations, but by decomposing $M$ as~\autoref{eq:M-decompose}, we can straightforwardly read off
\begin{align}
    M^{\mathrm{dual}}_a(v_+,v_-) &\simeq M_d(v_-^T, v_+)
\end{align}
up to an overall shift.

\subsubsec{Circuit classes}
Two automata are considered to belong to the same class if they are related by a point group transformation or change of basis. Equivalently, we define a class as all those related by just one point group transformation---left-right reflection---or by the transformation $X_1\leftrightarrow Y_1$, and/or $X_2 \leftrightarrow Y_2$, i.e. the similarity transformation $ M(X_i \rightarrow Y_i) = S_i M S_i^{-1}$ where
\begin{subequations}\label{eq:similar}
\begin{align}
    S_1 &= \begin{pmatrix}
     M_{R_Z[\pi/2]} & 0 \\
     0 & \mathbbm{1}
     \end{pmatrix} = S_1^{-1}, \\
   S_2 &= \begin{pmatrix}
     \mathbbm{1} & 0 \\
     0 & M_{R_Z[\pi/2]}
     \end{pmatrix} = S_2^{-1}
\end{align}
\end{subequations}
This change of basis preserves the iSWAP and SWAP cores while exchanging $R_X[\pi/2] \leftrightarrow R_Y[\pi/2]$ and $R_{(1,1,1)}[2\pi/3] \leftrightarrow R_{(1,1,1)}[-2\pi/3]$. The SWAP core is also preserved under transformations like $X \leftrightarrow Z$ and their compositions, implemented by replacing $R_Z$ with $R_X$ or $R_Y$ in the above expression.

To see that similarity transformations composed with left/right reflections generate all the automata (expressed in convention (a)) related by a point group transformation, note that taking the transpose of any single-qubit gate, followed optionally by a similarity transformation, yields the original gate up to a Pauli, i.e.
\begin{equation}\label{eq:cob}
    M_{v^T}=M_{v}^{-1} = \begin{cases}
    S M_v S^{-1} & v = R_{(1,1,1)}[\pm 2\pi/3] \\
    M_v & \mathrm{otherwise}
    \end{cases}
\end{equation}
This means that a square-lattice circuit is weakly self-dual under any point group transformation as long as it has (weak) invariance under left/right reflection. From the two cases in~\autoref{eq:cob}, we also see that one-site gates corresponding to automata of period 1 or 2 satisfy $u \simeq u^T$, whereas the period 3 automata have $u \not\simeq u^T$.

\section{Classes on the square lattice}\label{sect:classes}
We now apply the formalism in the previous section to classify the dual-unitary CQCA on the square lattice. The SWAP-core automata can be viewed as generalizations of the periodic class of $a=1$ automata. The iSWAP-core automata form six classes, which split into two groups: one group of "poor scramblers" is related to the $a=1$ glider class, while the "good scramblers" are related to the $a=1$ fractal class. {The CQCA formalism also provides another perspective on the trends in correlation functions, conserved quantities, and entanglement growth common to dual-unitary circuits, which we touch on throughout this section and further discuss in~\autoref{sect:dual-unitary}.}

\subsection{SWAP core}
Since the SWAP gate does not generate any entanglement, we already know that the STTI circuits with a SWAP core are non-entangling, with a dynamics that is in some sense "trivial." Nevertheless, writing out the 4x4 matrices that describe these circuits can elucidate their structure and situate them within the framework of $a=1$ CQCA.

Inserting~\autoref{eq:swap} for $M_{core}$,~\autoref{eq:M-decompose} simplifies to:
\begin{equation}
M(v_+,v_-) = \begin{pmatrix}
uM_{v_+} & 0 \\
0 & M_{v_-}
\end{pmatrix}
\end{equation}

$M$ is block-diagonal, where the 2x2 blocks on the diagonal describe the independent time evolution along the $+$ and $-$ diagonals of the lattice, determined by the single-qubit gates $v_+$ and $v_-$ respectively. Thus, the dynamics decompose into two $a=1$ automata in the periodic class, consistent with the fact that SWAP gates do not generate entanglement~\cite{Gutschow2010solo,Gutschow2010long}. Independently, the two automata have period 1, 2, or 3 depending on their trace (measured in units of $t=1/2$). But in any fixed frame, $\tau(m)$ is linear in $m$. This is because the top left block is symmetric with respect to the lattice point $d=1$, whereas the bottom right block is symmetric with respect to $d=0$ (i.e., is a centered SCA). The full automaton only appears periodic if we choose a "staggered frame" where in each time step, the odd sites are translated by one unit cell with respect to the even sites.

As written, $M$ contains an explicit dependence on the single-qubit gates $v_+$ and $v_-$, but we can always push the single-qubit gates through the SWAP core up to the top layer/boundary of the circuit. In this sense, all SWAP-core automata are equivalent to the bare SWAP circuit, which is self-octa-unitary. (Point group transformations would just change the boundary layers.) For this circuit, the recurrence time in units of $t=1/2$ on a system of $m$ unit cells is $m$. Any translation-invariant stabilizer state is invariant under the action of the circuit, so there is a large set of stationary states on a system of any size. 

Although the dynamics are fairly boring viewed through this lens, the SWAP class is actually "maximally entangling" from the perspective of Ref.~\cite{Berenstein2019}. Explicitly, starting from two pure subsystems $A$ and $B$, with some initial entanglement between the odd and even sites on each half, when a gate is introduced between $A$ and $B$, the SWAP model saturates the minimal cut bound on entropy production between $A$ and $B$. In fact, in generic dual-unitary circuits starting from a product state of $m$ nearest-neighbor Bell pairs on $2m$ sites, the entanglement entropy of a contiguous subregion $A$ saturates this bound~\cite{Piroli2019}, {which for a system of length $L$ with periodic boundary conditions, reads~\cite{Casini2016}:
 \begin{equation}\label{eq:min-cut}
     \lim_{L\rightarrow\infty} S_A(t) = \min(4t, |A|).
 \end{equation}}
For the SWAP circuit, the presence of initial entanglement already in the system is crucial, because the SWAP gate has zero entangling power~\cite{Aravinda2021}.

\subsection{iSWAP core}
Substituting $M_{\iSWAP}$ (\autoref{eq:iswap}) into~\autoref{eq:M-decompose} yields:
\begin{equation}\label{eq:M-iswap}
M(v_+,v_-) = \begin{pmatrix}
u \mathbf{b} M_{v_+} & u \mathbf{a} M_{v_-} \\
\mathbf{a} M_{v_+} & \mathbf{b} M_{v_-}
\end{pmatrix}.
\end{equation}

A key difference from the SWAP-core automata is that the "period 2" single-qubit gates are not all equivalent. Since the iSWAP gate has $Z$ as a special axis, a $Z$ rotation can be propagated through the core as in the case of a SWAP gate:
\begin{equation}\label{eq:type2-iswap}
    \iSWAP (R_Z[\pi/2] \otimes \mathbbm{1}) = (\mathbbm{1} \otimes R_{Z}[\pi/2])\iSWAP
\end{equation}
which tells us that after two layers, up to signs on Pauli operators, performing a $Z$ rotation is equivalent to acting with the identity~ \footnote{{At the level of one unit cell, however, placing a $Z$ rotation on only one leg breaks self-octa-unitarity.}}. On the other hand, $X$ and $Y$ rotations, when propagated through the iSWAP, change the core itself, as do the cyclic permutations~\footnote{This gives another view of the crystallography of the dense good scrambling class, which has $X$ rotations on each leg. Pushing $R_X[\pi/2]$ through the iSWAP core cancels out the decorations on all edges (up to signs on the stabilizers) but changes the core from $\exp(-i\pi/4(XX+YY))$ to $\exp(-i\pi/4(XX+ZZ))$. This leaves a rotated square lattice which is a checkerboard of the conventional iSWAP and the "XX+ZZ" iSWAP, with the enlarged unit cell $(T=1,a=2)$.}. This can be seen from~\autoref{eq:iswap} and~\autoref{eq:iswap-ab}: the only single-qubit CQCA that commute with both $\mathbf{a}$ and $\mathbf{b}$ are $\mathbbm{1}$ and $M_{R_Z[\pi/2]}$.

Thus, when considering the action of the automaton at integer times $t$, there are 3 distinct choices for each of $v_+$ and $v_-$: (1) $\mathbbm{1}$ and $R_Z[\pi/2]$, (2) $R_X[\pi/2]$ and $R_Y[\pi/2]$, (3) $R_{(1,1,1)}[\pm 2\pi/3]$. This implies that there are $_3C_2 = 6$ classes of iSWAP automata. Unlike with the SWAP core, these classes cannot be further combined by pushing single-qubit gates through to the boundary.

All six classes of automata generate volume-law entanglement, but they divide into two groups based on how much entanglement is generated for a random initial product state.  There is also a sharp distinction between the two groups with respect to the recurrence times on a finite system: "poor scramblers" have linear in $m$ recurrence times for all $m$, reminiscent of the $a=1$ glider class~\cite{Stephen2019}, whereas $\tau(m)$ grows superlinearly for $m\neq 2^k$ in the "good scrambling" classes.

\subsection{"Poor scramblers"}\label{sect:poor}
 In three classes, the "poor scramblers," the steady-state Page curve for a system starting in a random pure product state has a slope less than 1, i.e. the total entropy of a subsystem of length $|A|<L/2$ is $f|A|$, where $0<f<1$. {We emphasize that random product states do not belong to the class of solvable translation-invariant initial states defined in Ref.~\cite{Piroli2019}, hence the nonmaximal entanglement generation despite the dual-unitarity of the circuit.} All three classes have an identity (or $R_Z[\pi/2]$) on one or both legs. Choosing the identity gate to be $v_+$ without loss of generality, this yields:
 \begin{equation}\label{eq:poor}
     M(\mathbbm{1},v_-) = \begin{pmatrix}
     \begin{pmatrix}
     u & 0 \\ u & u
     \end{pmatrix} &
     u\mathbf{a} M_{v_-} \\
     \begin{pmatrix}
     0 & 0 \\ 1 & 0
     \end{pmatrix} & \mathbf{b} M_{v_-}
     \end{pmatrix}
 \end{equation}
Regardless of $v_-$, this automaton has a glider observable, $\vec{\xi}(Z_1)$, with eigenvalue $u$. In the "centered" frame (replacing $M_{shift}$ with $\tilde{M}_{shift}$ [\autoref{eq:center}]), the glider formally has eigenvalue $u^{1/2}$, so after two layers (one full time step) $Z_1^{(n)}$ shifts to $Z_1^{(n+1)}$, where $\sigma_j^{(n)}$ denotes the Pauli operator $\sigma$ on the $j$th site of the $n$th unit cell.

The presence of gliders provides some explanation for why the entanglement generated by these circuits is submaximal. Recall from~\autoref{sect:corr} that in any dual-unitary circuit, the two-point correlations at infinite temperature, which are nonvanishing only on the boundary of the light cone $x=\pm v t$, can be decomposed in terms of left and right quantum channels $\mathcal{M}_\pm$~\cite{Bertini2019}. All conserved charges are gliders, with eigenvalue $1$ for one of the channels, and since the product of gliders moving in the same direction is also a glider, the presence of one glider implies infinitely many~\cite{Borsi2022}. {Thus, a circuit for which some but not all of the eigenvalues are equal to 1 is generally interacting but non-ergodic, with some dynamical correlations remaining constant~\cite{Bertini2019}; see~\appref{app:corr} for more details. In fact, Ref.~\cite{Bertini2020soliton} proves that the \textit{only} square-lattice circuits supporting moving one-site gliders (referred to as "moving ultralocal solitons") are dual-unitary. Our poor scramblers are Clifford examples of the explicit formulas for glider-supporting gates in that work.}

In the present context, any $Z$ operator initialized on only odd sites gets shifted, but does not spread, under the action of the circuit. In particular, any initial product state with $Z$ stabilizers on all odd sites remains a product state at all times. On the other hand, if the initial product state is generated by only $X$ and $Y$ stabilizers, then it can become maximally entangled, but immediately starts to lose entanglement to return to a product state before the next recurrence.

The full details on the poor scramblers are provided in~\appref{app:poor}. Here we just introduce the simplest of the classes, the bare iSWAP:
\begin{equation}
    (v_+, v_-) = (\mathbbm{1}, \mathbbm{1}).
\end{equation}
The centered automaton after two layers is:
\begin{equation}\label{eq:centered-M-iswap}
    \tilde{M} = \ui M^2 = \begin{pmatrix}
    u & 0 & 0 & 0 \\
    0 & u & 1+u & 0 \\
    0 & 0 & \ui & 0 \\
    1+\ui & 0 & 0 & \ui \\
    \end{pmatrix}.
\end{equation}
Since the iSWAP gate preserves the symmetry between $X$ and $Y$, the changes of basis in~\autoref{eq:similar} exactly preserve the matrix, or in other words, there is only one unique circuit in this class. This is just a manifestation of the self-octa-unitarity of the bare iSWAP automaton.

Owing to this self-octa-unitarity, since the automaton is reflection-symmetric, not only is $Z_1$ a glider with eigenvalue $u$, but $Z_2$ is a glider with eigenvalue $\ui$. Moreover, this matrix can be made block diagonal, with $(X_1,Z_2)$ forming one block and $(Z_1, X_2)$ forming another block:
\begin{equation}\label{eq:block-iswap}
    \begin{pmatrix}
    \begin{pmatrix}
    u & 0 \\ 1+\ui & \ui
    \end{pmatrix} & {\bigzero} \\
    \bigzero & \begin{pmatrix}
    u & 1 + u \\ 0 & \ui
    \end{pmatrix}
    \end{pmatrix}
\end{equation}
Neither block is a symplectic matrix, so we cannot use the machinery for $a=1$ CQCA. However, it is worth noting that each block has the same trace, $u + \ui$, and determinant, $1$, as the class of one-step gliders with $a=1$, which can all be mapped to the "standard glider," $\mathbf{g}=\begin{pmatrix} 0 & 1 \\ 1 & u + \ui \end{pmatrix}$~\cite{Gutschow2010long}. Thus, the characteristic polynomial of $\tilde{M}$ is:
\begin{equation}\label{eq:char-poly-glider}
\chi_{\tilde{M}}(y) = (y^2 + (u + \ui) y + 1)^2 = \chi_{\mathbf{g}}(y)^2,
\end{equation}
and \gs{the two automata share the same minimal polynomial, $\mu_{\tilde{M}} = \mu_{\mathbf{g}} = \chi_{\mathbf{g}}$.} Thus, $\tilde{M}$ satisfies the same recursion relation as the standard glider automaton. {This leads to similarities in the operator spreading of initially local Pauli strings: some operators are gliders, while others fill the lightcone in a periodic pattern~\cite{Gutschow2010solo,Gutschow2010fractal}.}

{Another perspective on the iSWAP circuit is as implementing a free fermion Floquet operator, the massless Dirac QCA~\cite{DAriano2012}, via Jordan-Wigner transformation~\cite{Terhal2001}. Thus, the iSWAP automaton is in fact non-interacting and integrable. Free fermion QCA are discussed in more depth in~\appref{app:matchgate}.}

\subsection{"Good scramblers"}
The three remaining classes exhibit a nonlinear structure in $\tau(m)$, and generate Page curves with slope 1 on random initial product states in between the recurrences. Since neither $v_+$ nor $v_-$ is the identity gate, automata in these classes have no single-site gliders~\cite{Bertini2020soliton}. Instead, they have more in common with the fractal class of $a=1$ automata. A notable exception, however, is the dense good scrambling class introduced in~\autoref{sect:circuit8}, which we revisit before discussing the two classes with fractal structure.

\subsubsection{Nonfractal good scrambling class}
In~\autoref{sect:circuit8}, we highlighted the entanglement and error correction properties of the circuit with single-qubit gates:
\begin{equation}
    (v_+, v_-) = (R_X[\pi/2], R_X[\pi/2]).
\end{equation}
Referring to~\autoref{fig:point-group} confirms that this circuit is self-octa-unitary. The corresponding matrix is, after two layers,
    \begin{equation}
       \tilde{M} = \begin{pmatrix}
            0 & u & u & u \\
            u + 1 & u + 1 & 0 & 1 \\
            \ui & \ui & 0 & \ui \\
            0 & 1 & \ui + 1 & \ui + 1
        \end{pmatrix}.
    \end{equation}

Reflection and time reversal symmetry manifest in the characteristic polynomial (cf.~\autoref{eq:char-poly-LR} and~\autoref{eq:char-poly-TR}):
\begin{equation}
    \chi_{\tilde{M}}(y) = y^4 + (u + \ui) y^3 + (u^2 + 1 + u^{-2}) y^2 + (u + \ui) y + 1.
\end{equation}
Other members of this class, generated by the transformations~\autoref{eq:similar}, have $R_Y[\pi/2]$ instead of $R_X[\pi/2]$ on the left and/or right leg. 

\subsubsec{Fractal $d_f = 1.9$ class}
A second good scrambling class contains the circuit with single-qubit gates:
    \begin{equation}\label{eq:df-1.9}
        (v_+, v_-) = (R_X[\pi/2], R_{(1,1,1)}[-2\pi/3]).
    \end{equation}
After two layers,
\begin{equation}\label{eq:M5}
    \tilde{M} = \begin{pmatrix}
    0 & u & 0 & u \\
    u + 1 & u + 1 & 1 & 0 \\
    \ui & \ui & \ui & 0 \\
    0 & 1 & 0 & 1 + \ui
    \end{pmatrix}.
\end{equation}
Owing to left-right asymmetry, members of this class generate asymmetric fractal patterns. The similarity transformations amount to changing out $R_X[\pi/2]$ with $R_Y[\pi/2]$ and/or reversing the direction of the second-qubit cyclic permutation.

The characteristic polynomial, which is also the minimal polynomial, is:
\begin{equation}
    \chi_{\tilde{M}}(y) = y^4 + u y^3 + (u^2 + 1 + u^{-2})y^2 + \ui y + 1.
\end{equation}
{Since $\chi_{\tilde{M}}(y)$ is not invariant under either time reversal or left-right reflection, the automaton itself is not symmetric under these transformations. While the characteristic polynomial \textit{is} invariant under their composition (inversion), inversion is only a weak self-duality of the automaton itself (or the corresponding unitary), as is reflection through the $+$ diagonal, since $M_{v_-} = S M_{v_-^T} S^{-1}$ (\autoref{eq:cob}).} The only strong self-duality is under reflection through the downward-sloping ($-$) diagonal, which maps $(v_+, v_-)\rightarrow (v_+^T, v_-)$.

The fractal pattern of this class is not present in the $a=1$ automata; the inherent asymmetry of odd and even sites makes it fundamentally $a=2$. For example, the image of $Z_1 Z_2$ (\autoref{fig:ZZ-idx9}) is asymmetric even though the initial operator is reflection invariant with respect to the center of the unit cell. The cumulative number of $X$, $Y$, and $Z$ Paulis within the footprint of $Z_1Z_2(t)$ all scale with the same fractal dimension.

To determine the fractal dimension more precisely, we leverage one useful commonality with $a=1$, which is that much of this fractal structure can be seen just by studying the evolution of the trace. In Ref.~\cite{Berenstein2021}, the fractal structure of the CNOT automaton (which also has $a=2$, but is not dual-unitary) is deduced from the pattern of nonzero coefficients of powers of $u$ in the expansion of $\tr(\tilde{M}^t)$. Applying the same technique here, we find that the footprint of $\tr(\tilde{M}^t)$ appears as a "black-and-white" version of the colored spacetime diagram (\autoref{fig:trace-fractal}). Then, the fractal dimension $d_f$ can be inferred numerically from the scaling of the number of nonzero coefficients $N(t)$:
\begin{equation}
    \sum_{t'\leq t} N(t') \propto t^{d_f}.
\end{equation}
A fit up to $t=2^{14}$ yields:
\begin{equation}
    d_f = 1.90 \pm 0.01.
\end{equation}
\begin{figure}[t]

    \centering
    \subfloat[]{
    \includegraphics[width=\linewidth]{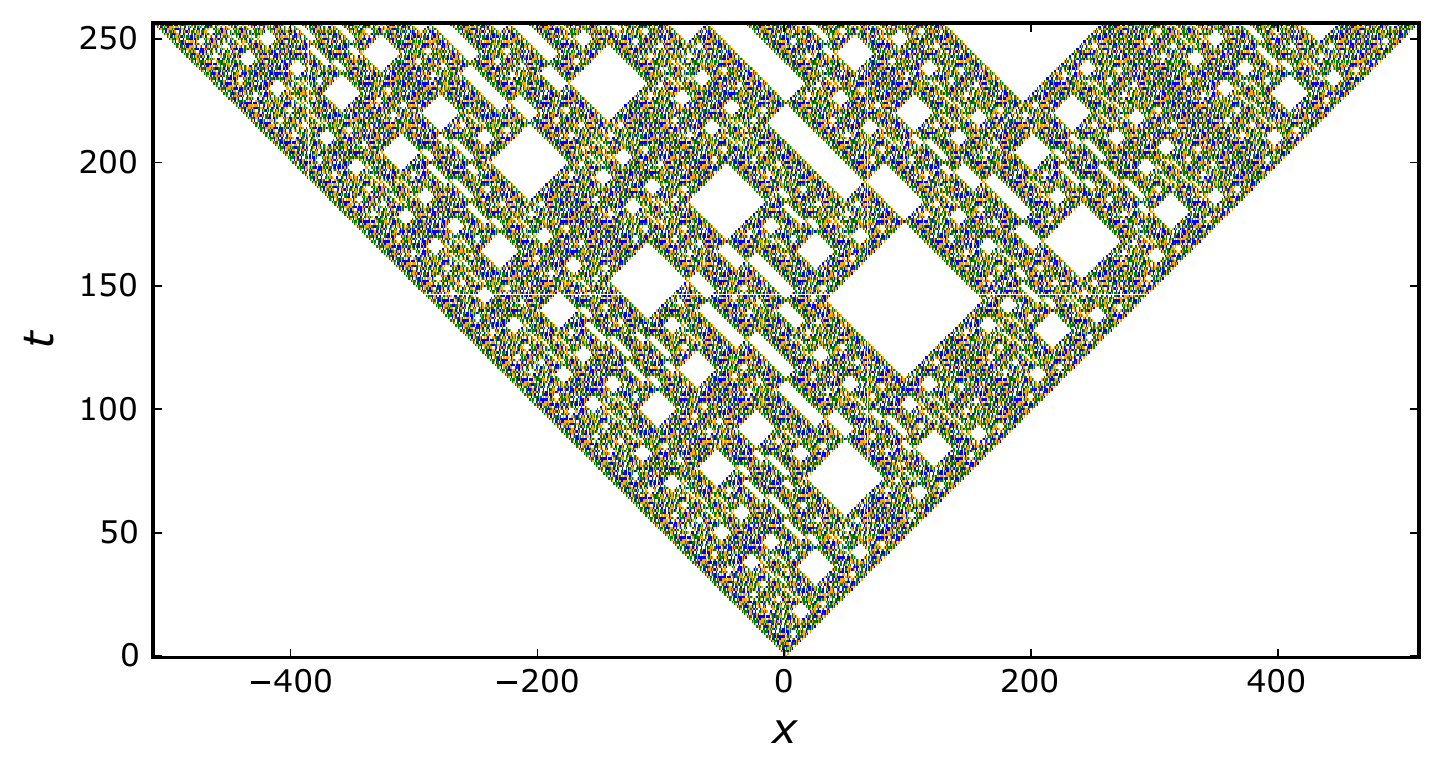}\label{fig:ZZ-idx9}
    } \\
    \subfloat[]{
    \includegraphics[width=\linewidth]{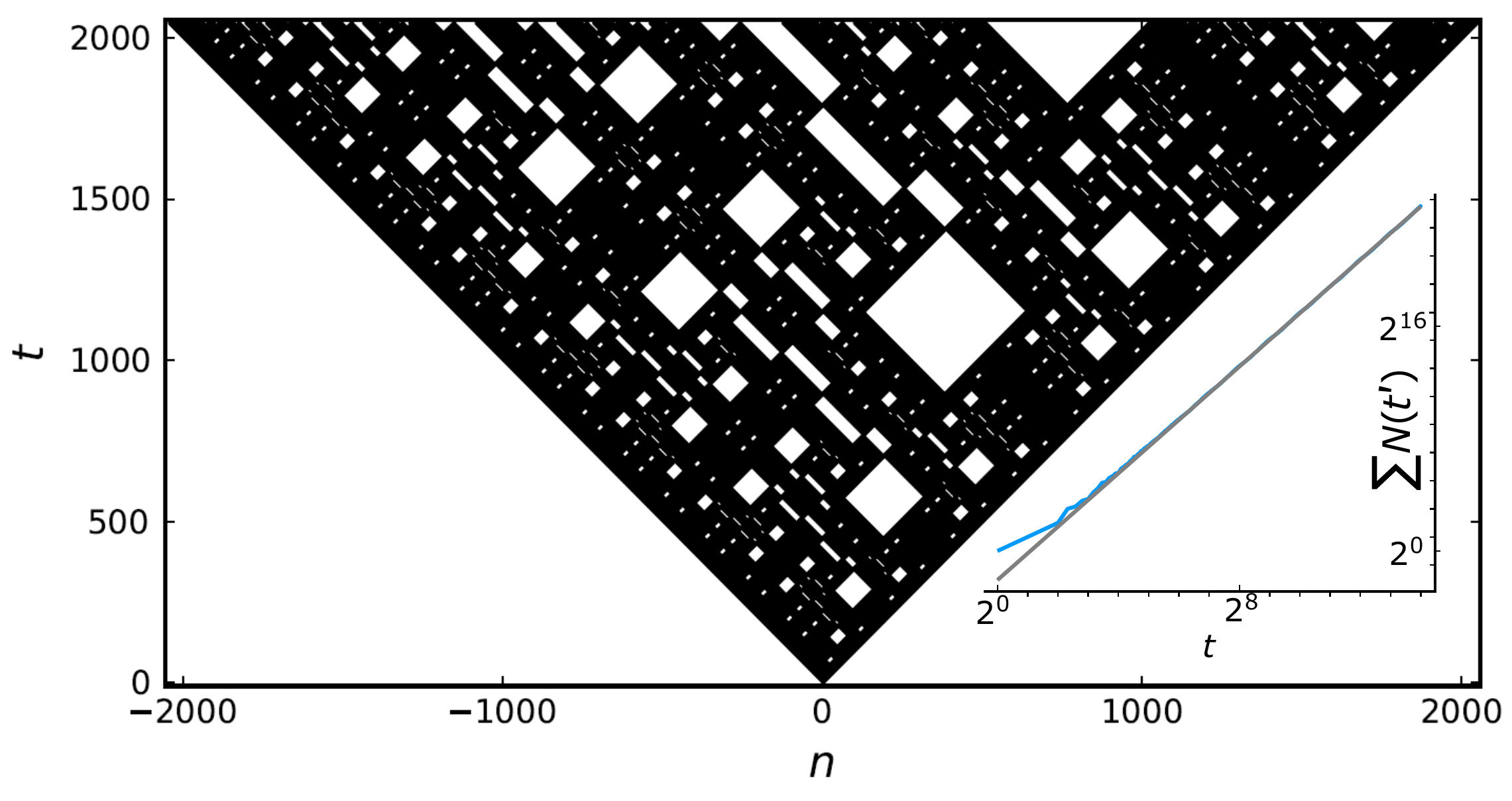}\label{fig:trace-fractal}}
    \caption{(a) Image of $Z_1 Z_2$ at integer time steps up to $t=256$ under the automaton~\autoref{eq:M5}. Blue, green, and orange pixels {cannot be individually distinguished but} correspond to $X$, $Y$, and $Z$, which follow similar fractal patterns. (b) Visual depiction of $\tr(\tilde{M}^t)$ up to $t=2048$. A black pixel at $(n,t)$ indicates that the coefficient of $u^n$ in $\tr(\tilde{M}^t)$ is 1. Inset: Power law fit to $\sum_{t\leq t'} N(t')$ yields $d_f=1.90(1)$.}
\end{figure}
  
\subsubsec{SDKI class}\label{sect:sdki}
The third good scrambling class has the deepest connections to $a=1$ CQCA,~as well as to a minimal model of maximal quantum chaos, the self-dual kicked Ising (SDKI) model~\cite{Akila2016,Bertini2018,Bertini2019entanglement,Gopalakrishnan2019}. The kicked Ising model is described by the Floquet unitary
\begin{equation}\label{eq:KI-floquet}
    \mathbbm{U}_{KI} = e^{-i b \sum_j X_j} e^{-i \sum_j J (Z_j Z_{j+1} + h_j Z_j)}
\end{equation}
 It is dual-unitary along the self-dual line $|J|=|b|=\pi/4$, and ergodic for any nonzero longitudinal field $h_j$~\cite{Bertini2018}.  Along the entire self-dual line, the entanglement velocity is maximal, implying a flat line tension in the membrane picture~\cite{Zhou2020}. 
 
Focusing on the Clifford point $h_j=h=\pi/4$, the SDKI model maps via a boundary circuit~\cite{Bertini2019} to
a representative automaton of this class, which has:
\begin{equation}\label{eq:sdki-U}
    (v_+, v_-) = \left(R_{(1,1,1)}[-2\pi/3], R_{(1,1,1)}[-2\pi/3]\right).
\end{equation}
Since $R_{(1,1,1)}[-2\pi/3]\neq R_{(1,1,1)}[-2\pi/3]^T$, this circuit is weakly self-dual under all point group transformations but is strongly invariant under only one, left-right reflection. {Indeed, all automata in this class---obtained from the representative~\autoref{eq:sdki-U} through the similarity transformations~\autoref{eq:similar}---are strongly symmetric under at most one kind of reflection, horizontal or vertical} \footnote{{Explicitly, since the similarity transformations~\autoref{eq:similar} reverse the sign of rotation on the single-qubit gates, which up to Paulis is equivalent to taking a transpose (\autoref{eq:cob}), another representative of this class has $(v_+, v_-) =R_{(1,1,1)}[2\pi/3],R_{(1,-1,1)}[2\pi/3]$. That circuit, for which $v_+ = v_-^T$, has symmetry under time reversal but no longer has left-right symmetry.}}.

It should be noted that there is a different way of decomposing the Floquet unitary from~\autoref{eq:KI-floquet} into a brickwork circuit ~\cite{Gopalakrishnan2019,Zhou2020}:
\begin{align} 
U_{KI} = &
e^{-i J Z_1Z_2 - i(h_1Z_1 + h_2Z_2)/2}e^{-i b (X_1 + X_2)} \notag \\
&e^{-i b Z_1Z_2 - i(h_1Z_1 + h_2Z_2)/2}.
\end{align}
This representation is strongly self-octa-unitary at the self-dual point with homogeneous $h$, as pointed out in~\cite{Mestyan2022}. However, this choice of gate is not Clifford~\footnote{Explicitly, $U_{KI}[J=\pi/4,b=\pi/4, h=\pi/4] = -(v \otimes v) U (Z v^\dag \otimes Z v^\dag)$, where $U=\iSWAP (R_{(1,1,1)}[-2\pi/3] \otimes R_-{(1,1,1)}[-2\pi/3])$ is the Clifford gate used as a representative of this class, and $v = R_Z[\pi/4] R_{(1,1,1)}[2\pi/3]$, which is not a Clifford gate.}.

As a 4x4 matrix, our chosen representative (\autoref{eq:sdki-U}) is:
\begin{equation}\label{eq:sdki-half}
M = \begin{pmatrix}
0 & u & 0 & 0 \\
u & 0 & 0 & u \\
0 & 0 & 0 & 1 \\
0 & 1 & 1 & 0
\end{pmatrix}.
\end{equation}
Once again, it is useful to consider the evolution of the centered automaton after two layers:
\begin{equation}\label{eq:sdki-class}
   \tilde{M} = \begin{pmatrix}
        u & 0 & 0 & u \\
        0 & u + 1 & 1 & 0 \\
        0 & \ui & \ui & 0 \\
        1 & 0 & 0 & 1 + \ui
    \end{pmatrix}.
\end{equation}
As with the bare iSWAP class, permuting rows and columns brings $\tilde{M}$ into block-diagonal form. Explicitly, $Z_1$ and $X_2$ form one block, and $X_1$ and $Z_2$ form another block, so an operator that starts with $Z$'s supported only on odd sites, for example, can only spread to a product of $Z$'s on odd sites and $X$'s on even sites. This is shown in~\autoref{fig:sdki} for the initial Pauli string $Z_1$. The block-diagonal matrix is:
\begin{equation}\label{eq:block-sdki}
    \begin{pmatrix}
    \begin{pmatrix}
    u & u \\ 1 & 1 + \ui
    \end{pmatrix} & {\bigzero} \\
    \bigzero & \begin{pmatrix}
    1 + u & 1 \\ \ui & \ui
    \end{pmatrix}
    \end{pmatrix}.
\end{equation}
Again, neither block is a valid CSCA, since they connect $X$'s and $Z$'s on opposite parity sites. However, writing $\tilde{M}$ in this form elucidates the connection to the SDKI model at the Clifford point, which as an $a=1$ automaton is~\cite{VonKeyserlingk2018}:
\begin{equation}\label{eq:M-SDKI}
    M_{SDKI} = \begin{pmatrix}
    \ui + u & 1 \\
    \ui + 1 + u  & 1
    \end{pmatrix}.
\end{equation}
This has the same characteristic polynomial as each block of~\autoref{eq:block-sdki}, and indeed 
\begin{equation}
    \chi_{\tilde{M}}(y) = (y^2 + (u + 1 + \ui)y + 1)^2 = \chi_{SDKI}(y)^2
\end{equation}
\gs{with $\tilde{M}$ and $M_{SDKI}$ sharing the same minimal polynomial, $\mu_{\tilde{M}}=\mu_{SDKI} = \chi_{SDKI}$.} \gs{Since SCA with the same minimal polynomial share a common fractal structure in their colored spacetime diagrams, i.e. the footprints of time-evolved initially local operators, the systems described by~\autoref{eq:sdki-class} and~\autoref{eq:M-SDKI} both have fractal dimension $d_f = \log_2[(3+\sqrt{17})/2]=1.8325...$, analytically determined in Ref.~\cite{Gutschow2010fractal} for another automaton with the same minimal polynomial.}

\begin{figure}[t]
    \centering
    \includegraphics[width=\linewidth]{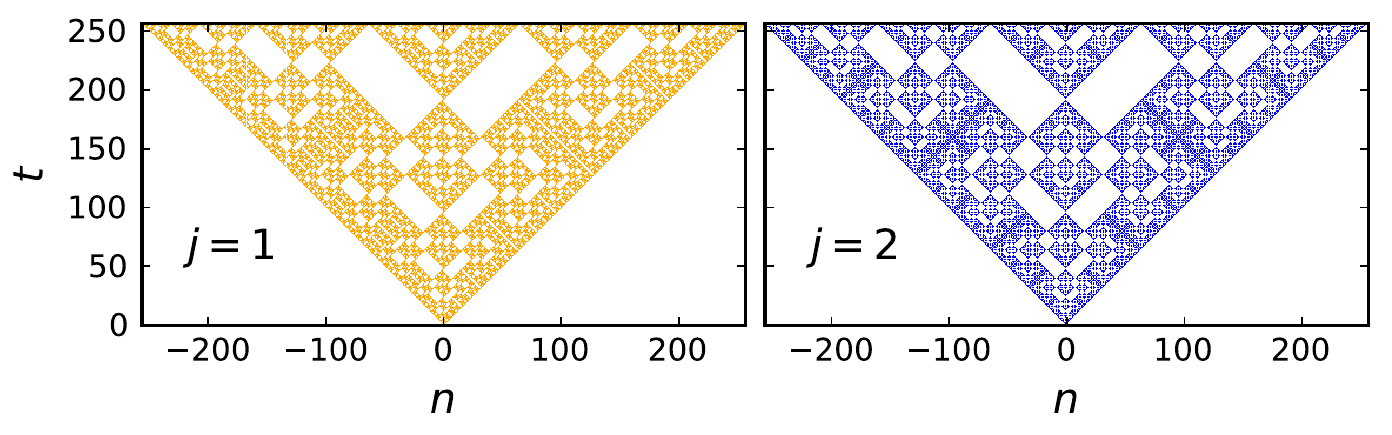}
    \caption{Time evolution from the initial operator $Z_1$ under~\autoref{eq:sdki-class}, split into odd (left) and even (right) sites. Owing to the block diagonal form of~\autoref{eq:block-sdki}, the image on odd sites is only $Z$'s (orange), while the image on even sites is only $X$'s (blue).}
    \label{fig:sdki}
\end{figure}

Therefore, just as the bare iSWAP class can be thought of as the natural $a=2$ descendant of the standard glider class, the SDKI class acts as the $a=2$ descendant of the simplest $a=1$ fractal class. {One remarkable feature of the standard fractal $a=1$ automaton examined in~\cite{Gutschow2010fractal} is that, if the Floquet operator for one step of the automaton is written as the exponential of a (non-unique) time-independent Hamiltonian $H$, then any choice of $H$ is non-local in a strict sense, i.e. the interactions do not decay with distance~\cite{Zimboras2022}. Consequently, conserved operators are also non-local. In contrast, unitary evolution of the bare iSWAP class, which maps onto free fermions, is generated by a time-independent Hamiltonian with algebraic decay of interactions~\cite{Zimboras2022}; see~\appref{app:matchgate}.} 

On the other hand, the $a=2$ SDKI class provides a case study for the ways in which $a=2$ automata can depart from the $a=1$ automata studied previously. Recall that for $a=1$ the only class that has either stationary translation-invariant stabilizer states or stationary product states (other than the fully mixed state) is the periodic class~\cite{Gutschow2010long}. In contrast, while a random pure product state becomes entangled when fed into the $a=2$ "SDKI-class" circuits, this class also has translation-invariant product stabilizer eigenstates. 
In particular, the state stabilized by $Z_1,X_2$ and all their translates, as well as its mirror image stabilized by $X_1^{(n)}, Z_2^{(n)}$, is stationary under two layers of the circuit.

To see this in the CQCA formalism, represent the translation-invariant stabilizer group $\mathcal{S}$ as a $4\times 2$ matrix $S$ over $\mathbbm{F}_2[u,\ui]$, where the $i$th column is the vector of polynomials corresponding to the $i$th generator. Under one step of the CQCA $\tilde{M}$, the generators evolve to $\tilde{M} S$. Individual generators can scramble while leaving the total group invariant, so to check for the invariance of the group, we perform row reduction on $(\tilde{M} S)^T$. For $\tilde{M}$ given by~\autoref{eq:sdki-class} and the initial group $\mathcal{S} = \langle X_1^{(n)}, Z_2^{(n)}\rangle$, this yields:
\begin{align}
    \tilde{M} \begin{pmatrix} 1 & 0 \\ 0 & 0 \\ 0 & 0 \\ 0 & 1 \end{pmatrix} = \begin{pmatrix} u & u \\ 0 & 0 \\ 0 & 0 \\ 1 & 1+ \ui \end{pmatrix} \Rightarrow \begin{pmatrix} 1 & 0 \\ 0 & 0 \\ 0 & 0 \\ 0 & 1 \end{pmatrix}.
\end{align}

\subsection{Good quasicyclic codes}\label{sect:good}
Two of the three good scrambling classes---the nonfractal class and the $d_f\cong 1.9$ class---are especially promising for quantum error correction. As demonstrated in~\autoref{fig:d1} and the surrounding discussion, the dense good scrambling class generates finite-rate codes with linear-in-$m$ code length $d_1$ for random initial product states. In fact, this property is enjoyed by all three good scrambling classes. 

We now make two further demands. First, rather than starting from a \textit{random} product state of some entropy density, consider the action of the circuit on translation-invariant product states of code rate $1/2$. The spatial periodicity of the automaton guarantees that such states remain translation-invariant at all times; for spatial period $a>1$, the resulting codes are known as quasicyclic codes~\cite{Lally2001,Guneri2020}. Existing decoding techniques for cyclic quantum codes~\cite{Grassl1999} and (quasi)cyclic classical codes~\cite{Feng1989,Semenov2012,Zeh2014,Mitchell2014} could prove useful for finding a decoder for our codes.

Restricting to translation-invariant product states gives 6 choices for the initial state, generated by $\sigma^{(n)}_j$ for $j=1,2$, $\sigma=X,Y,Z$. When any of these initial states is fed into a circuit in the SDKI class, the code length remains $O(1)$ to late time. On the other hand, members of the $d_f\cong 1.9$ class and nonfractal class are able to generate linear-in-$m$ code length, albeit with more frequent recurrences of short code length than for random initial states.

Second, for assessing the performance of the resulting codes under realistic noise models, the relevant metric is the code distance $d$, for which $d_1$ is only an upper bound. For a given circuit and initial state, consider the code defined by a snapshot of the system at the time when $d_1$ is maximized. While the distance $d$ of the resulting code is exponentially hard to compute, we can get a sense for its performance compared to random codes by subjecting it to erasures. For this simple error model, {an optimal decoder of cubic complexity is known~\cite{Delfosse2016,Delfosse2020}}, and the failure probability $P_F$ of the decoder can be efficiently computed~\cite{Gullans21}. Let $P_F(e,s,L)$ denote the failure probability for a code of rate $s$ on $L$ qubits, where the erasures are applied at random locations on a fixed fraction $e$ of the sites. For random codes, this quantity is well modeled by random matrix theory, and decays exponentially in $L$ for error rates far below threshold:
\begin{equation}
    P_F(e,s,L) \propto 2^{-2L(e_c-e)-1}, \quad e\ll e_c
\end{equation}
where the error threshold $e_c=(1-s)/2$~\cite{Gullans21}.

\begin{figure}[t]
    \centering
    \includegraphics[width=\linewidth]{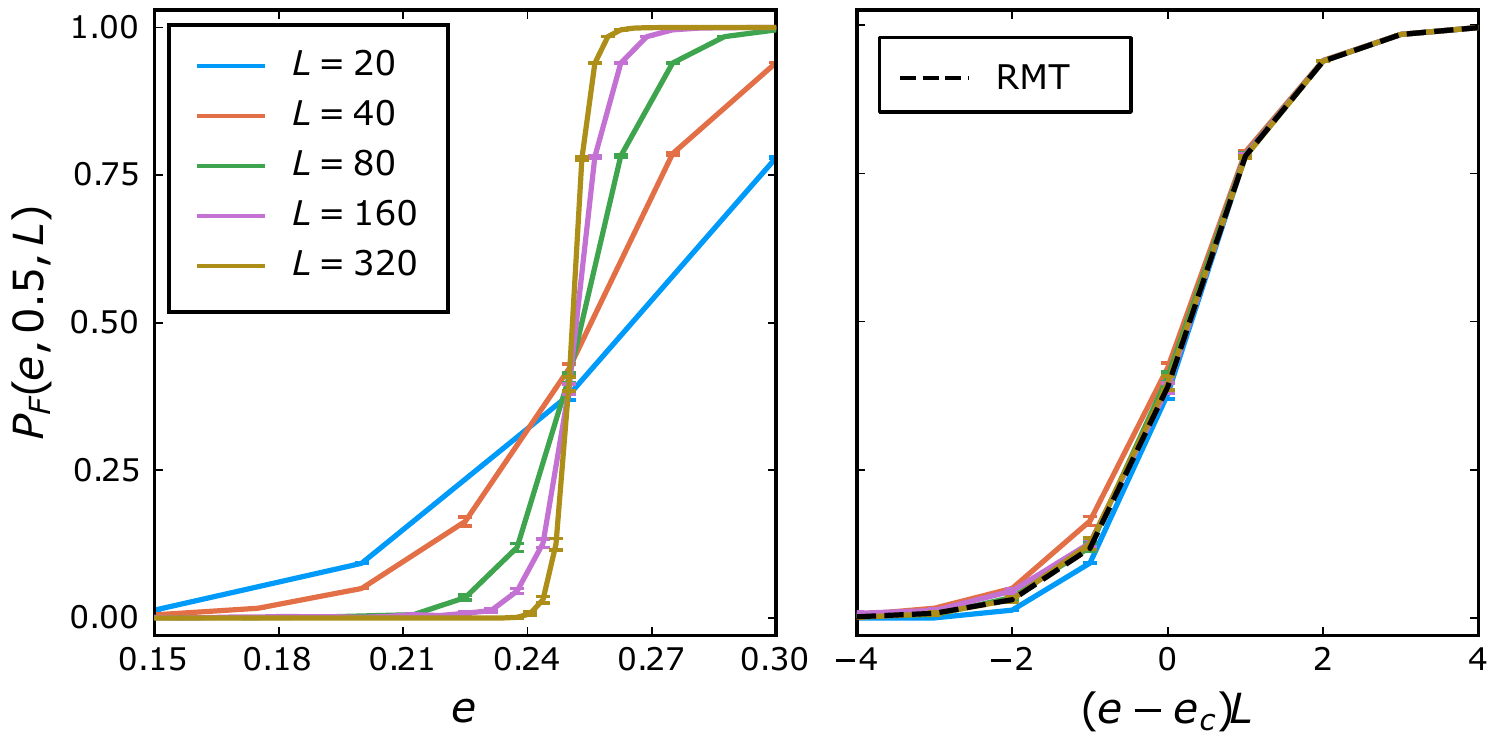}
    \caption{Failure probabilities for quasicyclic codes generated by a dense good scrambling circuit starting from the product state $\langle Z_1^{(n)}\rangle$ with code rate $1/2$, as a function of erasure rate, for $L=20,40,80,160,320$ qubits. Right: scaling collapse vs. $(e-e_c)L$. Black dashed line is the random matrix theory prediction for $L=320$. At least 1000 samples are taken at each erasure rate.}
    \label{fig:threshold}
\end{figure}

To evaluate the quasicyclic codes generated by good scrambling circuits, we first ask whether they achieve the optimal threshold.~\autoref{fig:threshold}, which shows the failure probabilities for quasicyclic codes produced by a dense good scrambling circuit for the initial $Z$ product state at code rate $1/2$, subject to randomly placed erasures, answers in the positive for the sequence of system sizes $L=10 \cdot 2^k$. Not only does the threshold saturate the bound $e_c=1/4$, but a scaling collapse of the form $P_F(e,s,L) = f((e-e_c)L)$ is consistent with random matrix theory (right panel). Similar results are obtained for other initial periodic states and for circuits in the $d_f\cong 1.9$ class.

Backing away from the threshold, we collect $\gtrsim 10^7$ samples at each system size to get a more precise estimate of the subthreshold failure probability at a fixed erasure rate of $e=0.75e_c$. As shown in~\autoref{fig:subthreshold}, the codes produced by the $d_f\cong 1.9$ class and nonfractal class are competitive with random codes for a wide range of $L$, but exhibit sharp peaks in $P_F$ for certain system sizes. Spikes in the failure probability are associated with system sizes for which the chosen snapshot of the system, despite having large code length, has poor code distance---an exception to the general trend that higher code length is correlated with lower failure rates. This poor performance can be avoided by restricting to certain system sizes (odd $m$ tend to fare better, and have fewer recurrences) or by monitoring the performance under erasures for the sequence of codes generated in time rather than just choosing the snapshot with maximum code length.

\begin{figure}[t]
    \centering
    \includegraphics[width=\linewidth]{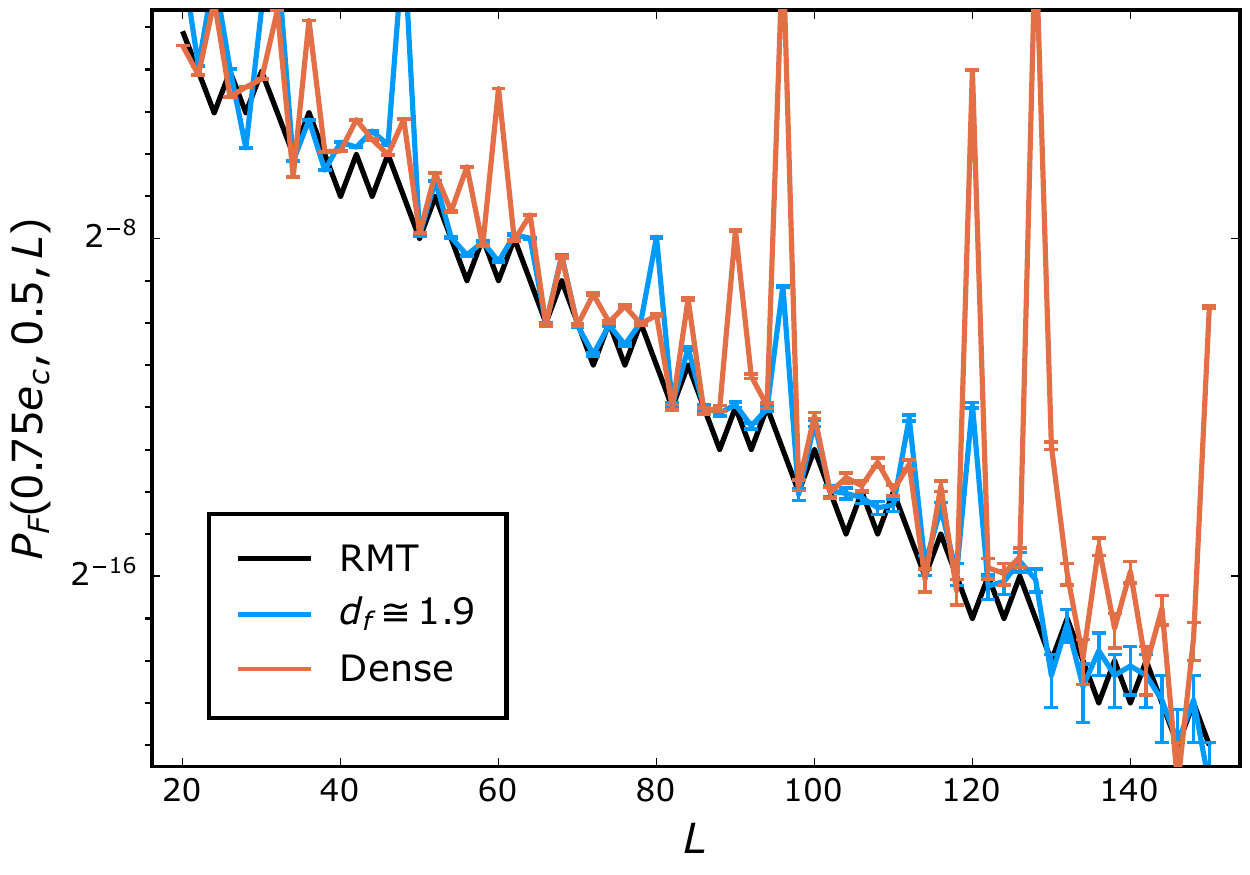}
    \caption{Failure probabilities at $e\approx 0.75e_c=0.1875$, for quasicyclic codes produced by a representative of the $d_f\cong 1.9$ class (blue) and dense good scrambling class (orange) compared to random codes (modeled by random matrix theory, black). Wiggles in the black curve come from rounding the number of erasures to the nearest integer.}
    \label{fig:subthreshold}
\end{figure}

{The astute reader may question our earlier emphasis on the nonfractal operator spreading in the dense good scrambling class, given that the $d_f \cong 1.9$ class appears to perform just as well, and in fact has less dramatic spikes in the failure probability. Thus, some clarifying points are in order. First, while we have defined the fractal dimension through the cumulative weight of Pauli operators spreading in spacetime, the code distance is concerned with the lowest weight of a logical operator at a specific time slice. If the fractal dimension is $d_f$, there must exist a sequence of time slices for which the Pauli weight grows at least as fast as $t^{d_f-1}$. For the $d_f\cong 1.9$ class, there is no sequence of times where the scaling is linear, but it is close enough that with the presently accessible system sizes we cannot distinguish the subthreshold scaling from that of a linear distance code. Moreover, the code distance for codes generated from a specific set of initial states is not necessarily monotonic in either the spacetime fractal dimension or the operator scaling along particular time slices. Even if $d_f$ is well above 1, the code distance may fail to grow at all, as is the case for circuits in the SDKI class when fed initial states with period $a=2$.}

\subsection{Dual-unitarity and beyond}\label{sect:dual-unitary}
{Throughout this section, we have noted several features of our automata that are general to dual-unitary circuits. Here we summarize these features and compare the iSWAP-core automata to those without dual-unitarity.}

{One key feature of dual-unitary circuits is their ability to saturate the minimal cut bound (\autoref{eq:min-cut}) on entanglement, and the existence of certain initial states for which this saturation is known to be exact in the limit of infinite system size at all times~\cite{Bertini2019entanglement,Piroli2019}. Numerically, we observe that the entanglement in our good scrambling circuits increases at a near-maximal rate starting from random pure product states. The significant suppression of entanglement growth in poor scrambling circuits acting on random product states, as well their complete failure to generate entanglement on certain translation-invariant Floquet eigenstates, does not violate any proven analytical results, since these initial states do not belong to the class of solvable initial states for which the bound is saturated. }

{Historically, the SDKI chain has served as a prototypical model within the broader realm of dual-unitary circuits, and the first for which the entanglement growth (among other quantities) was computed exactly~\cite{Bertini2019entanglement}. It is therefore striking that our $a=2$ CQCA include the closely related SDKI class. For the SDKI model, the class of initial states ("separating states") for which~\autoref{eq:min-cut} is exactly saturated includes product states in the computational basis. Again, the fact that our SDKI automaton admits Floquet product eigenstates is consistent with this result, since these eigenstates, when evolved under the boundary layer relating our automaton to the standard SDKI model (\autoref{eq:KI-floquet}), do not evolve into separating states.}

{Another special feature of dual-unitary circuits is the restriction of two-point correlations of one-site observables to the edges of the lightcone (\autoref{sect:corr}). As detailed in~\appref{app:corr}, the good scrambling classes of iSWAP-core CQCA enjoy an even stronger restriction: two-point correlations of nontrivial one-site operators vanish for all $t\geq 1$ (2 layers of gates). This is as close as we can get with two-qubit gates to the "maximally chaotic" behavior of quantum Bernoulli circuits, for which correlations of one- and even two-site operators vanish for all $t>0$. Such circuits arise when $U$ is a perfect tensor, which is possible for qudit dimension $q\geq 3$, and their ergodicity is robust to one-site gates dressing the legs~\cite{Aravinda2021}. Clearly, the iSWAP gate and its dressings lack this robustness, since the scrambling properties depend on $v_+$ and $v_-$. To wit, as already noted, in the poor scrambling classes (for which $v_+$ and/or $v_-$ is an identity gate), the presence of gliders results in some correlations that are constant in time.}

{Lifting the constraint of dual-unitarity, the only other Clifford gates that produce interacting dynamics are those with a CNOT core. Dividing all CNOT-core automata into classes as we did for the iSWAP-core automata, most classes are minor variations on those we have already encountered: an SDKI-like class and several glider classes, where now the gliders can have velocity other than $\pm 1$ owing to the lack of dual-unitarity (see~\appref{app:matchgate}). Another class contains the bare CNOT automaton, a brickwork circuit of CNOT gates also known as the Clifford East model~\cite{Gopalakrishnan2018}. This circuit preserves Calderbank-Shor-Steane (CSS) codes~\cite{Calderbank1996,Steane1996}---namely, it maps $X$'s to $X$'s and $Z$'s to $Z$'s---and thus its 4x4 matrix is block diagonal in the basis of $(X_1, X_2, Z_1, Z_2)$~\cite{Berenstein2021}:
\begin{equation}\label{eq:cnot-a2}
    M_{CNOT} = \begin{pmatrix}
    \begin{pmatrix} 
    u + 1 & u \\ 1 & 1 
    \end{pmatrix} & \bigzero \\
    \bigzero & \begin{pmatrix}
    1 & 1 \\ \ui & \ui + 1
    \end{pmatrix}
    \end{pmatrix}.
\end{equation}
Whereas the bare iSWAP class is a poor scrambler with gliders, under the action of a circuit in the bare CNOT class, an initially local Pauli string spreads fractally as a Sierpinski gasket, with fractal dimension $d_f=\log_2(3)=1.5824...$~\cite{Gopalakrishnan2018}. {A closely related CNOT-core class with the same minimal polynomial (and hence the same fractal dimension) as the Clifford East model generates quasicyclic codes with the optimal threshold under erasures, despite $d_f$ being lower than the $a=2$ SDKI class for which the code distance remains 1 at all times.} This fractal behavior is not present in our dual-unitary square lattice circuits, but it remarkably appears in the tri-unitary kagome lattice CQCA, to which we now turn.}

\section{Kagome lattice automata}\label{sect:kagome}
Turning to the kagome lattice, we consider three representative examples of the dynamics that occur when there are three (six including time-reversal) choices for the arrow of time.

Recall from~\autoref{fig:kagome} and the surrounding discussion that symmetry under three-fold rotations ("self-tri-unitarity"), imposes $1=3=5$ and $2=4=6$ on the six unique edges within the unit cell. We focus on a subset of self-tri-unitary circuits with an iSWAP core where, like on the square lattice, the single qubit gates on the same diagonals with respect to the core are identical. This corresponds to assigning identical gates to the edges of a common orientation on the kagome lattice, i.e. $3^T=6$, $1^T=4$, and $2=5^T$. When this is combined with $C_3$ symmetry, the resulting circuits are also invariant under the three reflections in ~\autoref{fig:kagome}c. The circuits fall into three classes: those with $\mathbbm{1}$ or $R_Z[\pi/2]$ on each leg, those with $R_X[\pi/2]$ or $R_Y[\pi/2]$ on each leg, and those with cyclic permutation gates on each leg.

Expressing the kagome lattice as a rectangular circuit (\autoref{fig:kagome-brickwork}), now $a=4$ and the corresponding SCA are 8x8 matrices. As on the square lattice, we could use a smaller unit cell by incorporating a shift, $(T=1, a=4, d=2)$, but the evolution is somewhat clearer if we just use $(T=2,a=4,d=0)$.  The three classes exhibit some notable similarities to automata with smaller $a$, indicating a latent connection to circuits with simpler geometries.

\begin{figure}[t]
    \centering
    \includegraphics[width=\linewidth]{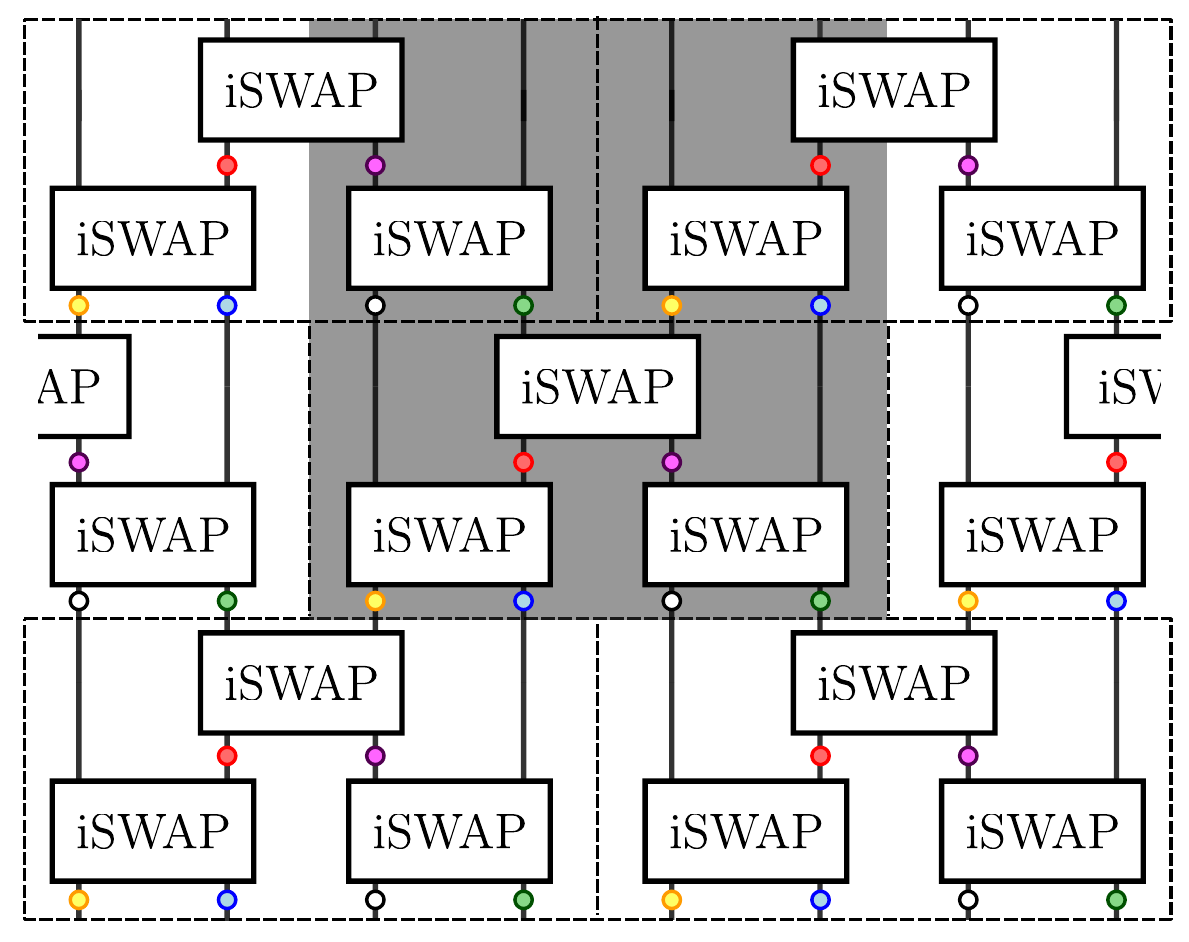}
    \caption{Kagome lattice expressed as a rectangular circuit. With time oriented in the vertical direction, the circuit is composed of $(T=1, a=4)$ "bricks" (dashed) with a shift by $d=2$ in between time steps. In constructing the automata, we instead use the enlarged unit cell (shaded gray) with $(T=2,a=4,d=0)$.}
    \label{fig:kagome-brickwork}
\end{figure}
\subsection{Bare iSWAP class}
The simplest example has identity gates on all the edges, and thus has the full $D_6$ symmetry. To elucidate the time evolution, we
permute the rows to be the image of $X_1,X_2,X_3,X_4,Z_1,Z_2,Z_3,Z_4$ respectively:
\begin{equation}\label{eq:iswap-kagome}
    T_1 = \begin{pmatrix}
    \mathbf{t_{ZZ}} & \mathbf{0} \\
    \mathbf{t_{ZX}} & \mathbf{t_{ZZ}}
    \end{pmatrix}
    \end{equation}
where
\begin{equation}
    \mathbf{t_{ZZ}} = \begin{pmatrix}
    u & 0 & 0 & 0 \\
    0 & 1 & 0 & 0 \\
    0 & 0 & 1 & 0 \\
    0 & 0 & 0 & \ui
    \end{pmatrix}
\end{equation}
and 
\begin{equation}
    \mathbf{t_{ZX}} = \begin{pmatrix}
0 &   u & u & u + 1 \\
1  &  0 &   0 &    1 \\
1 &   0 &   0 &    1 \\
1 + \ui & \ui & \ui & 0
\end{pmatrix}
\end{equation}
\begin{figure}[t]
    \centering
    \includegraphics[width=\linewidth]{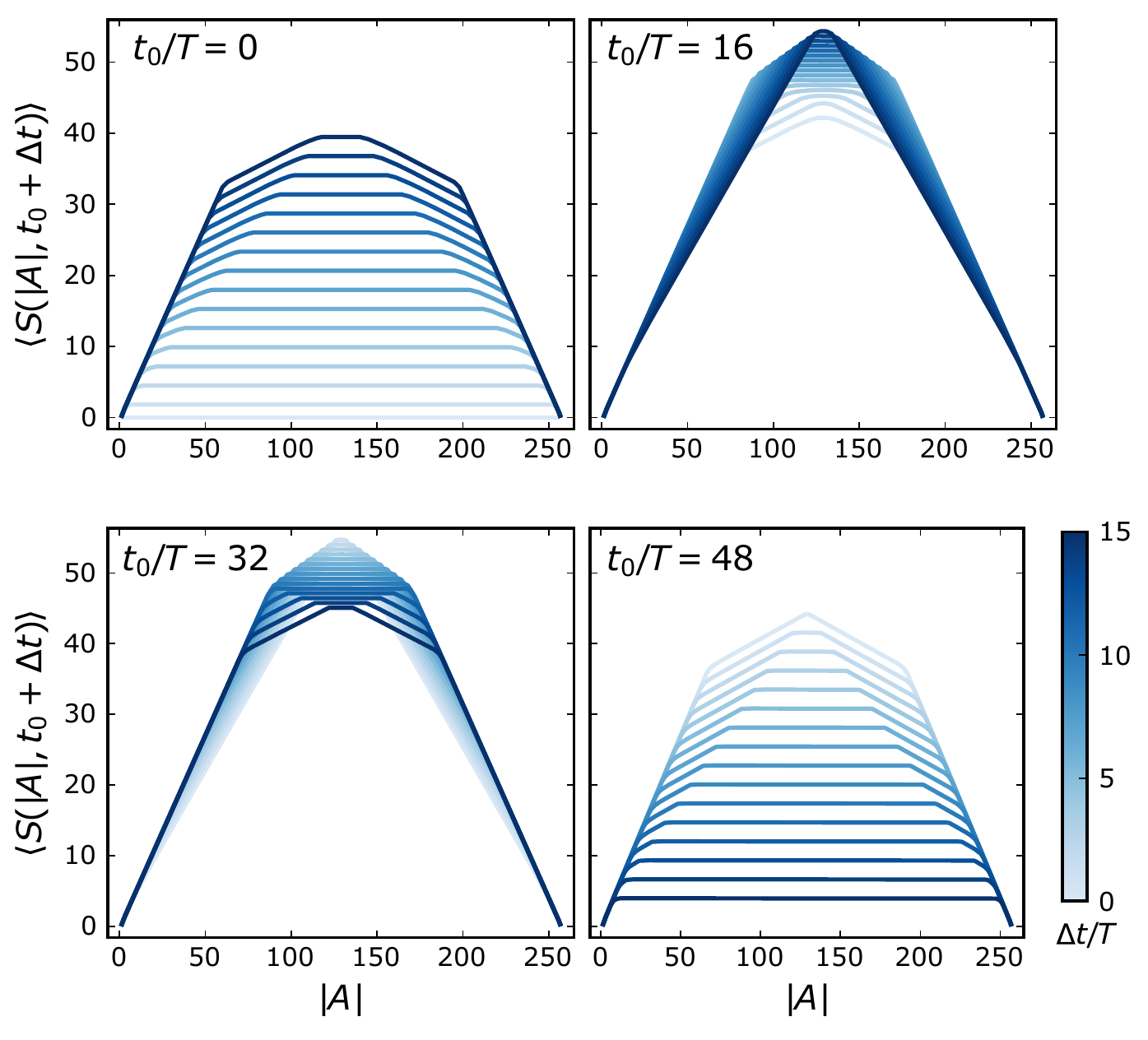}
    \caption{Subsystem entropy $\langle S(|A|,t_0+\Delta t) \rangle$ averaged over contiguous regions of length $|A|$ starting from a random pure product state on $L=256$ qubits, time-evolved under the bare iSWAP kagome circuit (\autoref{eq:iswap-kagome}). Each panel shows 16 time steps, where each time step consists of 4 layers ($T=2$), and darker (lighter) curves correspond to later (earlier) times $\Delta t$ with respect to $t_0$.}
    \label{fig:poor-kagome}
\end{figure}
From the form of $\mathbf{t_{ZZ}}$, we see that all $Z$ strings are (products of) gliders, just like in the iSWAP class on the square lattice. Indeed, this could have been anticipated by recalling that the iSWAP does not produce entanglement on $Z$ eigenstates. But owing to the modified geometry, instead of just left and right movers ($\vec{\xi}(Z_1)$ and $\vec{\xi}(Z_4)$ have eigenvalues $u$ and $\ui$ respectively), there are also "stationary gliders" ($\vec{\xi}(Z_2)$ and $\vec{\xi}(Z_3)$ both have eigenvalue 1). This reflects the different spacetime structure of two-point correlations at infinite temperature in tri-unitary circuits vis \'a vis dual-unitary brickwork circuits as discussed in~\autoref{sect:corr}, namely, the existence of nonvanishing correlations along the static worldline $x=0$. {Conserved charges of $Z$ strings, corresponding to nontrivial eigenvectors of the channels $\mathcal{M}_\pm$ and $\mathcal{M}_0$ with eigenvalue 1, thus place this bare iSWAP circuit in the nonergodic class of tri-unitary circuits~\cite{Jonay2021}.}

Despite the different geometry, this class is similar in spirit to the bare iSWAP class on the square lattice {in three regards. First,} as with all of the poor scrambling classes on the square lattice, random product states do not become maximally entangled. The entanglement generation on a system of $m=64$ unit cells ($L=256$) is shown in~\autoref{fig:poor-kagome}, where the maximum slope of the Page curve is well below 1. Second, the recurrence time is linear in $m$ for all $m$: $\tau(m)=2m$. Finally, {similarly to how the characteristic polynomial of the bare iSWAP class on the square lattice is the perfect square of that of the $a=1$ glider (\autoref{eq:char-poly-glider})}, the characteristic polynomial of the bare iSWAP class on the kagome lattice is also a perfect square:
\begin{equation}
    \chi_{T_1}(y) = (y^4 + (u + \ui) y^3 + (u + \ui) y^2 + 1)^2
\end{equation}
which means that although this matrix is 8x8, its minimal polynomial $\mu_{T_1}(y) = y^4 + (u + \ui) y^3 + (u + \ui) y^2 + 1$ is only degree 4. Note, however, that none of the $a=2$ automata considered in this paper have this as their characteristic polynomial.

\subsection{CNOT-like class}
A second class, which is symmetric under three-fold rotations and the three reflections in~\autoref{fig:kagome}c but none of the other transformations, contains the representative
\begin{align}
    1=3=5&=R_{(1,1,1)}[-2\pi/3], \notag \\
    2=4=6 &= R_{(1,-1,1)}[2\pi/3].
\end{align}
The corresponding automaton is:
\begin{equation}\label{eq:T-cnot}
T_2 = \begin{pmatrix}
\mathbf{t_{11}} & \mathbf{t_{12}} & \mathbf{t_{13}} & \mathbf{t_{14}} \\
\mathbf{t_{21}} & \mathbf{t_{22}} & \mathbf{0} & \mathbf{t_{24}} \\
\overline{\mathbf{t_{24}}} & \mathbf{0} & \overline{\mathbf{t_{22}}} & \overline{\mathbf{t_{21}}} \\
\overline{\mathbf{t_{14}}} & \overline{\mathbf{t_{13}}} & \overline{\mathbf{t_{12}}} & \overline{\mathbf{t_{11}}}
\end{pmatrix}
\end{equation}
where~\footnote{\gs{Although the circuit in~\autoref{eq:T-cnot} does not have gliders, it does have a simplifying feature that the matrices $\mathbf{t}_{nm}$ where $n\in \{2,3\}$ contain only 0's and 1's. This is in fact true of any kagome automaton, and reflects the fact that under one time step ($T=2$) of the automaton, operators that end up on the second and third sites of a given unit cell (shaded in~\autoref{fig:kagome-brickwork}) cannot have originated from beyond that cell.}}:
\begin{subequations}
\begin{align}
    \mathbf{t_{11}} &= \begin{pmatrix}
    u & 0 \\ 
    0 & u+1
    \end{pmatrix}, \quad \mathbf{t_{12}} = \begin{pmatrix}
    u & u \\ 1 & 0
    \end{pmatrix} \\
    \mathbf{t_{13}} &= \begin{pmatrix}
    u & 0 \\ u+1 & u+1 
    \end{pmatrix}, \quad \mathbf{t_{14}} = \begin{pmatrix}
    0 & u \\ u+1 & 0
    \end{pmatrix} \\
    \mathbf{t_{21}} &= \begin{pmatrix}
    0 & 1 \\ 1 & 1
    \end{pmatrix} = \overline{\mathbf{t_{21}}}, \quad \mathbf{t_{22}} = \begin{pmatrix}
    1 & 0 \\ 1 & 0
    \end{pmatrix} = \overline{\mathbf{t_{22}}} \\
    \mathbf{t_{24}} &= \begin{pmatrix}
    0 & 0 \\ 0 & 1
    \end{pmatrix} = \overline{\mathbf{t_{24}}}
\end{align}
\end{subequations}

The symmetry of this circuit under left-right reflection manifests in its automaton as invariance under~\autoref{eq:reflection}, which for $a=4$ reads:
\begin{equation}\label{eq:reflect-4}
    M_{1\leftrightarrow 4, 2\leftrightarrow 3} = \begin{pmatrix}
    0 & 0 & 0 & \mathbbm{1} \\
    0 & 0 & \mathbbm{1} & 0 \\
    0 & \mathbbm{1} & 0 & 0 \\
    \mathbbm{1} & 0 & 0 & 0
    \end{pmatrix}
    \overline{M}
    \begin{pmatrix}
    0 & 0 & 0 & \mathbbm{1} \\
    0 & 0 & \mathbbm{1} & 0 \\
    0 & \mathbbm{1} & 0 & 0 \\
    \mathbbm{1} & 0 & 0 & 0
    \end{pmatrix}.
\end{equation}

While the fractals are different, the behavior of this kagome class is reminiscent of the class on the square lattice with cyclic permutations on each edge. Recall that on the square lattice, the resulting $a=2$ SDKI class can roughly be thought of as decomposing into two copies of the $a=1$ SDKI automaton, in the sense that $\chi_{\tilde{M}}(y)=\chi_{SDKI}(y)^2$. Pauli strings spread with the same fractal dimension as the $a=1$ SDKI automaton, but these fractals are invisible in the footprint of $\tr(\tilde{M}^n)$, used to infer the fractal dimension in Ref.~\cite{Berenstein2021}. Similarly, for this CNOT kagome class, the trace is nonfractal---$\tr(T_2^n)=u^n + u^{-n}$ for all $n$---but the characteristic polynomial tells a more interesting tale: 
\begin{equation}
    \chi_{T_2}(y) = \mu_{T_2}(y)^2 = \mu_{CNOT}(y)^2
    \end{equation}
    where
\begin{align}
    \mu_{CNOT}(y) =  &= y^4 + (u+\ui)y^3 + y^2 + (u+\ui)y + 1 \notag \\
    &= (y^2 + uy + 1)(y^2 + \ui y + 1)
\end{align}
is the minimal polynomial for the bare CNOT automaton, a.k.a. the Clifford East model (\autoref{eq:cnot-a2}). 
\begin{figure}[t]
    \centering
    \includegraphics[width=\linewidth]{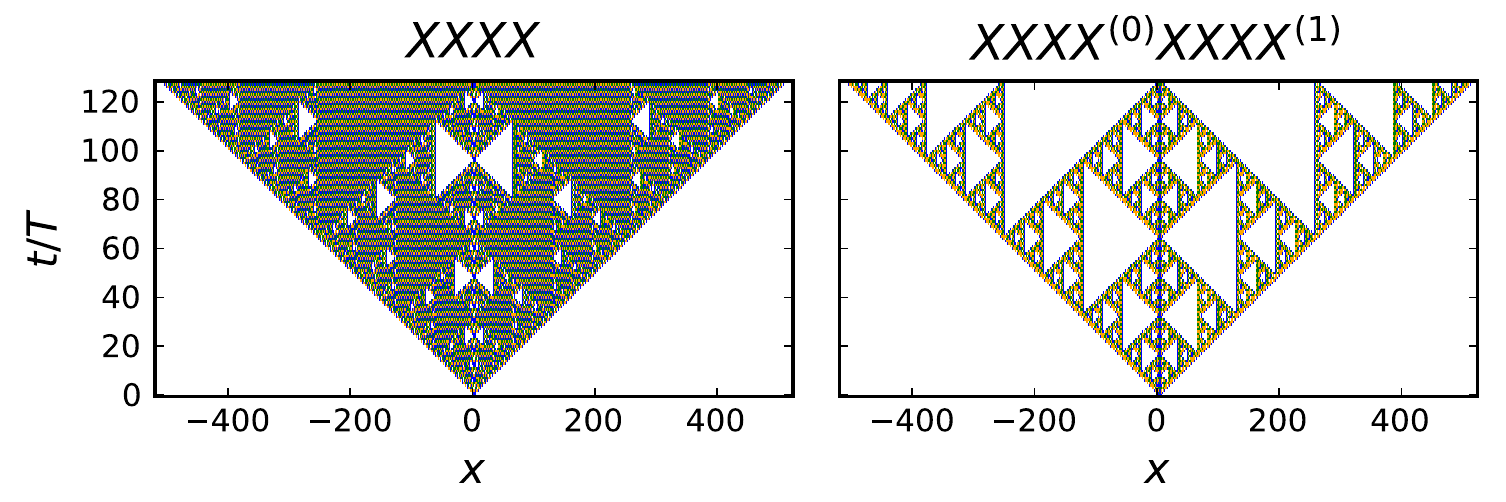}
    \caption{Time evolution under the automaton $T_2$ (\autoref{eq:T-cnot}) for the initial operator $P=XXXX$ (left) and $P^{(0)}P^{(1)}=XXXXXXXX$ (right), for $128$ time steps in units of $T=2$. Blue, orange, and green pixels correspond to $X$, $Y$, and $Z$ respectively.}
    \label{fig:XXXX}
\end{figure}

Pauli strings in the $a=4$ kagome CNOT class exhibit a fractal structure with the familiar Sierpinski motif, as presaged by the fact that its minimal polynomial is $\mu_{CNOT}(y)$. For a string initially localized to one unit cell, the non-identity part of the image is much less sparse than the standard Sierpinski gasket, with a fractal dimension near $2$. This can be seen in left panel of~\autoref{fig:XXXX} for the initial string $XXXX$, which remains reflection-invariant at all times owing to the left-right symmetry of the automaton. To recover the classic Sierpinski pattern, we note that each dense patch of $XXXX(t)$ contains a clear periodic structure. Thus, in the image of the product $XXXX^{(0)} XXXX^{(1)}$, where the superscript indexes the unit cell, the interior of each dense patch cancels out, and a fractal dimension of $\log_2(3)$ is recovered (right panel of \autoref{fig:XXXX}).

We leave the details of the origin of this relation to the CNOT automaton to future work but note that some insights can be gained by examining the footprint on every $a$th site. The time evolution of certain initial one-site operators particularly simple. For example, examining the footprint of $Z_1^{(0)}Z_1^{(1)}(t)$ on every fourth site reveals four monochrome Sierpinski gaskets: $Z$'s only live on for $x=na+1$ and $x=na+3$, while $x=na+2$ is all $Y$'s and $x=na+4$ is all $X$'s (\autoref{fig:Z1}).

\begin{figure}[t]
    \centering
    \includegraphics[width=\linewidth]{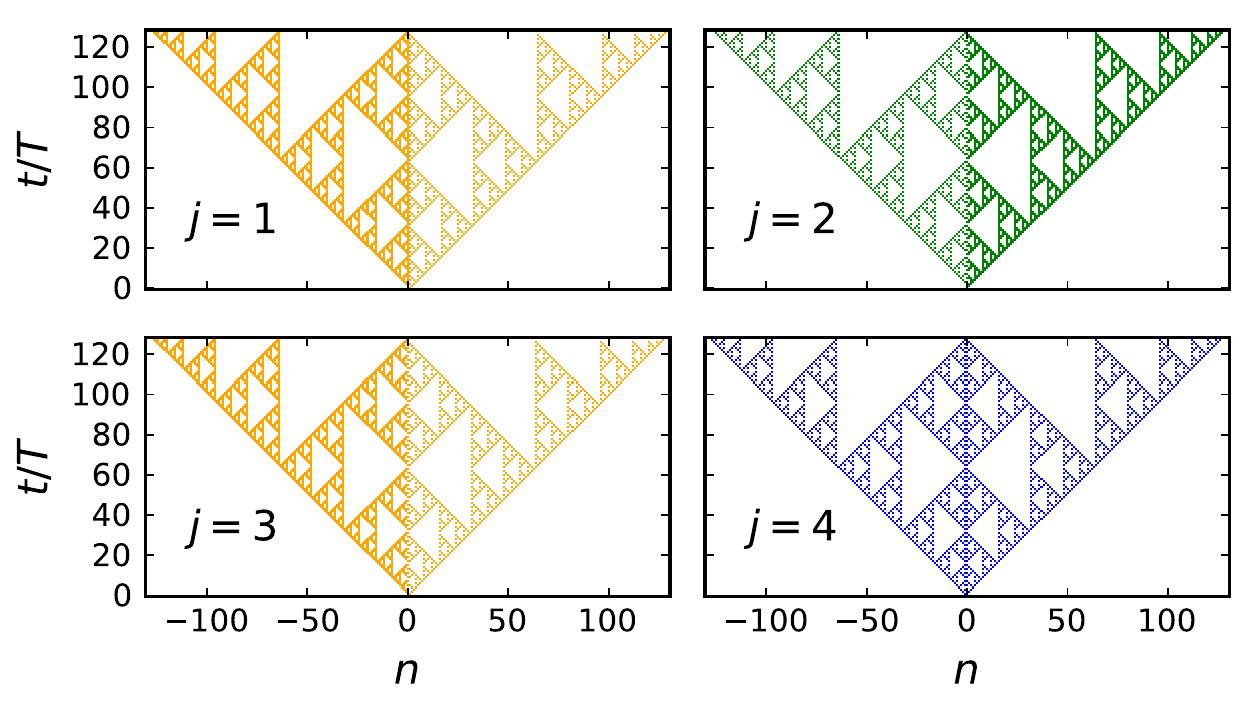}
    \caption{Time evolution under the automaton $T_2$ from the initial operator $Z_1^{(0)}Z_1^{(1)}$, for 128 time steps in units of $T=2$. Separate panels show the footprint on the four sites of the unit cell, i.e., $x=na+j$ for $j=1,2,3,4$ respectively. Blue, orange, and green pixels correspond to $X$, $Y$, and $Z$ respectively.}
    \label{fig:Z1}
\end{figure}
The connection between this kagome class and the Clifford East model has interesting implications for the entanglement growth and ergodicity, which have been analyzed for the latter model in several recent works. Ref.~\cite{Gopalakrishnan2018} finds that despite the absence of integrability, the half-chain entropy of typical many-body eigenstates only grows logarithmically with $L$ for $L=2^k$. A related "memory effect" is described in Ref.~\cite{Berenstein2021cnot} where for $t=2^k$, the single-qubit density matrix for an initial product state only converges to the fully mixed state polynomially in $t$.

\subsection{$D_6$-symmetric good scrambling class}
Finally, we decorate the kagome lattice with the one-site gates that on the square lattice produce the dense good scrambling class. As on the square lattice, placing $R_X[\pi/2] = R_X[\pi/2]^T$ on each edge maintains the full point group symmetry, which in this case is $D_6$. Unlike on the square lattice, however, this class exhibits a fractal structure, which is in fact quite similar to the CNOT-like class above.

\begin{figure}[t]
    \centering
    \includegraphics[width=\linewidth]{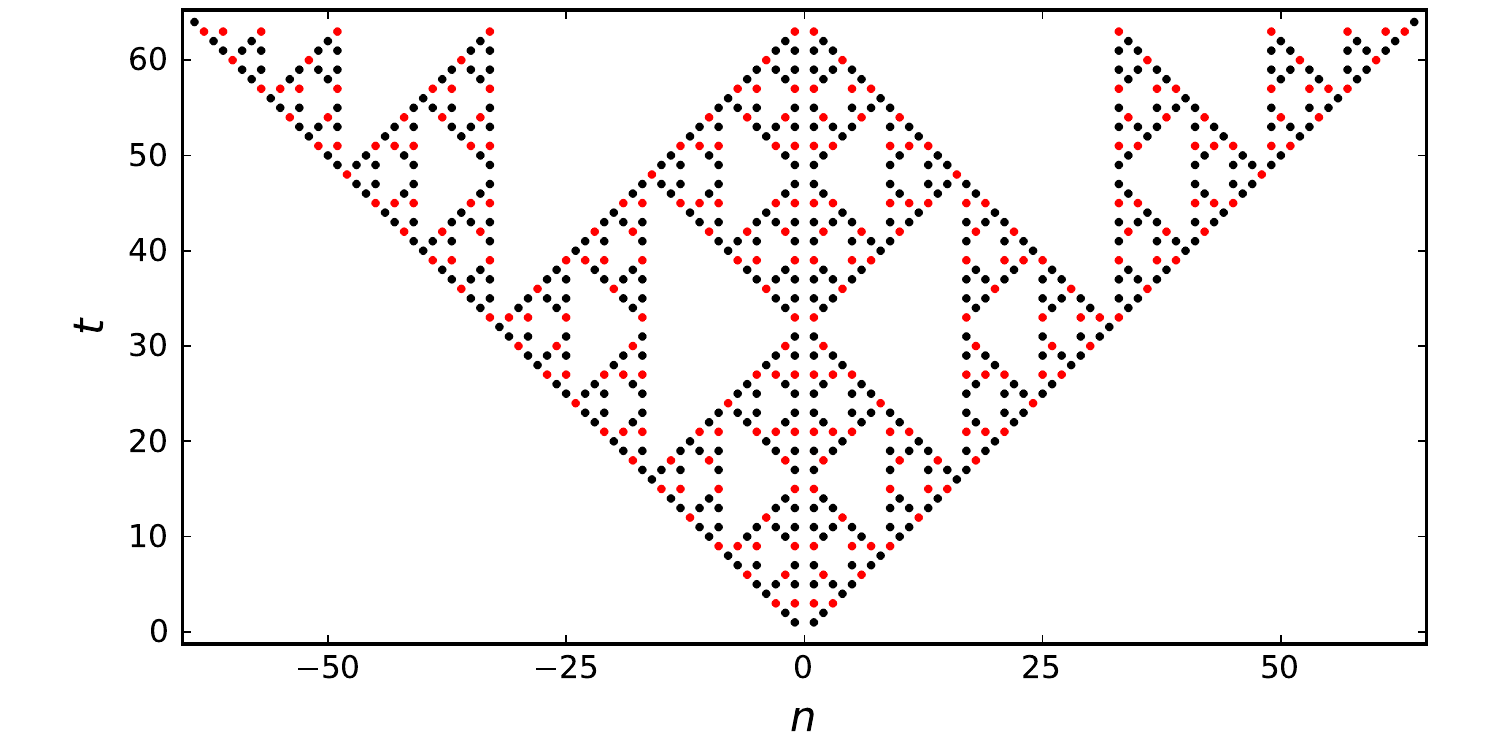}
    \caption{Visual depiction of $\tr(M^t)$ up to $t=64$. A black point at $(n,t)$ indicates that the coefficient of $u^n$ in the expansion of $\tr(M^t)$ is 1, for both $M=M_{CNOT}$ (\autoref{eq:cnot-a2}) and $M=T_3$  (\autoref{eq:T-d6}). Red points, which occur for $t \, \mod 3 = 0$, mark the location of nonzero coefficients in $\tr(M_{CNOT}^t)$, whereas $\tr(T_3^t)$ vanishes in every third time step.}
    \label{fig:trace}
\end{figure}
The automaton for this circuit is:
\begin{equation}\label{eq:T-d6}
T_3 = \begin{pmatrix}
\mathbf{s_{11}} & \mathbf{s_{12}} & \mathbf{s_{13}} & \mathbf{s_{14}} \\
\mathbf{s_{21}} & \mathbf{s_{22}} & \mathbf{s_{23}} & \mathbf{s_{24}} \\
\overline{\mathbf{s_{24}}} & \overline{\mathbf{s_{23}}} & \overline{\mathbf{s_{22}}} & \overline{\mathbf{s_{21}}} \\
\overline{\mathbf{s_{14}}} & \overline{\mathbf{s_{13}}} & \overline{\mathbf{s_{12}}} & \overline{\mathbf{s_{11}}}
\end{pmatrix}
\end{equation}
where:
\begin{subequations}
\begin{align}
    \mathbf{s_{11}} &= \begin{pmatrix}
    u & u \\ 
    u+1 & 0
    \end{pmatrix}, \quad \mathbf{s_{12}} = \begin{pmatrix}
    0 & 0 \\ u+1 & u
    \end{pmatrix} \\
    \mathbf{s_{13}} &= \begin{pmatrix}
    u & 0 \\ u & u 
    \end{pmatrix}, \quad \mathbf{s_{14}} = \begin{pmatrix}
    u & 0 \\ 1 & u+1
    \end{pmatrix} \\
    \mathbf{s_{21}} &= \begin{pmatrix}
    1 & 1 \\ 0 & 1
    \end{pmatrix} = \overline{\mathbf{s_{21}}}, \quad \mathbf{s_{22}} = \begin{pmatrix}
    0 & 1 \\ 0 & 0
    \end{pmatrix} = \overline{\mathbf{s_{22}}}\\
    \mathbf{s_{23}} &= \begin{pmatrix}
    0 & 0 \\ 1 & 1
    \end{pmatrix} = \overline{\mathbf{s_{23}}}, \quad
        \mathbf{s_{24}} = \begin{pmatrix}
    0 & 0 \\ 1 & 0
    \end{pmatrix} = \overline{\mathbf{s_{24}}}.
\end{align}
\end{subequations}
Again, \autoref{eq:T-d6} is explicitly invariant under the reflection implemented by~\autoref{eq:reflect-4}.

For this class, the characteristic polynomial does not factorize, but remarkably, the footprint of $\tr(T_3^t)$ is closely related to $\tr(M_{CNOT}^t)$. As shown in~\autoref{fig:trace}, 
\begin{equation}
    \tr(T_3^t) = \begin{cases}
    0 & t \, \mod 3 = 0 \\
    \tr(M_{CNOT}^t) & \mathrm{otherwise}.
    \end{cases} 
\end{equation}

Of course, the physical observable is not the trace (which can hide the true fractal structure of the operator spreading, as in the case of the CNOT-like class above), but the image of a spreading Pauli string. For strings initially localized on one unit cell, the operator spreading has far less white space than~\autoref{fig:trace}, and with a more intricate pattern of $X$, $Y$, $Z$ than in the CNOT-like class. But taking the product of two unit-cell-supported Paulis translated by $n=2$ with respect to each other, i.e. $P^{(0)}P^{(2)}$, yields the classic Sierpinski gasket with $d_f=\log_2(3)$. For example,~\autoref{fig:sierpinski-X} shows the time evolution for $P = XXXX$. Since $P$ is symmetric about the center of the unit cell, the image of $XXXX^{(0)}XXXX^{(2)}(t)$ on sites $x=an + 1$ is the mirror image of that on sites $x=an+4$, and $x=an+2$ is the mirror image of $x=an+3$. 

\begin{figure}[t]
    \centering
    \includegraphics[width=\linewidth]{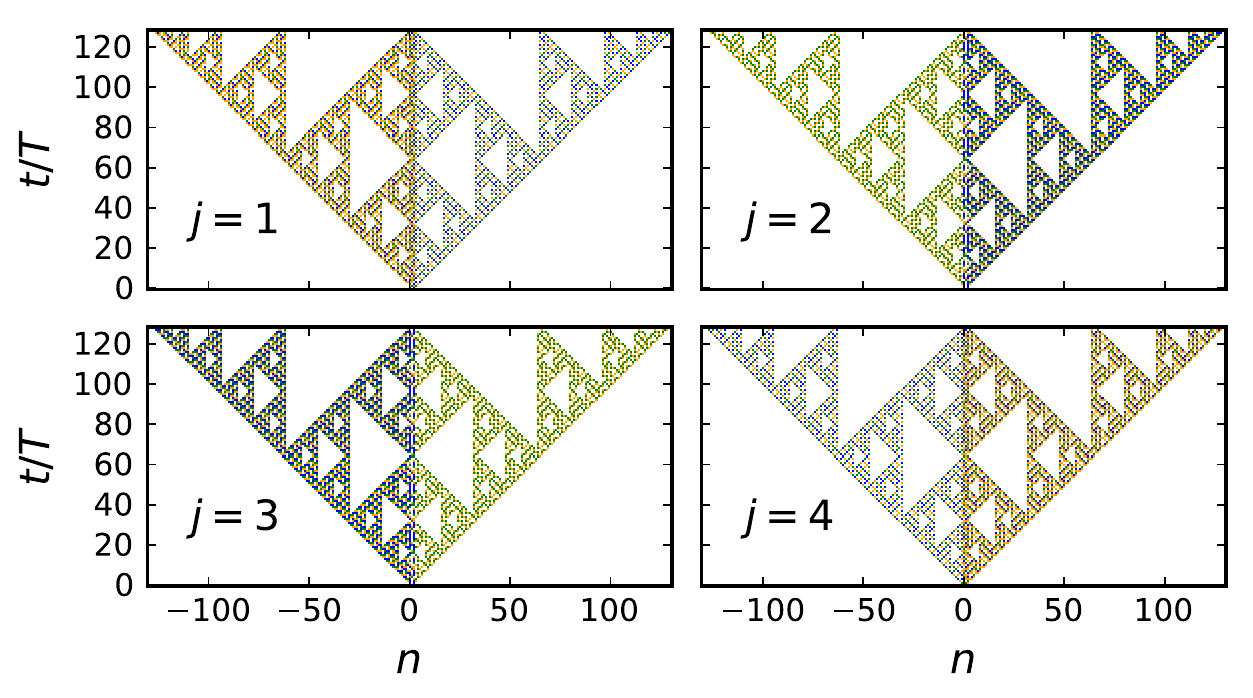}
    \caption{Time evolution under the automaton $T_3$ (\autoref{eq:T-d6}) from the initial operator $XXXX^{(0)}XXXX^{(2)}$, i.e. $XXXXIIIIXXXX$, up to 128 time steps in units of $T=2$. Separate panels show the footprint on the four sites of the unit cell, i.e., $x=na+j$ for $j=1,2,3,4$. $X$, $Y$, and $Z$ are shown in blue, green, and orange respectively.}
    \label{fig:sierpinski-X}
\end{figure}

In addition to both producing Sierpinski triangles in the operator spreading and generating Page curve with slope 1 on random initial product states, this class and the CNOT-like class above also have the same recurrence times $\tau(m)$ when applied to finite systems with periodic boundary conditions. As with the good scrambling classes on the square lattice, $\tau(m)$ is linear in $m$ for $m=2^k$ but grows superlinearly for generic $m$. But unlike on the square lattice, both the fractal dimension and the recurrence times are indifferent to whether the single-qubit gates are all cyclic permutations (as in the CNOT-like class) or $X$ or $Y$ rotations.

\section{Hybrid circuits}\label{sect:measurements}

Returning to the square lattice, we now break unitarity by adding projective measurements in a STTI fashion. While the measurement outcomes are random, for stabilizer circuits different quantum trajectories just differ with respect to signs on the stabilizers, so when considering the dynamics of stabilizer groups modulo signs, the spacetime translation invariance is preserved.

{The realm of possibilities for crystalline hybrid circuits is vast, and a more thorough treatment of the purification dynamics, steady state properties, and implications for quantum error correction is left to a forthcoming paper~\cite{Sommers2023a}. Here, we focus upon a minimal modification of the brickwork circuits studied in this paper (\autoref{fig:lattice}) in which one single-site measurement in the $\sigma$ basis is performed per doubled unit cell $(T=1, a=2)$ (\autoref{fig:hybrid-circuit}).} In addition to enlarging the unit cell of the lattice, the added measurements reduce the point group symmetry. {If the one-site gates along the diagonal containing the measurements (taken to be the $+$ diagonal in~\autoref{fig:hybrid-circuit}) are identity gates, then reflections about both diagonals (and thus inversion as well) preserve the relative positions of the gates and measurements. If the blue one-site gates are non-identities, then of the original point group transformations (\autoref{fig:point-group}), only reflection about the diagonal containing the measurements is a possible symmetry.}

Starting from a fully mixed initial state, the first layer of $m$ measurements performed on the $j$th site of each unit cell purifies the state by $m$ bits, to entropy $S(t=0) = m(a-1)$, since the measured operators are commuting and independent. Immediately after the measurements, the stabilizer group is generated by the measured operators:
\begin{equation}\label{eq:t1}
    \mathcal{S}(t=0) = \langle \sigma(na+j) \rangle \equiv \langle \sigma^{(n)}_j \rangle_{n=1,...,m}
\end{equation}
The stabilizer generators then spread under two layers of unitary gates. Subsequent measurement layers may or may not purify the state further; a given measurement causes a purification by 1 bit if and only if the measured operator commutes with all of $\mathcal{S}$, but does not already belong to $\mathcal{S}$, i.e. anticommutes with a logical operator. Once the state stops purifying, the stabilizer group (mixed or pure) is static, that is, invariant under one time step of the circuit; we call this the "plateau group."

\begin{figure}[t]
    \centering
    \includegraphics[width=\linewidth]{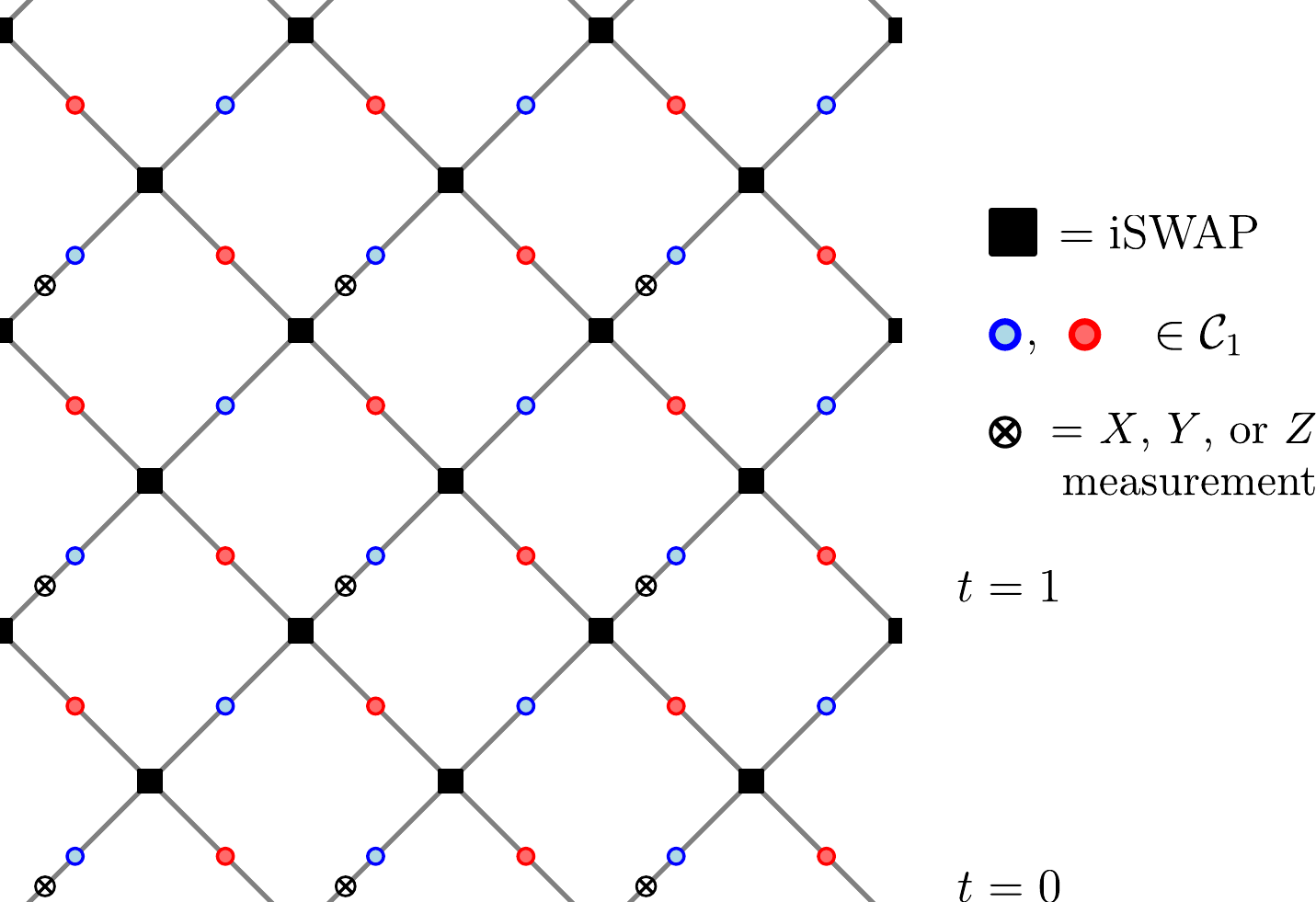}
    \caption{Square-lattice (brickwork) circuit with measurements (c.f.~\autoref{fig:lattice}). White x'ed circles represent measurements in a fixed Pauli basis.}
    \label{fig:hybrid-circuit}
\end{figure}

We have examined the dynamics for all choices of dual-unitary gates, measurement locations, and measurement bases with the geometry of~\autoref{fig:hybrid-circuit}. In most cases the plateau group is reached after $O(1)$ time steps, {which we refer to as "gapped purification" because the purification time does not scale with system size. But for circuits in the $d_f\cong 1.9$ class in the appropriate measurement basis, a fully mixed initial state purifies "gaplessly" in $m$ time steps to a pure product state for $m=2^k$.} This extensive purification time gives rise to nontrivial entanglement behavior and the appearance of Sierpinski fractals when the steady state is perturbed. {To give the reader a small taste of the rich dynamics that can arise in hybrid circuits, we now discuss this class of circuits in detail.}

\subsection{Purification dynamics}
Consider the representative circuit of the $d_f\cong 1.9$ class (\autoref{eq:df-1.9}), now with measurements in the $X$ basis at spacetime locations $(t,x)=(k,2n+1)$ with integer $k$. Each measurement immediately precedes $v_+=R_X[\pi/2]$ on the left incoming leg to the iSWAP core. The unitary circuit only has one strong point group symmetry---invariance under reflection through the downward-sloping diagonal---which is not present in the hybrid circuit. 

At $t=0$, the first round of measurements adds an $X$ stabilizer on the first site of each unit cell, i.e. $X^{(n)}_1$. From~\autoref{eq:M5}, we can read off the time-evolved stabilizer generators after the subsequent two layers:
\begin{equation}
    X^{(n)}_1\rightarrow X^{(n-1)}_2 Z^{(n)}_1 Z^{(n+1)}_1
\end{equation}
In the next round of measurements, we again measure $X^{(n)}_1$, for $n=1,2,...,m$. Since the measurements commute with each other, we can perform them in any order. The first $m-1$ measurements anticommute with a pair of stabilizer generators. But each measurement modifies $\mathcal{S}$ such that the final measurement commutes with the entire group, causing the state to purify by exactly one bit. To see why this is the case, note that once we have measured $X^{(n)}_1$ for all $n$, we will also have measured $\prod_{n=1}^m X^{(n)}_1$, which is a logical operator of the pre-measurement state. 

After the full round of measurements, the stabilizer group has $m+1$ generators:
\begin{equation}\label{eq:t2}
    \mathcal{S}(t=1) = \langle \{X^{(n)}_1\}_{n=1}^m, \prod_{n=1}^{m} X^{(n)}_2 \rangle
\end{equation}
This can be proven by noting that $\prod X^{(n)}_2$ is the only element of the pre-measurement stabilizer group that commutes with all the measurements.

Comparing~\autoref{eq:t2} to~\autoref{eq:t1}, we see that $\mathcal{S}(t=0)$ is a subgroup of $\mathcal{S}(t=1)$. Indeed, this is an example of a more general property of the purification dynamics in any Floquet Clifford circuit, with or without spatial translation invariance: for the fully mixed initial condition, or any state in the sequence of stabilizer groups from fully mixed to the steady state group, $\mathcal{S}(t-1)$ is a subgroup of $\mathcal{S}(t)$. A corollary is that the entropy $S(t)$ decreases at a non-increasing rate. {This gives us a nice way to partially fix the generators of the instantaneous stabilizer group: the "time-ordered" stabilizer tableau at time $t$ is defined so that for all $t'\leq t$, the first $L-S(t')$ stabilizers generate the group at time $t'$~\cite{Sommers2023a}.}

In the present example, for $m=2^k$, each subsequent time step induces exactly one purification event, until the state purifies completely at $t^*=m$. The final steady state is a product group:
\begin{equation}\label{eq:dark-state}
    \mathcal{S}^* = \langle X^{(n)}_j \rangle_{j=1,2; n=1,...,m}.
\end{equation}
Remarkably, although the plateau group has zero entanglement, since the time to reach this state scales linearly with $m$ for generic initial states (including the fully mixed state, as well as random product states of any entropy density), it is possible for the circuit to generate a volume-law transient despite the presence of measurements.
\begin{figure}[t]
    \centering
    \includegraphics[width=\linewidth]{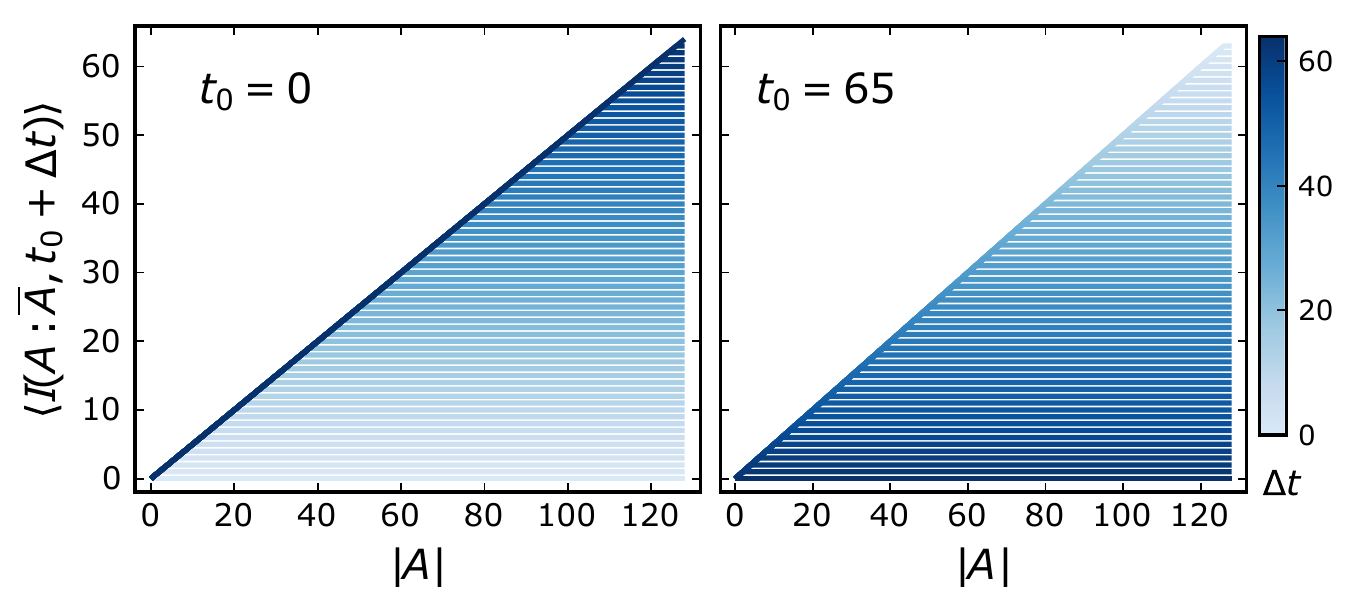}
    \caption{Mutual information $I(A:\overline{A}, t_0+\Delta t)$, averaged over all contiguous subregions $A$ of a given length $\leq L/2$, starting from a fully mixed state on $L=256$ qubits, under the circuit evolution depicted in~\autoref{fig:hybrid-circuit} with $U=\iSWAP (R_X[\pi/2] \otimes R_{(1,1,1)}[-2\pi/3])$ and measurements in the $X$ basis. Left panel shows mutual information increasing up to $t=64$ (darkest blue), while the right panel shows mutual information decreasing to $t=128=m$.}
    \label{fig:mutual}
\end{figure}
This can be seen from the growth of the mutual information from the fully mixed initial state (\autoref{fig:mutual}), defined as:
\begin{equation}
    I(A:\overline{A}, t) = S(\rho_A(t)) + S(\rho_{\overline{A}}(t)) - S(\rho(t))
\end{equation}
for contiguous regions $A$, where $\overline{A}$ is the complement of $A$ and $\rho(t)$ is the state of the full system of $L$ qubits at time $t$. Averaging over all contiguous regions (with periodic boundary conditions) of the same length $|A|\leq L/2$, at early times $\langle I(A:\overline{A})\rangle$ has a piecewise linear form: 
\begin{equation}
\langle I(A:\overline{A},t) \rangle = \begin{cases}
|A|/4 & |A|\leq 2t \\
t/2 & |A|>2t
\end{cases}
\end{equation}
Thus, the half-cut mutual information increases linearly until $t=m/2$ (left panel of \autoref{fig:mutual}). It then decreases linearly until $t=t^*$, at which point the steady state with zero entanglement is reached (right panel). An analogous trend is present in entanglement entropy starting from random pure product states.

{When $m$ is not a power of 2, the fully mixed initial state still purifies by one bit per time step, but does not purify completely, a phenomenon tied to an underlying fractality in the purification dynamics. Explicitly, for $m=p 2^k$ where $p$ is odd,
\begin{equation}
t^*=2^k, \qquad S(t^*) = m - t^* = (p-1) 2^k
\end{equation}
so the entropy density of the plateau group is $(p-1)/2p$, asymptoting toward 1/2 for large $p$. Thus, reminiscent of how the recurrence time of fractal or good scrambling CQCA is sensitive to the power of 2, when measurements are introduced, the \textit{purification time} can also be sensitive to powers of 2. All gapless circuits we have surveyed, across a wide range of unit cell dimensions and even when we populate each unit cell with random Clifford gates rather than dual-unitary gates, exhibit this sensitivity, indicating that gaplessness and fractality are intimately linked.}

{The fractal structure in our current example is a Sierpinski gasket, which can be seen from the time-ordered stabilizer tableau. The first $m$ generators in the tableau are the measured operators $X^{(n)}_1$. Thereafter, we extend the time-ordered tableau by one generator in each time step, and can further fix this generator such that its cycle length---the number of unit cells by which it must be translated before returning to itself---is minimized. At $t=1$, we obtain a fully translation-invariant generator $\prod_n X^{(n)}_2$, with cycle length 1. As time increases, the minimum cycle length increases, and a particular choice of translates of each generator produces a spacetime Sierpinski gasket in the non-identity entries of the tableau matrix. More closely related to the topic of operator spreading addressed throughout this paper, we also identify this Sierpinski gasket in the spreading of local perturbations, described next.} 
\subsection{Dark perturbations}
The pure group defined by~\autoref{eq:dark-state} can be viewed as an absorbing, or "dark", state of the purification dynamics: for $m=2^k$, any initial stabilizer group will evolve to this product group within $m$ time steps. Moreover, while the steady state group for $m\neq 2^k$ is mixed, ~\autoref{eq:dark-state} defines a stationary group within this mixed plateau, i.e. it has period 1 under the action of the circuit. We can then perturb this dark state in various ways and observe the fractal spreading of the perturbation. At time $t$, we mark the $n$th unit cell as dark if $X_2^{(n)}$ is contained in the group; otherwise, it is marked light. One choice of perturbation is a local perturbation where the entire state is dark, except for a contiguous region of $O(1)$ cells. 

\begin{figure}[t]
    \centering
    \includegraphics[width=0.9\linewidth]{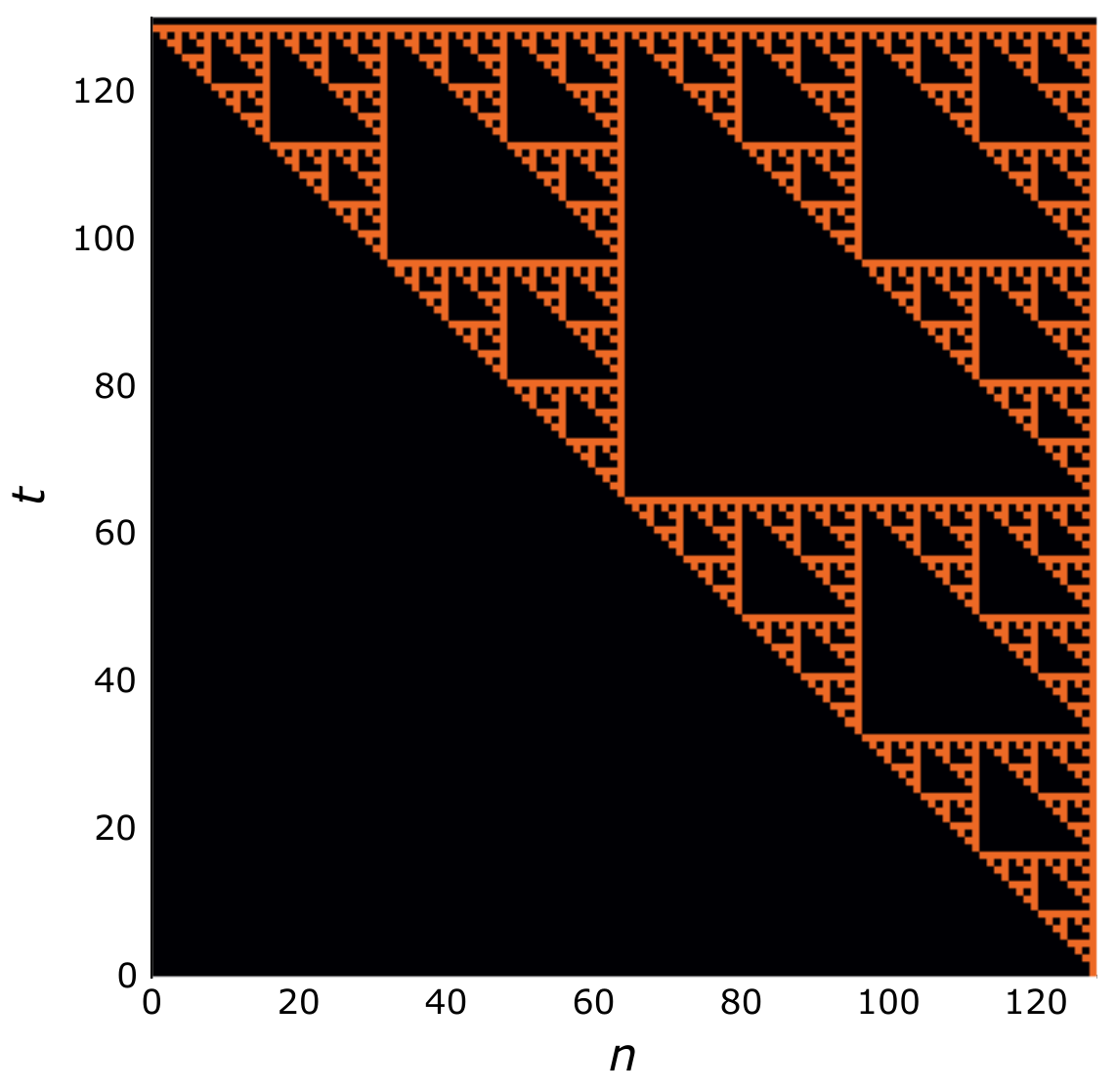}
    \caption{Pattern of light (orange) and dark (black) unit cells, for $m=128$ and a perturbation on the rightmost cell from $X\rightarrow Z$.}
    \label{fig:dark}
\end{figure}
As an example, consider a perturbation on the rightmost site, $X^{(m)}_2 \rightarrow Z^{(m)}_2$. This produces another product stabilizer group, with a single light cell. In the subsequent time evolution, shown in~\autoref{fig:dark}, the familiar Sierpinski gasket appears in the spacetime structure of the spreading "light" cells. Immediately after spreading through the entire system, the light is annihilated at $t = m$, upon return to the absorbing state. 

It should be emphasized that while the spacetime spreading from an initially local perturbation is fractal, this is a different fractal from that observed in the unitary circuits belonging to this class. Namely, while operator spreading in the unitary circuits is characterized by fractal dimension $d_f\cong 1.9$, the hybrid circuit produces Sierpinski gaskets, with $d_f=\log_2(3)= 1.5849...$, and a much starker asymmetry, as the light sites only spread left from the initial perturbation. {The strong asymmetry is tied to the fact that placing a measurement on odd sites only breaks the left/right symmetry more strongly than does the choice of different one-site gates.} 

\section{Discussion}\label{sect:conclude}
This work is the beginning of an investigation into the crystallography of quantum circuits, that is, the description of spacetime translation-invariant (STTI) quantum circuits defined on lattices with varying amounts of symmetries. {Just as randomness in certain limits imbues models of quantum many-body physics with analytic tractability, at the other end of the spectrum crystalline quantum circuits are also amenable to precise statements about operator spreading, entanglement growth, and purification dynamics. The analysis becomes particularly friendly when we restrict to Clifford gates, as we have done in this work, thus allowing our STTI circuits to be represented as Clifford quantum cellular automata (CQCA) with unit cell $a$. Leveraging this formalism,} we have classified all dual-unitary Clifford circuits with one gate per unit cell on the square lattice ($a=2$), which roughly separate into periodic, glider, and fractal classes like the $a=1$ CQCA studied previously. Strikingly, we also find a class of circuits, {a representative of which is composed by applying the gate $\iSWAP (R_X[\pi/2]\otimes R_X[\pi/2])$ in a brickwork fashion, which possesses the full symmetry of the square lattice while also acting as a "good scrambler" with nonfractal operator spreading.} We have moreover examined the effect of translation-invariant measurements on square-lattice CQCA, as well as analyzing a subset of tri-unitary Clifford circuits on the kagome lattice.

{The two main features of the "dense good scrambling class"---symmetry under point group transformations and nonfractal operator spreading---have served as overarching themes of this work. The latter theme points to our aim to bring the tools of crystallography well-known to condensed matter physicists to bear on the study of quantum circuits, while building on the current understanding of dual-unitary~\cite{Bertini2019}, tri-unitary~\cite{Jonay2021}, and, broadly, multidirectional-unitary~\cite{Milbradt2023,Mestyan2022} gates. When the constituent gates remain unitary under all point-group transformations, we can then ask how that transformed unitary circuit relates to the original circuit. A circuit left invariant under a given transformation is said to be strongly "self-dual", and on the square lattice, the dense good scrambling circuit is self-octa-unitary---invariant under all elements of the $D_4$ point group.} 
A broad question is how the presence or absence of certain point group symmetries manifests in the circuit dynamics, and whether imposing these symmetries bears any relation to desirable coding features. Left/right reflection invariance clearly manifests in whether initially reflection-symmetric operators remain so under time evolution, but the interpretation of invariance under other point group transformations, such as rotations of the spacetime axes, is less clear. One observable which is sensitive to all point group transformations is the two-point function of one-site operators, discussed in~\appref{app:corr}, but for good-scrambling circuits these correlations are non-vanishing only at very early times. Thus, follow-up work is needed to identify probes of symmetry in the late-time dynamics of both unitary and hybrid circuits.

{Meanwhile, fractality in operator spreading provides an important point of contrast between random quantum circuits, in which the operators become scrambled and spread densely within the lightcone, and most good scrambling CQCA, which generate state entanglement but where operators only spread on a spacetime region of fractal dimension $d_f<2$. In this work, we have found fractal motifs in good scrambling iSWAP-core CQCA on the square and kagome lattices and used the minimal polynomial to relate them to previously studied automata such as the SDKI automaton (\autoref{eq:M-SDKI}) and Clifford East model (\autoref{eq:cnot-a2}). We have also discovered a new class of fractal CQCA with $d_f\cong 1.9$ and asymmetric operator spreading. Fractals arise in hybrid STTI circuits as well, as exemplified by the Sierpinski gasket in the purification dynamics and response to dark perturbations when measurements are added to a $d_f\cong 1.9$ circuit. In light of the prevalence of fractals which result in weak ergodicity breaking of otherwise chaotic CQCA~\cite{Kent2023}, the dense good scrambling class with $d_f=2$ is particularly interesting. It approaches the sort of mixing behavior seen in random Clifford circuits, yet the underlying structure of the circuit is still present in the nonuniformity of Pauli strings within the bulk of a spreading operator and the linear recurrence time on finite systems with $m=2^k$ unit cells.} We did not discover such a nonfractal class on the kagome lattice when we imposed the full $D_6$ symmetry, but it remains open whether there exist nonfractal good scramblers in kagome circuits with less symmetry, or in ($D_6$-symmetric) triangular lattice automata containing irreducible three-qubit interactions.

{The question of whether any lattices, besides the square lattice, support a nonfractal good scrambling class is relevant to an additional aspect of our work, the application to quantum error correction. Time evolution under an STTI circuit starting from a translation-invariant initial mixed state produces a quasicyclic stabilizer code, and while we have not yet developed optimal decoders for realistic noise models, the possible utility of their quasicyclic structure for decoding was one motivation for the project of "derandomization" embarked upon in this work. A larger fractal dimension does not necessarily imply the ability to encode better finite-rate codes, but insofar as the spreading of an individual local Pauli places a bound on the achievable code distance---the lowest weight of a logical representative---a linear code distance can only be achieved if there exists a sequence of time slices and system sizes for which local operators spread to a finite fraction of the system. While we cannot compute the code distance efficiently,} on the square lattice, both the dense good scrambling class and the $d_f\cong 1.9$ class generate codes that are competitive with random codes under erasures for certain system sizes. Future work is also needed to clarify the relation between contiguous code length and code distance for fractal and nonfractal automata, {as the naive approach of choosing a snapshot in time with the maximal code length to define one's quantum-error-correcting code can result in suboptimal codes (\autoref{fig:subthreshold}).}

Beyond spacetime translation invariance, the circuits considered in this paper are restricted to those comprised of dual-unitary (so the circuit produced by any point group transformation is also unitary), Clifford gates (allowing their representation as CQCA). Lifting dual-unitarity, the spacetime dual remains a useful construct even when the spatial evolution is nonunitary. Mapping to the spatial direction is both a valuable analytical tool, e.g. for computing the spectral form factor~\cite{Sonner2022,Garratt2021feynman,Garratt2021}, and an asset to certain experimental protocols~\cite{Foss-Feig2021,Ippoliti2021}. Nontrivial phases and phase transitions in the dual can be related to those in the unitary circuit~\cite{Basu2022,Ippoliti2022}, with measurement-induced phase transitions being just one example~\cite{Lu2021}. With two-qubit Clifford gates, the only interacting gates that are not dual-unitary are those with a CNOT core, discussed briefly in~\autoref{sect:dual-unitary}. Surprisingly, the fractal classes on the kagome lattice (which are self-tri-unitary) exhibit the same Sierpinski gasket as the bare CNOT automaton on the square lattice (which is not even dual-unitary). Further investigation should elucidate this connection. In addition to this Sierpinski fractal class, a complete classification of all square-lattice CNOT-core automata with one gate per unit cell yields several glider classes and an SDKI-like class, but again, no dense good scrambling class.

{We now elaborate on some future avenues for research.}

While the dynamics of Clifford circuits can be quite rich, they are not universal, and a natural next step would therefore be to go beyond Clifford. A first step in this direction is to consider matchgate (free fermion) circuits, which are also classically simulatable~\cite{Valiant2001,Terhal2001}. The subset of free fermion circuits which are also Clifford are discussed in~\appref{app:matchgate}.

In addition, we can generalize beyond the square and kagome lattices, both in $1+1$D and in higher dimensions. Hyperbolic lattices, considered either as $1+1$D spacetime or as the 2D space of a $2+1$D circuit, offer particularly rich crystallography~\cite{Kollar2020,Boettcher2022} realizable in experiment~\cite{Kollar2019}. {Quantum circuits can also be defined on general graphs, including trees~\cite{Nahum2020,Feng2022}, which are amenable to tensor network methods for analytic computation of the code distance and more general noise models~\cite{Cao2022}. Preliminary investigation of tree circuits in which every gate is identical reveals promising classes of circuits for which the code distance grows exponentially in the tree depth~\cite{Sommers2023a}.}

Moving to $2+1$D makes available a greater variety symmetry groups while still being relevant to near-term quantum computing devices~\cite{Arute2019,Andersen2020,Semeghini2021,Zhao2022,Krinner2022}; Floquet codes such as the honeycomb code are one example~\cite{Hastings2021,Gidney2021,Paetznick2022,Haah2022,Aasen2022}. Here we have restricted ourselves to lattices with coordination number 4, such that each vertex is a SWAP or iSWAP core, but the broad project of classifying STTI circuits and their symmetries can also be applied to lattices with higher coordination number. It would be interesting to connect these crystallographic classifications to the broader topological and group theoretic characterization of (non)trivial QCA in higher dimensions~\cite{Haah2018,Haah2019,Freedman2020,Freedman2022,Shirley2022}. {QCA can be used to define subsystem symmetry-protected topological (SPT) phases, characterized by line-like and fractal symmetries for the $a=1$ glider and fractal classes respectively~\cite{Stephen2019}. How, then, should we interpret the phase defined by our nonfractal good scrambling class, and might it be useful as a resource state for universal measurement-based quantum computation~\cite{Devakul2018,Stephen2019}?}

{Possibilities also abound when we increase the local Hilbert space dimension $q$. Two-qudit gates with $q\geq 3$ can be not only dual-unitary, but also unitary along the diagonal, making them "perfect tensors" with maximal entanglement power~\cite{Aravinda2021,Borsi2022,Rather2022}. In the operator-state correspondence, these operators are absolutely maximally entangled (AME) states, which are maximally entangled with respect to all bipartitions of the legs~\cite{Goyeneche2015}. The symplectic cellular automaton formalism can be used for general (composite) $q$, so we can also test whether the "washing-out" of fractal structure observed for the $a=1$ CQCA in the limit of large $q$ in Ref.~\cite{Kent2023} is also found for crystalline circuits with $a>1$.}

The research directions for hybrid STTI circuits, with or without dual-unitarity of the gates, are also numerous. Enlarging the unit cell to reduce the density of measurements allows for circuits with a mixed, volume-law-entangled steady-state group with linear code length and high performance under erasure errors, of interest for quantum error correction~\cite{Sommers2023a}. 
To develop an analytic understanding of these hybrid quantum circuits en route to the steady state, it would be useful to adapt the techniques of cellular automata to nonunitary dynamics, an area of research still in its infancy ~\cite{Richter1996,Brennen2003,Piroli2020}. In considering circuits with measurements and/or noise, it would also be fruitful to leverage recent work generalizing dual-unitary circuits to 3- and 4-way-unital open quantum channels~\cite{Kos2022}.

In this paper, we considered ideal circuits without noise.  Adapting techniques from fault-tolerance to either make the circuits robust to noise or the codes generated by the circuit useful for quantum computation is an interesting direction of research.  As an intermediate goal, one can design quantum cellular automata that are robust to small amounts of randomness in the choice of Clifford gates or measurement locations.  Partial progress in this direction has recently been reported for two-dimensional Floquet codes \cite{Davydova2022}. Fault-tolerant constructions for reliable computation with classical cellular automata have a rich history \cite{Gacs01}.

Finally, another way to make nonrandom circuits is by adding measurements or deforming gates in a deterministic, \textit{quasiperiodic} manner. This motivated our recent work on a model of self-dual quasiperiodic percolation on the square lattice~\cite{Sommers2023}. When quasiperiodic projective measurements are added to a good scrambling dual-unitary circuit, we find that there is a measurement-induced phase transition which falls outside the universality class of the random Clifford transition~\cite{Sommers2023a,Li2018,Li2019,Gullans2020,Zabalo2022}.

\begin{acknowledgments}
We wish to acknowledge helpful conversations with Arpit Dua, Sarang Gopalakrishnan, Stefan Krastanov, and Jon Nelson. We thank Jeongwan Haah and Matthew Hastings for helpful comments on the manuscript, and Matteo Ippoliti for pointing out the relation between the kagome and triangular lattice constructions of tri-unitary circuits. We also credit Suhail Rather with alerting us to the vanishing correlations between one-site operators for $t\geq 1$ in the good scrambling square-lattice circuits. GMS is supported by the Department of Defense (DoD) through the National Defense Science \& Engineering Graduate (NDSEG) Fellowship.  We also acknowledge support from the National Science Foundation (QLCI grant OMA-2120757). Numerical work was completed using computational resources managed and supported by Princeton Research Computing, a consortium of groups including the Princeton Institute for Computational Science and Engineering (PICSciE) and the Office of Information Technology's High Performance Computing Center and Visualization Laboratory at Princeton University. The open-source \texttt{QuantumClifford.jl} package was used to simulate Clifford circuits~\cite{QuantumClifford}. 
\end{acknowledgments}
\section*{Code Availability}
Select code used for this study, including a demo Jupyter notebook, is available at \url{https://github.com/gsommers/clifford-QCA}. Further code is available upon request.

\section*{Data Availability}
Data on iSWAP-core automata and recurrence times is available at \url{https://github.com/gsommers/clifford-QCA}. Additional data is available upon request.
\appendix
\counterwithin{figure}{section}
\section{Two-point correlations}\label{app:corr}
In this appendix, we more carefully define the two-point correlation functions of one-site observables at infinite temperature and discuss their connection to strong/weak symmetries of crystalline circuits. While we focus on dual-unitary brickwork circuits here, the broad concepts generalize to tri-unitary circuits and beyond.

\subsection{General formalism}
~\autoref{fig:corr} shows a close-up of the square lattice circuit. Each unit cell is labeled by a time $t$, where $\Delta t=1$ corresponds to two layers (one full time step), and a spatial coordinate $y$. Within each unit cell are four distinct spacetime locations, $\tau_\mu = \pm 1/2_\pm$. Here $\tau=-1/2$ marks the time before the one-site gates, $\tau=+1/2$ marks the time after the one-site gates but before the core, and $\mu=\pm$ lie along the diagonals with slope $\pm 1$~\footnote{The spatial coordinate $x$ used in, e.g.,~\autoref{fig:paulis-circuit8}, is $x(y,\mu)=2y-\mu$ up to an even integer.}.

\begin{figure}[t]
\centering
\includegraphics[width=0.9\linewidth]{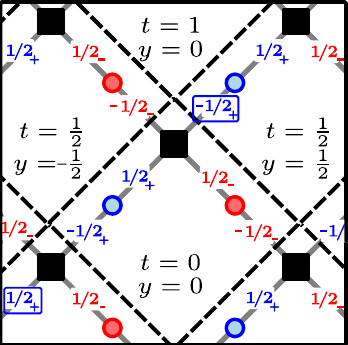}
\caption{Spacetime locations in finer detail on the brickwork circuit of~\autoref{fig:lattice}. Dashed lines indicate unit cells of the rotated square lattice, labeled by a spatial coordinate $y$ and time coordinate $t$. Within each unit cell are four distinct spacetime locations: $-1/2_\pm$ before the one-site gates, and $+1/2_\pm$ after the one-site gates but before the two-site core, associated with the $\pm$ light cones. Blue boxes indicate a pair of spacetime locations for which all correlations vanish in good scrambling iSWAP-core circuits (\autoref{eq:depol}). ~\label{fig:corr}}
\end{figure}
Now consider the correlation functions:
\begin{equation}\label{eq:Dab}
D^{\alpha\beta}_{\mu \mu'}(t,t',y,y',\tau,\tau') = \langle \sigma_{\alpha} (t,y,\tau_\mu) \sigma_{\beta} (t',y',\tau'_{\mu'}) \rangle. 
\end{equation}
Here $\{\sigma_\alpha\}_{\alpha=0}^{q^2-1}$ is a complete orthonormal basis of operators on q-dimensional qudits, $\tr(\sigma_\alpha^\dag \sigma_\beta) = q \delta_{\alpha\beta}$, where $\sigma_0=\mathbbm{1}$ and the remaining $q^2-1$ operators are traceless~\cite{Bertini2019}. In our case, $q=2$, so $\sigma_1,\sigma_2,\sigma_3$ are the usual Pauli operators $X, Y, Z$. As has been the convention throughout this paper, time evolution is in the Schrodinger picture. We work at infinite temperature, i.e. the expectation value is taken in the fully mixed state $\rho = \mathbbm{1}/q^L$ where $L$ is the number of qudits.

For a dual-unitary circuit, the correlations are nonvanishing only for~\cite{Bertini2019}:
\begin{equation}\label{eq:du}
\mu = \mu', \quad t'-t=\mu (y'-y)
\end{equation}
i.e., for operators along the same diagonal.

Inserting~\autoref{eq:du} into~\autoref{eq:Dab} and exploiting translation invariance, we arrive at:
\begin{equation}
D^{\alpha\beta}_{\mu \mu'}(t,t',y,y',\tau,\tau') \equiv \delta_{\mu \mu'} \delta_{t'-t, \mu(y'-y)} C^{\alpha\beta}_\mu (t'-t, \tau, \tau')
\end{equation}
where:
\begin{equation}\label{eq:ti-corr}
C^{\alpha\beta}_\mu (t',\tau,\tau') = \langle \sigma_{\alpha} (0,0,\tau_\mu) \sigma_{\beta} (t',\mu t',\tau'_{\mu})\rangle.
\end{equation}
This expression differs from works such as~\cite{Bertini2019} in two respects: it uses the Schrodinger picture rather than the Heisenberg picture, and it distinguishes between two times with each layer, labeled by $\tau,\tau'$. 

Let us denote the blue and red gates along the diagonals by $v_+, v_-$ respectively, and the two-site core by $U_{core}$. Then~\autoref{eq:ti-corr} can be decomposed into the following two functions:
\begin{subequations}
\begin{align}\label{eq:corr-2}
C^{\alpha\beta}_\pm \left(0, -\frac{1}{2},\frac{1}{2}\right) &= \frac{1}{q} \tr [\sigma_\alpha (v_\pm \sigma_\beta v_\pm^\dag)] \\
C^{\alpha\beta}_\pm \left(\frac{1}{2}, \frac{1}{2}, -\frac{1}{2}\right) &= \frac{1}{q} \tr [\sigma_\alpha M_\pm(\sigma_\beta; U_{core})]
\end{align}
\end{subequations}
where the $+$ and $-$ quantum channels are~\cite{Bertini2019}:
\begin{subequations}\label{sect:channels}
\begin{align}
M_+(\sigma; U) &= \frac{1}{q} \tr_1 [U (\sigma \otimes \mathbbm{1}) U^\dag] \\
M_-(\sigma; U) &= \frac{1}{q} \tr_2 [U (\mathbbm{1} \otimes \sigma) U^\dag].
\end{align}
\end{subequations}
As is standard in the literature, we can encode the correlations after an integer number of layers of the brickwork circuit with $U = U_{core} (v_+ \otimes v_-)$ in the pair of $q^2 \times q^2$ matrices~\cite{Bertini2019,Aravinda2021,Jonay2021}:
\begin{align}\label{eq:curly-M}
\mathcal{M}^{\alpha\gamma}_\pm[U] &\equiv \frac{1}{q} \tr [\sigma_\alpha M_\pm (\sigma_\gamma; U)] = C^{\alpha\gamma}_\pm \left(\frac{1}{2},-\frac{1}{2},-\frac{1}{2}\right) \notag \\
&= \sum_\beta C^{\alpha\beta}_\pm \left(0, -\frac{1}{2}, \frac{1}{2}\right) C^{\beta\gamma}_\pm \left(\frac{1}{2}, \frac{1}{2}, -\frac{1}{2}\right).
\end{align}
Both channels preserve the identity operator, i.e. $M_\pm(\sigma_0; U) = \sigma_0$. The remaining $2(q^2-1)$ nontrivial eigenvalues $\{\lambda_i\}_{i=1}^{2(q^2-1)}$ of $\mathcal{M}_\pm$ determine whether the associated circuit is (1) non-interacting (all $\{\lambda\}$ are $1$, all correlations are constant), (2) non-ergodic ($n$ unit eigenvalues, where $1 \leq n < 2(
q^2-1)$, so some correlations are constant), (3) ergodic but non-mixing (no unit eigenvalues, but at least one has $|\lambda|=1$, resulting in persistent oscillations but vanishing of time-averaged correlations at large $t$), and (4) ergodic and mixing (all $|\lambda|<1$, so correlations vanish at large $t$ even before time-averaging)~\cite{Bertini2019,Jonay2021}. To this hierarchy Ref.~\cite{Aravinda2021} adds a special case of (4), quantum Bernoulli circuits, for which $\mathcal{M}_\pm$ is the perfectly depolarizing channel (diagonalizable, and all nontrivial eigenvalues are zero). This requires $U$ to be a perfect tensor, which for 2-qudit gates is only possible when $q>2$.

\subsection{Symmetries}
If the circuit is strongly self-dual under a given point group transformation, then the correlation functions must also be invariant in the following sense:
\begin{enumerate}
\item Left/right reflection: $C^{\alpha\beta}_+(t,\tau,\tau') = C^{\alpha\beta}_- (t,\tau,\tau')$
\item Time-reversal: $C^{\alpha\beta}_\pm (t,\tau,\tau') = C^{\tilde{\alpha}\tilde{\beta}}_\mp (-t, -\tau,-\tau')$
\item Symmetry under reflection about:
\begin{enumerate} 
\item $+$ diagonal: $C^{\alpha\beta}_-(t,\tau,\tau') = C^{\tilde{\alpha}\tilde{\beta}}_- (-t, -\tau, -\tau')$
\item $-$ diagonal: $C^{\alpha\beta}_+(t,\tau,\tau') = C^{\tilde{\alpha}\tilde{\beta}}_+ (-t, -\tau, -\tau')$
\end{enumerate}
\item Symmetry under inversion: both (3a) and (3b)
\item Symmetry under $\pi/2$ rotation: both (1) and (2)
\end{enumerate}
where the tilde indicates transposition of the operator basis:
\begin{equation}
\sigma_{\tilde{\alpha}} = \sigma_\alpha^T.
\end{equation}
That is, if the point group transformation changes the sign of time along the given diagonal, then we must also reverse time in the basis of operators, which corresponds to taking the transpose.

Some remarks are in order:

First, not surprisingly, the correlations along the $+$ diagonal do not depend on $v_-$, and vice versa. One consequence of this is that, in the iSWAP-core circuits, the existence of gliders with velocity $+1$ and $-1$ depends only on the one-site gates $v_+$ and $v_-$ respectively.

Second, since we distinguish between $\tau=-1/2$ and $\tau=+1/2$ within each layer, in order to guarantee the equality of certain correlation functions we require the one-site gates and core to be individually invariant. Equality for $t=1/2,\tau=1/2,\tau'=-1/2$ imposes symmetry on the core, while equality for $t=0,\tau=-1/2,\tau'=1/2$ imposes symmetry on the one-site gates. If the circuit is only weakly self-dual, then the correlations are invariant up to a change of basis, and the required change of basis can depend on $\tau$ and $\tau'$.

\subsection{Correlations in iSWAP-core automata}
Specializing to Clifford gates, the two-point correlations of one-site operators can take only three values: $-1$, 0 or $1$. For the iSWAP-core CQCA studied in~\autoref{sect:classes}, the quantum channels corresponding to $U_{core}=\iSWAP$ are manifestly symmetric:
\begin{equation}\label{eq:channel-iswap}
\mathcal{M}_+[\iSWAP] = \begin{pmatrix} \begin{pmatrix} 1 & 0 \\ 0 & 0 \end{pmatrix}
& \bigzero \\
\bigzero & \begin{pmatrix} 0 & 0 \\ 0 & 1 \end{pmatrix}
\end{pmatrix} = \mathcal{M}_-[\iSWAP]
\end{equation}
This is diagonalizable, and the two nontrivial nonzero eigenvalues are the $Z$ gliders discussed in~\autoref{sect:poor}. 

In the main text, we claimed that the good scrambling classes have vanishing correlations for $t\geq 1$. The exact statement is that for these circuits, for any one-site operators $\sigma_\alpha$ and $\sigma_\beta$ at spacetime locations along the diagonal with more than one iSWAP core between them (and hence at least one intervening one-site gate), the correlation between them is zero. A pair of such locations is indicated with blue boxes in~\autoref{fig:corr}. To wit, 
\begin{align}\label{eq:depol}
C^{\alpha\gamma}_\pm (1,1/2,-1/2) &= \mathcal{M}_\pm^{\alpha\beta}[\iSWAP] \mathcal{M}_\pm^{\beta\gamma}[U] \notag \\
&= \begin{pmatrix}
1 & 0 & 0 & 0 \\
0 & 0 & 0 & 0 \\
0 & 0 & 0 & 0 \\
0 & 0 & 0 & 0
\end{pmatrix} = \ket{0}\bra{0}
\end{align}
where $U = \iSWAP (v_+ \otimes v_-)$. This is the completely depolarizing channel. As noted in the main text, it is impossible for our circuits to satisfy the stronger condition $\mathcal{M}_\pm[U] = \ket{0}\bra{0}$, because this would indicate that $U$ is a perfect tensor~\cite{Aravinda2021}. Instead, $\mathcal{M}_\pm[U]$ are nondiagonalizable: for each channel, there is one pair $\alpha\neq \beta$ such that $\mathcal{M}[U]^{\alpha\beta} = \pm 1$. 

As an example, consider the representative circuit of the dense good scrambling class (\autoref{eq:circuit8-gate}). From~\autoref{eq:corr-2} we can read off:
\begin{equation}\label{eq:C-8}
C_\pm\left(0,-\frac{1}{2},\frac{1}{2}\right) = \begin{pmatrix} 
\begin{pmatrix} 1 & 0 \\ 0 & 1 \end{pmatrix} & \bigzero \\
\bigzero & \begin{pmatrix} 0 & 1 \\ -1 & 0 \end{pmatrix}
\end{pmatrix}.
\end{equation}
{The symmetry under left/right reflection is manifest, while time reversal and self-duality are more subtle, since $\sigma_2^T = -\sigma_2$.}

Then, after the iSWAP core,
\begin{equation}
\mathcal{M}_\pm[U] = \begin{pmatrix} 
\begin{pmatrix} 1 & 0 \\ 0 & 0 \end{pmatrix} & \bigzero \\
\bigzero & \begin{pmatrix} 0 & 1 \\ 0 & 0 \end{pmatrix}
\end{pmatrix}
\end{equation}
Note that the only surviving nontrivial correlation after one layer is:
\begin{equation}
\langle \sigma_2 (0, 0, -1/2_{\pm}) \sigma_3 (1/2, 1/2, -1/2_{\pm}) \rangle = 1.
\end{equation}
That is, $Y_1^{(1)} \rightarrow Z_2^{(1)}$ after one full layer, while all other one-site Paulis spread to two sites. $Z_2^{(1)}$ then spreads to two sites in the next layer, hence the vanishing of all nontrivial two-point correlations of one-site operators. 

\section{Free fermion circuits}\label{app:matchgate}
In this appendix, we review matchgate circuits and their mapping to free fermions, then specialize to free fermion translation-invariant Clifford circuits and draw connections to CQCA glider classes, both with and without dual-unitarity.

\subsection{Classical simulation of matchgate circuits}
Matchgate circuits are composed of nearest-neighbor gates of the form~\cite{Terhal2001}:
\begin{align}\label{eq:U-matchgate}
U &= e^{i\phi} \begin{pmatrix}
U_{11}^{(1)} & 0 & 0 & U_{12}^{(1)} \\
0 & U_{11}^{(2)} & U_{12}^{(2)} & 0 \\
0 & U_{21}^{(2)} & U_{21}^{(2)} & 0 \\
U_{21}^{(1)} & 0 & 0 & U_{22}^{(1)}
\end{pmatrix}
\end{align}
where $U^{(1)}, U^{(2)} \in \mathrm{SU}(2)$ and $\phi$ is an arbitrary phase. Loosely speaking, circuits of this form can be classically simulated in polynomial time, a statement that can take on different meanings. In Ref.~\cite{Terhal2001}, it is proven that, given an initial state in the computational basis, the probability distribution of measurement outcomes on any subsystem (one or more qubits) can be efficiently computed. Ref.~\cite{Jozsa2008} proves efficient classical computation of a slightly different quantity: the probability of a measurement outcome on \textit{one} qubit, given \textit{any} initial product state. The circuits in question can be supplemented by special gates on the first two qubits only~\cite{Valiant2001}, classical conditioning on projective measurements in the computational basis~\cite{Terhal2001}, and conjugation by Clifford gates~\cite{Jozsa2008}, while preserving classical simulatability. On the other hand, just by adding a SWAP gate, or by adding arbitrary single-qubit gates, the circuits become universal for quantum computation~\cite{Terhal2001}.

Focusing on nearest-neighbor gates of the form~\autoref{eq:U-matchgate}, efficient classical simulation rests on the ability to express $U$ as evolution under a free fermion Hamiltonian:
\begin{align}
U = e^{i\phi} e^{i (H_1 + H_2 + H_3)}
\end{align}
where
\begin{subequations}
\begin{align}
H_1 &= \alpha_1 ZI + \beta_1 IZ = 2(\alpha_1 c_1^\dag c_1 + c_2^\dag c_2) \\
H_2 &= \alpha_2 XX + \beta_2 YY \notag \\
&= \alpha_2 (c_1^\dag - c_1)(c_2^\dag + c_2) - \beta_2 (c_1^\dag + c_1)(c_2^\dag - c_2) \\
H_3 &= \alpha_3 XY + \beta_3 YX \notag \\
&= -i\alpha_3 (c_1^\dag - c_1)(c_2^\dag - c_2) - i\beta_3 (c_1^\dag + c_1)(c_2^\dag + c_2)
\end{align}
\end{subequations}
with $c_i, c_i^\dag$ the fermionic creation and annihilation operators, obtained via a Jordan-Wigner transformation ~\cite{Terhal2001}.

\subsection{Free fermion (C)QCA}
When nearest-neighbor gates of the form~\autoref{eq:U-matchgate} are arranged on a crystalline lattice, the resulting circuits are less interesting than the "good scramblers" discussed in the main text, because they can be cast in terms of noninteracting fermions. Nevertheless, they can exhibit nontrivial topological phases which can be classified according to Floquet band theory~\cite{Roy2016,Harper2020,Rudner2020}. To our knowledge, it remains an open question which of the 10 Floquet topological classes~\cite{Roy2016}---the Floquet versions of the Altland-Zirnbauer classes for time-independent Hamiltonians~\cite{Altland1997,Kitaev2009}---can be realized in infinite systems with nearest-neighbor gates and a finite Floquet period~\footnote{Ref.~\cite{Jozsa2008} proves that, on a system of $N$ qubits, evolution under \textit{any} free fermion Hamiltonian can be expressed in terms of $O(N^3)$ layers of nearest-neighbor free fermion gates; however, this would necessitate an infinite Floquet period as $N\rightarrow \infty$.}.

While free fermion gates were obtained from a Jordan-Wigner transformation in the previous subsection, we can also \textit{start} from fermions and define a quantum cellular automaton in terms of how it transforms the creation and annihilation operators (or, equivalently, Majorana fermions satisfying $a_i=a_i^\dag$)~\cite{Arrighi2019}. In this context, a quasi-free fermionic QCA is one that acts as a linear transformation on the Majorana operators, i.e. transforming each Majorana fermion into a linear combination of Majoranas~\cite{Zimboras2022}.

Restricting to Clifford quantum cellular automata, spacetime translation-invariant circuits of free fermion gates either are periodic or host gliders. Quasi-free stationary states of the $a=1$ standard glider class are discussed in terms of the Araki-Jordan-Wigner construction---an extension of the Jordan-Wigner transformation to infinite spin chains---in Ref.~\cite{Gutschow2010long}. 

For $a>1$, we have already seen an example of a matchgate circuit: the bare iSWAP circuit. In fact, the bare iSWAP class (which includes circuits dressed by $Z$ rotations on the edges) is the only class of iSWAP-core automata that contains matchgate circuits. 

As we know from the CQCA representation (\autoref{eq:centered-M-iswap}), after two layers, $Z_1^{(n)} \rightarrow Z_1^{(n+1)}$ while $Z_2^{(n)}\rightarrow Z_2^{(n-1)}$. These gliders manifest in the free fermion mapping through the momentum eigenoperators on odd and even sites, $c(k,1)$ and $c(k,2)$, satisfying:
\begin{subequations}
\begin{align}
\mathbbm{U}_F c(k,1) \mathbbm{U}_F^\dag &= -e^{ik} c(k,1) \\
\mathbbm{U}_F c(k,2) \mathbbm{U}_F^\dag &= -e^{-ik} c(k,2)
\end{align}
\end{subequations}
where $\mathbbm{U}_F$ is the Floquet operator corresponding to one time step (two layers) of the brickwork circuit.
The quasienergy spectrum in terms of Dirac fermions therefore has two bands,
with quasienergy $\epsilon(k) = \pi \pm k$, in a Brillouin zone of $[-\pi,\pi]$ and Floquet zone of $[-\pi,\pi]$. This dispersion relation is shown in the left panel~\autoref{fig:dispersion}. Here we use units of $x/a$ in the spatial direction, so the gliders have velocities $\pm 1$.

\begin{figure}[t]
\includegraphics[width=\linewidth]{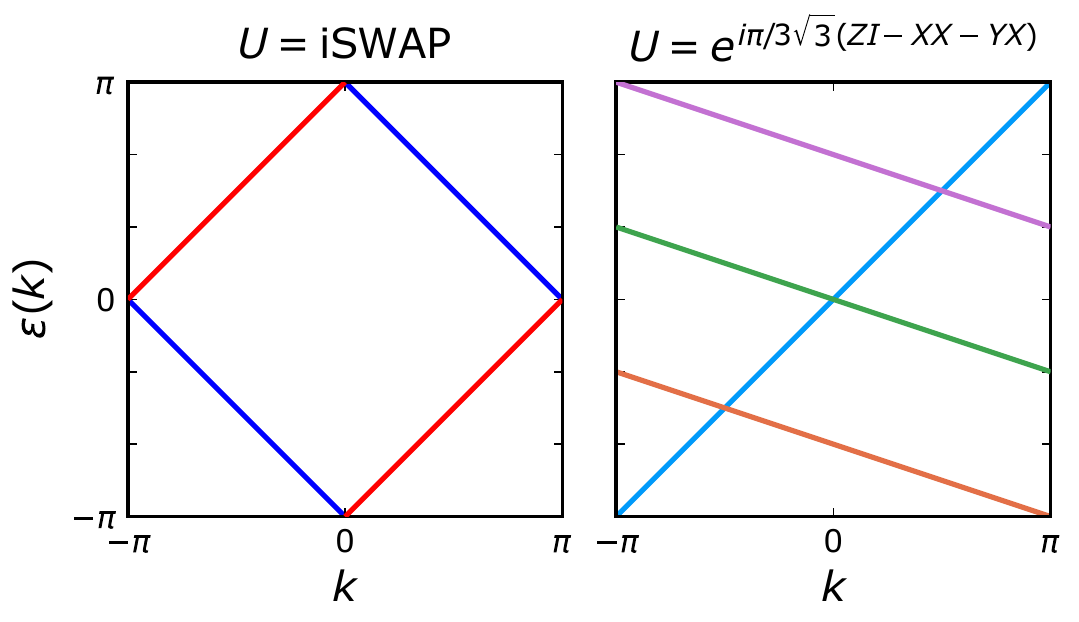}
\caption{Dispersion relation of (left) two momentum bands of Dirac fermions in the bare iSWAP automaton and (right) four momentum bands of Majorana fermions in the CNOT-core automaton with the unitary gate given by~\autoref{eq:cnot-matchgate}.\label{fig:dispersion}}
\end{figure}

This illustrates an important feature of dual-unitary free fermionic QCA: in order for the single-particle dispersion for the dual Floquet operator $\tilde{\mathbbm{U}}_F$ to be well-defined, the Floquet bands must have a nontrivial winding in the quasienergy. This has implications for the locality of the (non-unique) time-independent Hamiltonian that generates $\mathbbm{U}_F$, i.e. $\mathbbm{U}_F = e^{-iH}$. Ref.~\cite{Zimboras2022} proves that if $\mathbbm{U}_F$ is the unitary associated to a quasi-free fermionic QCA, it can be generated by a time-independent Hamiltonian whose interactions decay with distance. However, whereas the interactions decay exponentially if all bands have zero winding, the decay is only a power law in the case of nonzero winding. Indeed, $\mathbbm{U}_F$ for the iSWAP circuit is none other than the so-called Dirac QCA at the massless point~\cite{DAriano2012}, which is generated by a Hamiltonian with $1/r$ interactions~\cite{Zimboras2022,Farrelly2019}. When the mass term is restored, the corresponding quantum circuit is no longer Clifford but remains a free fermion circuit, an interesting direction for future work.

\subsection{CNOT-core free fermion automata}

Lifting the constraint of dual-unitarity, there are three matchgate classes of square-lattice automata with a CNOT core, where members of the same class are related by a Clifford change of basis. Since the CNOT gate is not dual-unitary, the single-particle Floquet bands can have windings other than $\pm 1$ in the quasienergy. Two of these classes have "stationary gliders," and thus one time step can be generated by a exponentially localized Hamiltonian~\cite{Zimboras2022}~\footnote{Another class of circuits, containing the circuit with $U = \mathrm{CNOT}(R_{(1,1,1)}[-2\pi/3] \otimes R_Y[\pi/2])$, has stationary gliders as well, but is not Clifford-equivalent to a matchgate circuit. We leave open the question of whether under a general change of basis, i.e. $U\rightarrow U' = (u_1 \otimes u_2) U (u_2^\dag \otimes u_1^\dag)$, $U'$ can be made to satisfy the matchgate condition.}.

The third class of circuit is more subtle. It is described by the gate:
\begin{align}\label{eq:cnot-matchgate}
U &= (R_Y[\pi/2] \otimes R_X[\pi/2]) \mathrm{CNOT} (R_X[\pi/2] \otimes I) \notag \\
&= e^{-i\pi/4} e^{i \pi/(3 \sqrt{3}) (ZI - XX - YX)}.
\end{align}

Unlike the Dirac QCA to which the iSWAP circuit maps,~\autoref{eq:cnot-matchgate} maps onto a free fermion QCA in which only the fermion parity, and not the fermion number, is conserved. The time evolution is best understood in terms of Majorana fermions. Then, there are four Majorana modes per unit cell, three with velocity $-1/3$ and one with velocity $+1$. Thus, as with the iSWAP circuit, this CNOT-core circuit described by~\autoref{eq:cnot-matchgate} has nontrivial winding (right panel of~\autoref{fig:dispersion}), unique among the CNOT-core matchgate automata.

\begin{figure}[t]
\includegraphics[width=\linewidth]{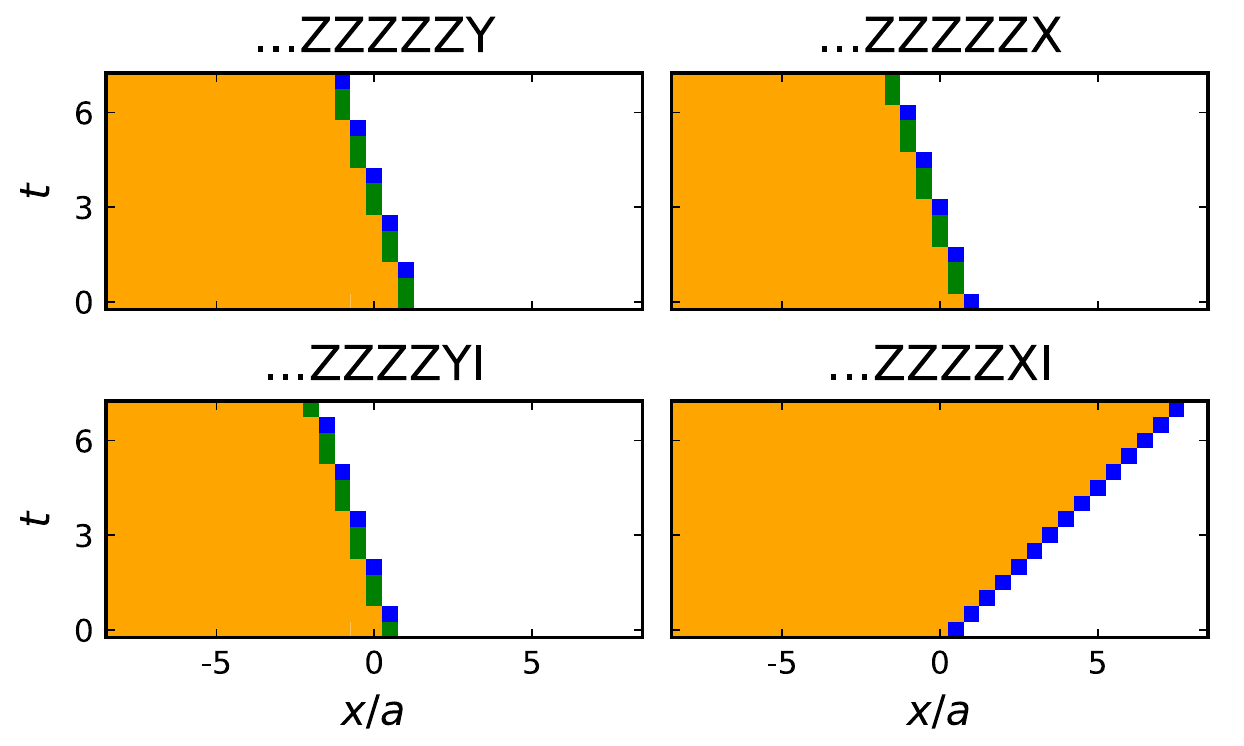}
\caption{Image of odd-parity Pauli strings under the CNOT-core matchgate automaton given by~\autoref{eq:cnot-matchgate}. In the top (bottom) row, the terminating $X$ and $Y$ are on even (odd) sites. The image of each string is shown after each layer, i.e. in steps of $t=1/2$.\label{fig:infinite-gliders}}
\end{figure}

To understand these eigenmodes in terms of the original circuit of qubits, consider the four semi-infinite strings $...ZZZZX$ and $...ZZZZY$ where the terminating $X$ or $Y$ can be on the first or second site of the unit cell. These Pauli strings all have odd fermion parity and square to 1, thus acting as Majorana fermions. As shown in~\autoref{fig:infinite-gliders}, three of these strings glide to the left with speed $1/3$ and a period of 3 layers (returning to the same point in the unit cell, with the same sign, after 6 layers/3 time steps), while the fourth glides to the right with a period of 1 layer (and returns the same point in the unit cell after 1 full time step). In contrast, in the bare iSWAP circuit, the same Pauli strings come in pairs, two with velocity $+1$ and two with velocity $-1$, and (taking the signs on the operators into account, which are not included in the CQCA representation) all acquire a sign of $-1$ after one full time step. {This minus sign is the reason why $\epsilon(k=0)=\pi$ for both Dirac fermion modes in the left panel of~\autoref{fig:dispersion}.}

As proven in Ref.~\cite{Bertini2020soliton}, only dual-unitary circuits can host moving one-site gliders. These "ultralocal solitons" $\sigma$ are preserved under one time step up to a phase and shift by one unit cell: $\mathbbm{U}_F \sigma_x \mathbbm{U}_F^\dag = \sigma_{x\pm a}$. This is perfectly consistent with~\autoref{fig:infinite-gliders}, because while $...ZZZZXI$ does move with velocity $+1$, it relies upon the semi-infinite string of $Z$'s to keep from spreading in the backward direction. As for the $v=-1/3$ gliders, multiplying $....ZZZY$ with $...ZZZX$ (top row of~\autoref{fig:infinite-gliders} does yield a one-site "glider" $Z_2^{(n)}$ of even fermion parity, but this is a soliton only on stroboscopic time scales: after 3 time steps $\mathbbm{U}_F^3 Z_2^{(n)} (\mathbbm{U}_F^\dag)^3 = Z_2^{(n-1)}$, but in the intervening layers it transforms as $Z_2^{(n)}\rightarrow -X_1^{(n)}Y_2^{(n)}\rightarrow -X_1^{(n)}X_1^{(n)}\rightarrow...$.

\section{Poor Scramblers}\label{app:poor}
In~\autoref{sect:poor} we introduced the group of iSWAP-core automata on the square lattice which have glider observables. Here we provide more detail on the three classes in this group, in order of increasing complexity.

\subsection{Bare iSWAP class}
In the main text, we found that the bare iSWAP class is described by the automaton~\autoref{eq:block-iswap} after two steps when written in the basis $(X_1, Z_2, Z_1, X_2)$. In that basis, $\tilde{M}$ is block diagonal, and
an analytic expression for $\tilde{M}^n$ can be proven by induction:
\begin{equation}
    \tilde{M}^n = 
       \begin{pmatrix}
       \begin{pmatrix}
       u^n & 0 \\
      f(n) & u^{-n}
       \end{pmatrix} & \bigzero \\
       \bigzero & \begin{pmatrix}
   u^n &  \overline{f}(n) \\
   0 & u^{-n} \end{pmatrix}
    \end{pmatrix}
\end{equation}
where
\begin{equation}
    f(n) = (u^{n} + 1) \sum_{j=1}^n u^{-j}.
\end{equation}
\comment{
\begin{equation}
    \tilde{M}^n = 
       \begin{pmatrix}
    u^n & 0 & 0 & 0 \\ 0 & u^n &  (u^{-n} + 1) \sum_{j=1}^n u^j & 0 \\
    0 & 0 & u^{-n} & 0 \\ (u^n + 1)\sum_{j=1}^n u^{-j} & 0 & 0 & u^{-n}
    \end{pmatrix}
\end{equation}
}
From this we can read off the time evolution of any initial Pauli string after an integer number of time steps. It is also clear that $\tau(m)=m$ for all $m$. While a typical pure stabilizer state will recur with period $\tau(m)$, this class also has several stationary states: any translation-invariant product stabilizer state with $Z_1$ and/or $Z_2$ as a stabilizer generator is an eigenstate under $\tilde{M}$. This is consistent with the fact, noted above, that the iSWAP gate alone generates no entanglement on a separable state for which one of the two qubits is in a $Z$ eigenstate.

The failure to generate entanglement on $Z$ eigenstates results in the "poor scrambling" behavior described in the main text: starting from a random pure product state, all three poor scrambling classes succeed in generating some entanglement, but do not saturate their Page curves. This is shown in~\autoref{fig:poor-scrambling} for the bare iSWAP class, on a system of $m=63$ unit cells. The entropy reaches a maximum at $t\cong m/4$, and periodically thereafter, but with a slope of $\cong 0.4$. Immediately after reaching a maximum, the entanglement begins to decrease, returning to an area law twice per period. For all $m$ and all three poor scrambling classes, the system returns to area law entanglement every $m/2$ time steps.

\begin{figure}[t]
    \centering
    \subfloat[]{\includegraphics[width=\linewidth]{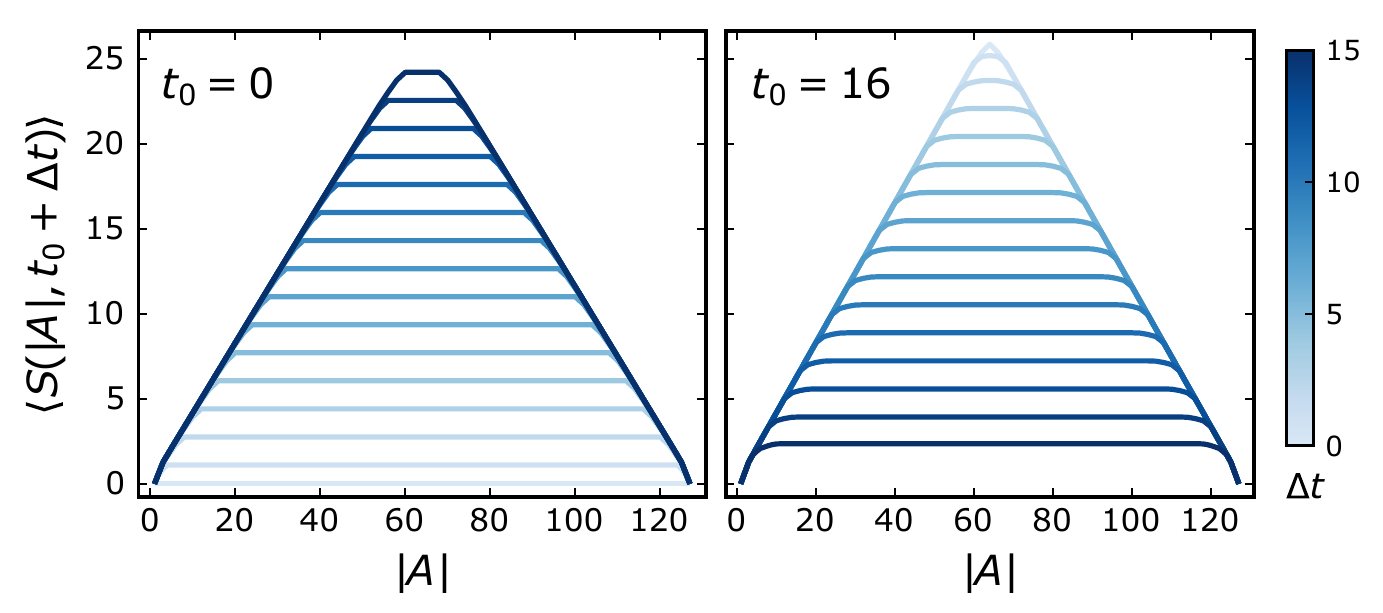}\label{fig:poor-scrambling}} \\
    \subfloat[]{\includegraphics[width=\linewidth]{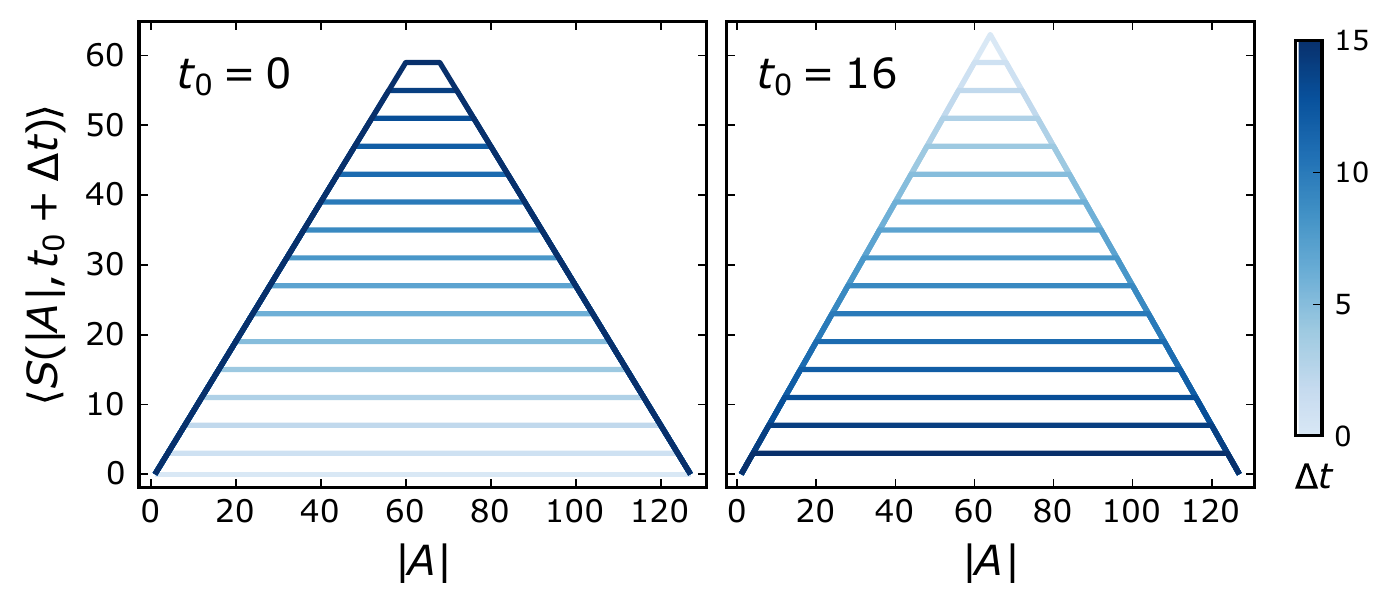}\label{fig:ok-scrambling}}
    \caption{Entanglement generation under the bare iSWAP circuit on $L=126$ qubits, or $m=63$ unit cells, for (a) a random pure product initial state and (b) a pure product state with randomly chosen $X$ and $Y$ stabilizers on each site. In both cases, the subsystem entropy, averaged over all contiguous regions of length $|A|$, increases linearly until $t\cong 16$ (left), then immediately starts to decrease for $t\geq 16$ (right). Darker (lighter) curves correspond to later (earlier) times $\Delta t$ with respect to $t_0$.}
\end{figure}

Since the presence of $Z$ gliders suppresses the entanglement, more entanglement can be produced if we start in a product state of only $X$ and $Y$ gliders. Indeed, in that case a finite system saturates to a slope 1 Page curve (\autoref{fig:ok-scrambling}). However, there is still a recurrence of area law entanglement at $t=m/2$, and since the total entropy can only increase by $\leq 2$ bits per layer, or 4 per time step (\autoref{eq:min-cut}), the earliest it can saturate is at $t=m/4$. Thus, as with the random initial state, the entropy immediately starts to decrease after reaching a maximum. This is another key distinction from the good scramblers (e.g., \autoref{fig:page}), where the slope-1 Page curve survives for $O(m)$ time steps if $m=2^k$, and much longer for generic $m$ ($O(\tau(m)$ steps).

\subsection{Traceless glider class}
The second poor scrambling class has
\begin{equation}\label{eq:M2}
    M = \begin{pmatrix}
    u & 0 & 0 & 0 \\
    u & u & 0 & u \\
    0 & 0 & 0 & 1 \\
    1 & 0 & 1 & 0
    \end{pmatrix}
\end{equation}
corresponding to the pair of single-qubit gates:
\begin{equation}
    (v_+,v_-) = (\mathbbm{1}, R_{(1,1,1)}[-2\pi/3]).
\end{equation}
The 4x4 matrix for this automaton can in fact be inferred from the bare iSWAP and SDKI classes, since the first two column vectors, determined by $v_+$, are the same for all poor scramblers (\autoref{eq:poor}), and the last two column vectors, determined by $v_-$, are the same as in~\autoref{eq:sdki-half}.

Consulting~\autoref{fig:point-group}, the only strict point group symmetry for this automaton is under reflection about the $-$ diagonal, which takes $M_a(v_+,v_-)\rightarrow M_d(v_+^T,v_-)$. In particular, automata in this class evidently lack left-right reflection symmetry, nor can we massage away this asymmetry through a similarity transformation. One consequence of this is that the set of gliders is "chiral": $Z_1$ is a glider with eigenvalue $u$ but $Z_2$ is not.
After two layers,
\begin{equation}\label{eq:tilde-M2}
    \tilde{M} = \ui M^2 = \begin{pmatrix}
    u & 0 & 0 & 0 \\
    1 & u & 1 & u \\
    \ui & 0 & \ui & 0 \\
    1 & 0 & 0 & \ui
    \end{pmatrix}.
\end{equation}
While not immediately obvious from~\autoref{eq:M2}, $\tilde{M}$ has the same characteristic polynomial, and indeed the same minimal polynomial $\mu_{\mathbf{g}}(y)$, as the bare iSWAP class. Thus, the characteristic polynomial is invariant under all point group transformations, even though $M$ itself is only invariant under one: invariance of the characteristic polynomial is a necessary, but not sufficient, condition for the invariance of the corresponding automaton. \gs{Although the bare iSWAP class and this traceless glider class have different symmetries and are not related by a point group transformation, their common minimal polynomial points to the similar structure of their $t\rightarrow\infty$ "spacetime diagrams" ~\cite{Gutschow2010fractal}, as initially local Pauli strings either fill the lightcone or travel along the boundary, with no nontrivial fractal pattern.}  

Permuting rows and columns corresponding to $X_2$ and $Z_1$, as we did for the bare iSWAP class, simplifies the matrix a bit:
\begin{equation}\label{eq:M2-permute}
    \tilde{M}' = \begin{pmatrix}
    \begin{pmatrix} 
    u & 0 \\
    \ui & \ui
    \end{pmatrix} & \bigzero \\
    \begin{pmatrix}
    1 & 1 \\ 1 & 0 \end{pmatrix}
    & \begin{pmatrix}  u & u \\
    0 & \ui
    \end{pmatrix}
    \end{pmatrix}.
\end{equation}
In this basis it is clear that, like in the bare iSWAP class, Pauli strings of only $Z$'s evolve into products of only $Z$'s.

Again using induction, we find:
\begin{equation}
    \tilde{M'}^n = 
\begin{pmatrix}
\begin{pmatrix}
u^n & 0 \\
\ui g(n) & u^{-n} 
\end{pmatrix} & \bigzero \\
\begin{pmatrix}
g(n) & g(n) \\
g(n) & 0
\end{pmatrix}
& \begin{pmatrix}
u^n & u g(n) \\
0 & u^{-n}
\end{pmatrix}
    \end{pmatrix}
\end{equation}
where
\begin{align}
    g(n) &= \sum_{j=0}^{n-1} u^{n-2j-1}.
\end{align}
When $m$ is even, this factors as
\begin{equation}
    g(n) = (u^n + 1)\sum_{j=0}^{n/2-1} u^{-2j-1} = \overline{f}(n).
\end{equation}
Thus $g(m)$ vanishes modulo $u^m+1$ for even $m$, from which we deduce the recurrence time:
\begin{equation}
    \tau(m) = \begin{cases}
    m & m \, \mod 2 = 0 \\
    2m & \mathrm{otherwise}.
    \end{cases}
\end{equation}

\subsection{Poor scramblers with nonzero trace}
The final class of "poor scramblers" is also reflection-asymmetric, with the pair of gates:
\begin{equation}
    (v_+,v_-) = (\mathbbm{1}, R_X[\pi/2])
\end{equation}
corresponding, after two layers, to the automaton:
\begin{equation}
    \tilde{M} = \begin{pmatrix}
    u & 0 & 0 & 0 \\
    1 & u & u & u + 1 \\
    \ui & 0 & 0 & \ui \\
    1 & 0 & \ui & \ui
    \end{pmatrix}.
\end{equation}

Unlike the first two classes, not all products of $Z$'s remain $Z$'s. $Z_1$ is a glider, as anticipated from the general form of~\autoref{eq:poor}, but $Z_2$ evolves into a tensor product of $Z_1$ spreading on odd sites, and a "periodic glider" alternating between $X$, $Y$, and $Z$ on even sites. The characteristic polynomial is
\begin{equation}
    \chi_{\tilde{M}}(y) = (y^2 + u^{2})(y^2 + \ui y + u^{-2}) = \mu_{\tilde{M}}(y).
\end{equation}

Unlike the previous class, the asymmetry under left-right reflection is manifest in the characteristic polynomial, since $\chi_{\tilde{M}}(y) \neq \chi_{\tilde{M}}(y; u\rightarrow \ui)$. $\chi_{\tilde{M}}(y)$ is also asymmetric under time reversal. To wit,
\begin{align}
    \chi_{\tilde{M}^{-1}}(y) &= (y^2 + u^{-2})(y^2 + u y + u^2) \notag \\
    &= \chi_{\tilde{M}_{1\leftrightarrow 2}}(y).
\end{align}
That is, reflections in time have the same effect on the characteristic polynomial as reflections in space. In fact, since $v_-=v_-^T$, they have the same effect on the automaton itself (up to a change in convention for the placement of the single-qubit gates relative to the core). This means that together, time + space reflection---which is just inversion, or rotation by $\pi$---is a symmetry of the automaton. $M$ is also invariant under reflection through either diagonal, which in some works (see, for example, Refs.~\cite{Gopalakrishnan2019,Bertini2019,Aravinda2021}) is used as the definition of the spacetime dual.

We empirically observe that
\begin{equation}
    \tau(m) = \begin{cases}
    m & m \, \mod 6 = 0 \\
    3m/2 & m \, \mod 2 =0, m\, \mod 3 \neq 0 \\
    2m & m \, \mod 2 = 1, m \, \mod 3 = 0 \\
    3m & \mathrm{otherwise}.
    \end{cases}
\end{equation}
Note that in all cases, $\tau(m)$ is divisible by 3. This can be traced to the existence of translation-invariant product states, stabilized by $\langle Z_1^{(n)}, \sigma_2^{(n)} \rangle$ on each unit cell $n$, which cycle through $\sigma=X,Y,Z$ with period 3.

\end{document}